\documentclass[12pt]{article}
\usepackage{cite}

\renewcommand{\theequation}
{\arabic{section}.\arabic{equation}}

\makeatletter
\def\eqnarray{ \stepcounter{equation} \let\@currentlabel=\theequation
 \global\@eqnswtrue
 \global\@eqcnt\z@
 \tabskip\@centering
 \let\\=\@eqncr
 $$\halign to \displaywidth\bgroup\@eqnsel\hskip\@centering
 $\displaystyle\tabskip\z@{##}$&\global\@eqcnt\@ne
 \hfil$\displaystyle{{}##{}}$\hfil
 &\global\@eqcnt\tw@$\displaystyle\tabskip\z@{##}$\hfil
 \tabskip\@centering&\llap{##}\tabskip\z@\cr}
\makeatother

\makeatletter
\def\@arrayacol{\edef\@preamble{\@preamble \hskip .5\arraycolsep}}
\def\array{\let\@acol\@arrayacol \let\@classz\@arrayclassz
\let\@classiv\@arrayclassiv \let\\\@arraycr\def\@halignto{}\@tabarray}
\makeatother

\renewcommand{\arraystretch}{1.6}

\makeatletter                                                              
\newcounter{subeqncnt} 
\def\thesubeqncnt{\alph{subeqncnt}}
\def\subequations{\begingroup%
   \stepcounter{equation}\edef\@tempa{\theequation}%
   \let\c@equation\c@subeqncnt\c@subeqncnt\z@
   \edef\theequation{\@tempa\noexpand\thesubeqncnt}}

\makeatother

\newcommand{\del}{\partial}
\newcommand{\ep}{\varepsilon}

\newcommand{\dd}{{\rm d}}
\newcommand{\UvUa}{${\rm U}(1)_{\rm V} \!\times\! {\rm U}(1)_{\rm A}$}

\begin{document}

\setlength{\baselineskip}{7mm}
\begin{titlepage}
\begin{flushright}
{\tt DIAS-STP-04-18} \\
{\tt TIT/HEP-533} \\
December, 2004
\end{flushright}
 
\vspace{2cm}

\begin{center} 
{\Large Quantization of Neveu-Schwarz-Ramond Superstring Model \\
in 10+2-dimensional Spacetime} 

\vspace{1cm}

{\sc{Takuya Tsukioka}}$^*$
and 
{\sc{Yoshiyuki Watabiki}}$^\dagger$\\
$*${\it{School of Theoretical Physics,}} 
{\it{Dublin Institute for Advanced Studies,}} \\
{\it{10 Burlington Road, Dublin 4, Ireland}} \\
{\sf{tsukioka@synge.stp.dias.ie}} \\
and \\
$\dagger${\it{Department of Physics,}} 
{\it{Tokyo Institute of Technology,}} \\
{\it{Oh-okayama, Meguro, Tokyo 152-8551, Japan}} \\
{\sf{watabiki@th.phys.titech.ac.jp}}
\end{center}

\vspace{1cm}

\begin{abstract}
We construct a Neveu-Schwarz-Ramond superstring model 
which is invariant under supersymmetric {\UvUa} gauge 
transformations as well as the super-general coordinate,  
the super local Lorentz and the super-Weyl transformations 
on the string world-sheet.  
We quantize the superstring model by covariant BRST formulation 
{\it \'a la} Batalin and Vilkovisky and noncovariant light-cone 
gauge formulation. 
Upon the quantizations the model turns out to be formulated 
consistently in 10+2-dimensional background spacetime involving 
two time dimensions.       
\end{abstract}

\end{titlepage}

\section{Introduction}
\setcounter{equation}{0}
\setcounter{footnote}{0}

It is the purpose of this paper to cast some further light upon 
constructions of theories involving two or more time dimensions. 
We propose an explicit Lagrangian description to the end 
from the viewpoint of string theory. 
It might be a clue for understanding the origin of time and 
spacetime itself to consider the physics in which two or more time 
coordinates are introduced.  

From the point of view of the string unification, 
the relations between string theories in various 
dimensions have been studied and  
it was also conjectured that all of these string theories 
were regarded as different phases of an underlying theory in 
higher-dimensional spacetime. 
Meanwhile, the idea of extra time dimensions, 
which might be hidden dimensions, was suggested and studied.  
In this context, 
several unitary theories formulated 
in spacetime with extra time coordinates were 
investigated from various 
viewpoints~\cite{bd, ov, v, t, b, bdk, br, rz, ns, hpnn, rssu, dg, k},
such as super $p$-brane scanning~\cite{bd}, 
$N=2$ heterotic string theories~\cite{ov},  
the perspective for F-theory~\cite{v,t}, 
two-time physics~\cite{b, bdk, br, rz},  
12-dimensional super Yang-Mills and supergravity 
theories~\cite{ns}, super (2,2)-brane~\cite{hpnn} and 
superalgebraic analysis~\cite{rssu}. 

The study in this paper is focused on a model which is constructed 
in spacetime involving two time dimensions, 
although our idea for introducing extra time dimensions 
might be applied to formulate other theories involving more than 
two time dimensions. 
In particular, we would like to investigate a 
superstring model which is consistently formulated in 
10+2-dimensional background spacetime. 
Our approach might make some connections to other 
models~\cite{bd, ov, v, t, b, bdk, br, rz, ns, hpnn, rssu, dg, k}  
from more fundamental and unified point of view.  

Some years ago, one of the authors (Y.W.) had proposed a model
which has a {\UvUa} gauge symmetry in two-dimensional
spacetime and also applied the idea to 
string theories~\cite{wata1, wata2}.  
The striking feature of these models is that 
extra negative norm states appear besides 
usual ones 
and these are removed by the quantization procedure
as the same as string theories. 
This fact suggests two time coordinates might be introduced 
in the background spacetime. 
For the {\UvUa} bosonic string model, we explicitly carried out 
the quantization by covariant BRST and noncovariant light-cone 
gauge formulations and showed the critical dimension was  
26+2 including two time dimensions~\cite{tw1}.         

This paper is a continuation of our work~\cite{tw1}.  
We wish to introduce supersymmetry into our previous 
{\UvUa} bosonic string model. 
The extensions are considered by two ways {\it i.e.}\ 
by introducing the supersymmetry on the string world-sheet 
(Neveu-Schwarz-Ramond model) and on the background spacetime 
(Green-Schwarz model).  
In this paper we focus our attentions on the {\UvUa} NSR superstring 
model. 
We propose an explicit Lagrangian description of the supersymmetric 
model by using the superspace formulation~\cite{howe} 
and study the quantization. 
A subject for the Green-Schwarz superstring model based 
on our framework will be discussed in an additional work 
elsewhere~\cite{tw2}. 

The {\UvUa} superstring model is constructed as gauge field theory
on two-dimensional world-sheet. 
Although the similar models were investigated in refs.~\cite{bdk,dg},
an advantage of the formulation of our model is 
its manifest covariant expression in the background spacetime
by using the {\UvUa} gauge symmetry, 
so that we can easily carry out the quantization
with preserving the covariance. 
An obtained gauge-fixed action might be useful for the 
perturbation theory.    
That is the {\UvUa} gauge symmetry is essential in our model.
In the formulation, 
the generalized Chern-Simons action~\cite{kawata} 
proposed by Kawamoto and one of the authors (Y.W.) 
as a new type of topological action
plays an important key role. 
In fact, the generalized Chern-Simons action is introduced for the 
action to be covariant. 

As we mentioned in the previous paper~\cite{tw1}, 
there are two remarks for the quantization. 
These are also inherited to our superstring model.   
Firstly the action has a reducible symmetry 
which originally arises from gauge structures 
of the generalized Chern-Simons action~\cite{kostu}. 
Secondly the gauge algebra is open. 
In the covariant BRST quantization of the system 
including reducible and open gauge symmetry,
we need to use the formulation developed by
Batalin and Vilkovisky~\cite{bv}.  
By adopting this method  
we explicitly show the covariant quantization is successfully
carried out in the Lagrangian formulation. 

In order to treat the dynamics of our model more directly,
we also quantize the same model in noncovariant light-cone gauges.
The suitable noncovariant gauge conditions can be imposed by
residual symmetries of the supersymmetric {\UvUa} gauge symmetry
and we can then solve all of the gauge constraints explicitly.
We can also confirm that the existence of two time coordinates
is not in conflict with the unitarity of the theory,
since these are required 
by our ``gauge'' symmetry. 

As an important feature of quantum string models,
one can argue the critical dimension of the background 
spacetime~\cite{koh, ggrt, oimnu}. 
In usual superstring theories, the critical dimension is 
9+1~\cite{gswbhp}. 
For our superstring model, the critical dimension turns out to be 10+2.
We obtain this result directly from both the BRST and the noncovariant
light-cone gauge formulations. 

This paper is organized as follows: 
The brief review of the {\UvUa} bosonic string model is 
provided in Section 2.
Then, we introduce the {\UvUa} superstring model involving $N=1$ 
supersymmetry on the world-sheet in Section 3. 
In this section, symmetries and semiclassical aspects of the {\UvUa} 
superstring model are explained.  
The covariant quantization for the model based 
on the Lagrangian formulation is presented in Section 4. 
In this section we investigate perturbative aspects of the 
quantized model and determine the critical dimension of 
our {\UvUa} superstring model. 
In Section 5, the quantization under noncovariant light-cone gauge 
fixing conditions is carried out. 
We then study the symmetry of the background spacetime and 
obtain the same critical dimension by direct computation 
of the full quantum Poincar\'e algebra. 
We also present a mass-shell relation of the model and 
discuss low energy quantum states. 
Conclusions and discussions are given in the final section. 
Appendixes A, B and C contain our notational conventions. 

\section{{\UvUa} bosonic string model}
\setcounter{equation}{0}
\setcounter{footnote}{0}

The {\UvUa} bosonic string model~\cite{wata1, wata2, tw1} 
described by two-dimensional field theory 
consists of scalar fields $\xi^I(x)$, $\phi^I(x)$ and $\bar{\phi}^I(x)$, 
gauge fields $A_m(x)$, $B_m^I(x)$ and $\tilde{C}(x)$ and 
the metric $g_{mn}(x)$. 
We shall consider closed string theories throughout this paper.  
The $D$ scalar fields $\xi^I(x)$ are considered to be 
string coordinates in $D$-dimensional flat background spacetime 
with the metric: 
\begin{equation}
\label{FlatmetricDefinitionGravity}
\eta_{IJ}  \ = \  \eta^{IJ}  \ = \ 
   \left\{\begin{array}{cl}
       -1   &  \hspace{36pt}\hbox{($I=J=0$)}           \\
        1   &  \hspace{36pt}\hbox{($I=J=i$, \ $i=1$, $2$, \ldots, $D-3$)} \\
       -1   &  \hspace{36pt}\hbox{($I=J=\widehat{0}$)} \\
        1   &  \hspace{36pt}\hbox{($I=J=\widehat{1}$)} \\
        0   &  \hspace{36pt}\hbox{(otherwise)}
   \end{array}\right.
\end{equation}
The indices $I$ and $J$ run through 
$0$, $1$, $2$, \ldots, $D-3$, $\widehat{0}$, $\widehat{1}$. 
As we will explain, 
the unitarity of the theory requires two negative signatures to 
the background metric $\eta_{IJ}$ (\ref{FlatmetricDefinitionGravity}),  
because the ${\rm U}(1)_{\rm A}$ gauge transformation 
as well as the general coordinate transformations
removes a negative norm state. 

The covariant action of the bosonic {\UvUa} string model~\cite{tw1} is  
\begin{eqnarray}
S=\int\!\dd^2x \, \sqrt{-g} \, 
\bigg\{
&-& 
\frac{1}{2} g^{mn} \partial_m \xi^I \partial_n \xi_I 
-g^{mn} \partial_m \bar{\phi}^I \partial_n \phi_I \nonumber \\ 
&+& 
\tilde{A}^m \phi_I \partial_m \xi^I 
+\tilde{B}^{mI} \partial_m \phi_I
-\frac{1}{2} \tilde{C} \phi^I \phi_I
\,\bigg\}, 
\label{classical_action_01}
\end{eqnarray}
where we denote 
$$
\sqrt{-g}\tilde{A}^m = \ep^{mn}A_n, 
\quad
\sqrt{-g}\tilde{B}^{mI}
= \ep^{mn}B^I_n, 
\quad   
\sqrt{-g}\tilde{C} 
= \frac{1}{2}\ep^{mn}C_{mn}, 
$$
and $g(x)=\det g_{mn}(x)$.  
The last two terms in (\ref{classical_action_01}) arise from   
the generalized Chern-Simons action~\cite{kawata} formulated 
in two-dimensional spacetime.  
The action (\ref{classical_action_01}) is invariant 
under the following gauge transformations including 
{\UvUa} gauge transformations~\cite{tw1}, 
\begin{eqnarray}
\delta\xi^I 
&=& v'\phi^I, 
\nonumber 
\\
\delta\tilde{A}^m 
&=& 
\frac{\ep^{mn}}{\sqrt{-g}}\del_nv+g^{mn} \del_n v', 
\nonumber 
\\ 
\delta\phi^I
&=&
0, 
\nonumber 
\\
\delta\bar{\phi}^I 
&=& 
u'^I, 
\label{u1u1}
\\
\delta\tilde{B}^{mI} 
&=& 
-v\frac{\ep^{mn}}{\sqrt{-g}}\del_n\xi^I 
+v' g^{mn} \del_n\xi^I 
+\frac{\ep^{mn}}{\sqrt{-g}}\del_nu^I 
+g^{mn}\del_nu'^I -\tilde{w}^m\phi^I, 
\nonumber 
\\
\delta\tilde{C} 
&=& 
\del_m v' \tilde{A}^m - v' \nabla_m \tilde{A}^m 
+\nabla_m\tilde{w}^m, 
\nonumber 
\\ 
\delta g_{mn} 
&=& 
0. 
\nonumber 
\end{eqnarray}
The parameters $\Big(v(x), u^I(x)\Big)$ and 
$\Big(v'(x), u'^I(x)\Big)$ correspond 
to the vector ${\rm U}(1)$ transformations ``${\rm U}(1)_{\rm V}$'' and the 
axial vector ${\rm U}(1)$ transformations ``${\rm U}(1)_{\rm A}$'', 
respectively. 
Although the scalar field $\bar{\phi}^I(x)$ 
might be gauged away by using the gauge degree of freedom for $u'^I(x)$, 
we leave this gauge degree of freedom in order to keep  
the {\UvUa} gauge structure.    
The gauge transformations corresponding to the gauge parameters 
$u^I(x)$ and $\tilde{w}^m(x)=\ep^{mn}w_n(x)/\sqrt{-g(x)}$  
originally come from the generalized Chern-Simons theory~\cite{kawata}. 
The action (\ref{classical_action_01}) is also invariant under 
the general coordinate and the Weyl transformations   
\begin{eqnarray}
\delta\xi^I 
&=& 
k^n \del_n \xi^I,  
\nonumber 
\\
\delta\tilde{A}^m 
&=& 
k^n \del_n \tilde{A}^m 
-\del_n k^m \tilde{A}^n 
+2s\tilde{A}^m,  
\nonumber 
\\
\delta\phi^I 
&=& 
k^n \del_n \phi^I, 
\nonumber 
\\
\delta\bar{\phi}^I 
&=& 
k^n\del_n\bar{\phi}^I, 
\label{general_coordinate_Weyl_trans} 
\\
\delta\tilde{B}^{mI} 
&=& 
k^n\del_n\tilde{B}^{mI}
-\del_n k^m \tilde{B}^{nI} 
+2s\tilde{B}^{mI}, 
\nonumber 
\\
\delta\tilde{C} 
&=& 
k^n\del_n\tilde{C}+2s\tilde{C}, 
\nonumber 
\\
\delta g_{mn} 
&=& 
k^l\del_l g_{mn} 
+\del_m k^l g_{ln} 
+\del_n k^l g_{ml}
-2sg_{mn}, 
\nonumber
\end{eqnarray}
where $k^n(x)$ is a parameter for the general coordinate 
transformation and $s(x)$ is a scaling parameter for the 
Weyl transformation. 
The transformations (\ref{u1u1}) and 
(\ref{general_coordinate_Weyl_trans}) are all local.

It is worth to mention about some algebraic structures of the symmetry.
The first is a reducibility of the symmetry.  
The system is on-shell reducible
because the gauge transformations (\ref{u1u1})
have on-shell invariance under the following transformations of
the gauge parameters with a reducible parameter $w'(x)$,
\begin{equation}
\begin{array}{rcl}
\delta' u^I
&=& w' \phi^I, 
\\
\delta' \tilde{w}^m 
&=& \displaystyle
\frac{\ep^{mn}}{\sqrt{-g}} \del_n w'.  
\end{array}
\label{reducibility_condition_00}
\end{equation}
Since the transformations (\ref{reducibility_condition_00}) are not
reducible anymore, 
the action (\ref{classical_action_01}) is called a first-stage
reducible system.
The second is that the gauge algebra is open.
This means that the gauge algebra closes only when
the equations of motion are satisfied.
Actually, a direct calculation of the commutator of
two gauge transformations on $\tilde{B}^{mI}(x)$ leads to
\begin{eqnarray*}
[\delta_1, \delta_2]\tilde{B}^{mI} 
&=& \cdots -(v'_1v_2-v'_2v_1)\frac{\ep^{mn}}{\sqrt{-g}}\del_n\phi^I,
\end{eqnarray*}
where the dots $(\cdots)$ contain terms of the
usual ``structure constants'' of the gauge algebra. 

In addition to the gauge symmetries 
(\ref{u1u1}) and (\ref{general_coordinate_Weyl_trans}), 
the action (\ref{classical_action_01}) is invariant 
under the following global transformations,  
\begin{eqnarray}
\delta \xi^I  
&=& \omega^I{}_J \xi^J + a^I, 
\nonumber\\
\delta \tilde{A}^m  
&=& r \tilde{A}^m + \sum_{i=1}^{2g} \alpha_i h^{(i)m}, 
\nonumber \\
\delta \phi^I 
&=& - r \phi^I + \omega^I{}_J \phi^J, 
\nonumber \\
\delta \bar{\phi}^I 
&=& r \bar{\phi}^I + \omega^I{}_J \bar{\phi}^J, 
\label{global_trans} \\
\delta \tilde{B}^{mI} 
&=& r \tilde{B}^{mI} + \omega^I{}_J \tilde{B}^{mJ}  
    + \sum_{i=1}^{2g} ( \beta_i^I + \alpha_i \xi^I) h^{(i)m}, 
\nonumber \\
\delta \tilde{C} 
&=& 2 r \tilde{C}, 
\nonumber \\
\delta g_{mn} 
&=& 0, 
\nonumber 
\end{eqnarray}
where the parameters $\omega_{IJ} = -\omega_{JI}$, 
$a^I$ and $r$ are global parameters for the $D$-dimensional Lorentz
transformation, the translation and the scale transformation, 
respectively.  
The functions $h^{(i)m}(x)$ 
are harmonic functions which satisfy 
$\nabla_m h^{(i)m}(x) = \ep^{mn}\nabla_mh^{(i)}_n(x) = 0$ 
($i$ $=$ $1$, $2$, $\dots$, $2g$; $g=$ genus of two-dimensional 
spacetime) and $\alpha_i$ and $\beta_i^I$ are global parameters. 

As we have shown in ref. \cite{tw1}, 
the critical dimension of our bosonic string model 
(\ref{classical_action_01}) is
\begin{equation}
D=28. 
\end{equation}
This means the quantum {\UvUa} bosonic string theory is consistently 
formulated in 26$+$2-dimensional spacetime involving two 
time coordinates. 
The observation was directly obtained from both BRST quantization 
based on Batalin-Fradkin-Vilkovisky formulation and non-covariant 
light-cone quantization~\cite{tw1}.    
It would be interesting to extend the model by introducing
world-sheet supersymmetry as we will explain in this paper.  

\section{{\UvUa} NSR superstring model}
\setcounter{equation}{0}
\setcounter{footnote}{0}

In this section we construct a {\UvUa} string model which 
holds $N=1$ supersymmetry on the world-sheet {\it i.e.}\  
Neveu-Schwarz-Ramond (NSR) type superstring model.  
In order to formulate supersymmetric theory, 
we use the $(1,1)$ type superspace with coordinates 
$z^M=(x^m, \ \theta^\mu)$, \ $(m=0, 1; \ \mu=1, 2)$ where 
$\theta^\mu$ are fermionic coordinates~\cite{howe}. 
The geometry of the superspace is given in Appendix B. 
Field variables of two-dimensional supergravity are 
a vielbein $E_M{}^A(z)$ and a connection $\Omega_M(z)$. 
In particular, we impose kinematic constraints on 
torsion components which are also explained in Appendix B. 

We begin by introducing superfields 
$\Xi^I(z)$, $\Phi^I(z)$, $\bar{\Phi}^I(z)$, 
$\tilde{\Psi}^\alpha(z)\equiv(\bar{\sigma}\Psi(z))^\alpha$, 
${\tilde{\Pi}}^{\alpha I}(z)\equiv(\bar{\sigma}\Pi^I(z))^\alpha$ and 
$\tilde{\Lambda}(z)\equiv
-\frac{1}{2}(\bar{\sigma})^{\alpha\beta}\Lambda_{\alpha\beta}(z)$, 
instead of the fields $\xi^I(x)$, $\phi^I(x)$, $\bar{\phi}^I(x)$, 
$\tilde{A}^m(x)$, $\tilde{B}^{mI}(x)$ and $\tilde{C}(x)$, respectively. 
The superfields $\Xi^I(z)$, $\Phi^I(z)$, $\bar{\Phi}^I(z)$ and 
$\tilde{\Lambda}(z)$ are bosonic scalar superfields, 
while $\tilde{\Psi}^\alpha(z)$ and
$\tilde{\Pi}^{\alpha I}(z)$ are fermionic spinor superfields on the
world-sheet. 
A covariant action for the {\UvUa} NSR superstring model 
is then given by the similar form to the 
bosonic string action (\ref{classical_action_01}),  
\begin{eqnarray}
S=
\int\dd^2x\dd^2\theta 
E\bigg\{
&&
-\frac{1}{2}{\cal D}^\alpha\Xi^I{\cal D}_\alpha\Xi_I 
      -{\cal D}^\alpha\bar{\Phi}^I{\cal D}_\alpha\Phi_I  
\nonumber 
\\
&&
+\tilde{\Psi}^\alpha\Phi_I{\cal D}_\alpha\Xi^I
      +\tilde{\Pi}^{\alpha I}{\cal D}_\alpha\Phi_I 
      -\frac{1}{2}\tilde{\Lambda}\Phi^I\Phi_I
 \bigg\}.
\label{classical_super_action_01}
\end{eqnarray}
The last two terms in (\ref{classical_super_action_01}) are 
constructed from the supersymmetric generalized Chern-Simons 
action~\cite{wata2}. 

The action (\ref{classical_super_action_01}) is invariant under the 
following local supersymmetric {\UvUa} transformations,   
\begin{eqnarray}
\delta\Xi^I 
&=& V'\Phi^I, 
\nonumber 
\\
\delta\tilde{\Psi}_\alpha
&=& (\bar{\sigma}{\cal D})_\alpha V
   +{\cal D}_\alpha V', 
\nonumber 
\\
\delta\Phi^I
&=&
0, 
\nonumber 
\\
\delta\bar{\Phi}^I
&=& U'^I,  
\label{u1u1_super_trans} 
\\
\delta\tilde{\Pi}_\alpha^I 
&=&-V(\bar{\sigma}{\cal D})_\alpha\Xi^I 
   +V'{\cal D}_\alpha\Xi^I
   +(\bar{\sigma}{\cal D})_\alpha U^I  
   +{\cal D}_\alpha U'^I 
   -\tilde{W}_\alpha\Phi^I, 
\nonumber 
\\
\delta\tilde{\Lambda}
&=& {\cal D}^\alpha V'\tilde{\Psi}_\alpha 
   -V'{\cal D}^\alpha\tilde{\Psi}_\alpha
   +{\cal D}^\alpha\tilde{W}_\alpha, 
\nonumber 
\\
\delta E_M{}^A
&=&
\delta \Omega_M=0, 
\nonumber 
\end{eqnarray}
where the gauge parameters $V(z)$, $V'(z)$, $U^I(z)$ and 
$U'^I(z)$ are bosonic scalar superfields and $\tilde{W}_\alpha(z)$ is 
a fermionic spinor superfield.    
The parameters $\Big(V(z), U^I(z)\Big)$ and 
$\Big(V'(z), U'^I(z)\Big)$ 
correspond to the supersymmetric versions of the vector 
U(1) transformations ``${\mbox{U}}(1)_{\rm V}$'' and the axial vector 
U(1) transformations ``${\mbox{U}}(1)_{\rm A}$'', respectively. 
Again, we leave the gauge degree of freedom for the parameter 
$U'^I(z)$ as well as we did in the previous section. 
If this gauge degree of freedom is gauged away, 
the model turns to be the same one which we have discussed 
in the previous work~\cite{wata2}.  
The parameters $U^I(z)$ and 
$\tilde{W}_\alpha(z)\equiv(\bar{\sigma}W(z))_\alpha$ are related with 
the symmetries of the supersymmetric generalized Chern-Simons 
action~\cite{wata2}.   
   
The action (\ref{classical_super_action_01}) is also invariant 
under the super-general coordinate, the super local Lorentz 
and the super-Weyl scaling transformations, 
\begin{eqnarray} 
\delta\Xi^I 
&=& 
K^N\del_N\Xi^I, 
\nonumber 
\\
\delta\tilde{\Psi}_\alpha 
&=& 
K^N\del_N\tilde{\Psi}_\alpha 
-\frac{1}{2}L(\bar{\sigma})_\alpha{}^\beta\tilde{\Psi}_\beta 
+\frac{1}{2}S\tilde{\Psi}_\alpha, 
\nonumber 
\\
\delta\Phi^I 
&=& 
K^N\del_N\Phi^I, 
\nonumber 
\\
\delta\bar{\Phi}^I 
&=& 
K^N\del_N\bar{\Phi}^I, 
\nonumber  
\\
\delta\tilde{\Pi}_\alpha^I 
&=& 
K^N\del_N\tilde{\Pi}_\alpha^I 
-\frac{1}{2}L(\bar{\sigma})_\alpha{}^\beta\tilde{\Pi}_\beta^I 
+\frac{1}{2}S\tilde{\Pi}_\alpha^I, 
\label{supergeneral_lorentz_weyl_trans}
\\
\delta\tilde{\Lambda} 
&=& 
K^N\del_N\tilde{\Lambda} 
+S\tilde{\Lambda}, 
\nonumber 
\\
\delta E_M{}^a
&=& 
K^N\del_NE_M{}^a 
+\del_MK^NE_N{}^a
+E_M{}^bL\ep_b{}^a 
-SE_M{}^a, 
\nonumber 
\\
\delta E_M{}^\alpha 
&=& 
K^N\del_NE_M{}^\alpha 
+\del_MK^NE_N{}^\alpha 
+\frac{1}{2}E_M{}^\beta L(\bar{\sigma})_\beta{}^\alpha 
-\frac{1}{2}SE_M{}^\alpha 
+\frac{i}{2}E_M{}^a(\sigma_a)^{\alpha\beta}{\cal D}_\beta S, 
\nonumber 
\\
\delta\Omega_M 
&=& 
K^N\del_N\Omega_M 
+\del_MK^N\Omega_N 
+\del_ML 
+E_M{}^a\ep_a{}^b{\cal D}_b S 
+E_M{}^\alpha(\bar{\sigma})_\alpha{}^\beta{\cal D}_\beta S,
\nonumber 
\end{eqnarray}
where the superfields $K^N(z)$, $L(z)$ and $S(z)$ are gauge  
parameters for the super-general coordinate, the super local Lorentz and 
the super-Weyl scaling transformations, respectively. 

Some algebraic structures of the gauge symmetry 
which we mentioned in the bosonic string model   
are also inherited to the superstring model. 
The gauge transformations (\ref{u1u1_super_trans})
have on-shell invariance under the following transformations of 
the gauge parameters with a reducible scalar superfield 
parameter $W'(z)$, 
\begin{equation}
\begin{array}{rcl}
\delta' U^I
&=& W'\Phi^I, 
\\
\delta' \tilde{W}_\alpha 
&=& (\bar{\sigma})_\alpha{}^\beta{\cal D}_\beta W'.  
\label{reducibility_condition_01} 
\end{array}
\end{equation}
The gauge algebra is also open in the superstring model. 
Actually, a direct calculation of the commutator of 
two gauge transformations on $\tilde{\Pi}^I_\alpha(z)$ leads to 
\begin{eqnarray*}
[\delta_1, \delta_2]\tilde{\Pi}^I_\alpha 
&=& \cdots -(V'_1V_2-V'_2V_1)
   (\bar{\sigma})_\alpha{}^\beta{\cal D}_\beta\Phi^I,  
\end{eqnarray*}
where the dots $(\cdots)$ contain terms of the 
usual ``structure constants'' of the gauge algebra. 
From the points of view of these structures of the gauge symmetry  
we may adopt the Batalin-Vilkovisky formulation~\cite{bv} 
which allows us to deal with reducible and open gauge symmetries 
to obtain covariant gauge-fixed theories. 
The on-shell reducibility is 
the characteristic feature of the gauge symmetry 
for the generalized Chern-Simons action and the quantization of such a system 
has been discussed in the previous works~\cite{kostu}. 

In addition to these gauge symmetries (\ref{u1u1_super_trans}) and 
(\ref{supergeneral_lorentz_weyl_trans}),  
the action (\ref{classical_super_action_01}) is invariant under  
the following global transformations, 
\begin{eqnarray}
\delta\Xi^I 
&=& \omega^I{}_J\Xi^J + a^I, 
\nonumber \\
\delta\tilde{\Psi}_\alpha 
&=& \displaystyle 
    r\tilde{\Psi}_\alpha 
   +\sum_{i=1}^{4g}\alpha_iH_\alpha^{(i)}, 
\nonumber \\
\delta\Phi^I 
&=& -r\Phi^I 
    +\omega^I{}_J\Phi^J, 
\nonumber \\
\delta\bar{\Phi}^I
&=& r\bar{\Phi}^I 
   +\omega^I{}_J\bar{\Phi}^J, 
\\
\delta\tilde{\Pi}_\alpha^I 
&=& \displaystyle 
    r\tilde{\Pi}_\alpha^I 
   +\omega^I{}_J\tilde{\Pi}_\alpha^J 
   +\sum_{i=1}^{4g}(\beta_i^I+\alpha_i\Xi^I)H_\alpha^{(i)}, 
\nonumber \\ 
\delta\tilde{\Lambda} 
&=& 2r\tilde{\Lambda}, 
\nonumber \\
\delta E_M{}^A 
&=& \delta \Omega_M=0, 
\nonumber 
\end{eqnarray}
where $\omega_{IJ}=-\omega_{JI}$, $a^I$, $r$, $\alpha_i$ and 
$\beta_i^I$ are all constant parameters. 
Poincar\'e symmetry is ISO($D-2, 2$) as the same as that of the 
bosonic model in the previous section. 
The functions $H_\alpha^{(i)}(z)$ result in harmonic functions 
on two-dimensional superspace which satisfy
${\cal D}H^{(i)} = {\cal D}\bar{\sigma}H^{(i)} = 0$
($i=1,2,\ldots,4g$; $g=$ genus of two-dimensional spacetime). 

Now we introduce component fields for the superfields. 
In two-dimensional supergravity, we impose Wess-Zumino gauge for 
the vielbein $E_M{}^A(z)$ and the connection $\Omega_M(z)$ 
whose explicit forms are given 
in Appendix B. 
The other superfields are expressed as  
\begin{eqnarray}
\Xi^I 
&=&
     \xi^I 
   +i(\theta\lambda^I) 
   +\frac{i}{2}(\theta\theta)F^I, 
\nonumber 
\\
\tilde{\Psi}_\alpha
&=&
   i\hat{\psi}_\alpha 
   +i\theta_\alpha X' 
   +i(\bar{\sigma}\theta)_\alpha X 
   +i(\sigma^m\theta)_\alpha\tilde{A}_m 
   +(\theta\theta)\psi_\alpha,
\nonumber 
\\
\Phi^I 
&=&
    \phi^I
   +i(\theta\kappa^I) 
   +\frac{i}{2}(\theta\theta)G^I, 
\nonumber 
\\
\bar{\Phi}^I 
&=&
    \bar{\phi}^I 
   +i(\theta\bar{\kappa}^I) 
   +\frac{i}{2}(\theta\theta)\bar{G}^I,
\\
\tilde{\Pi}_\alpha^I 
&=&
    i\hat{\rho}^I_\alpha 
   +i\theta_\alpha Y'^I 
   +i(\bar{\sigma}\theta)_\alpha Y^I 
   +i(\sigma^m\theta)_\alpha\tilde{B}^I_m 
   +(\theta\theta)\rho^I_\alpha, 
\nonumber 
\\
\tilde{\Lambda}
&=&\displaystyle
    -2i\Big(H+i(\theta\pi)+\frac{i}{2}(\theta\theta)\tilde{C}
       \Big). 
\nonumber 
\end{eqnarray}
The {\UvUa} gauge parameters are also expressed as 
\begin{eqnarray}
V 
&=& v+i(\theta\mu)+\frac{i}{2}(\theta\theta)M, 
\nonumber \\ 
V' 
&=& v'+i(\theta\mu')+\frac{i}{2}(\theta\theta)M', 
\nonumber \\
U^I 
&=& u^I+i(\theta\nu^I)+\frac{i}{2}(\theta\theta)N^I, 
\\
U'^I 
&=& u'^I+i(\theta\nu'^I)+\frac{i}{2}(\theta\theta)N'^I, 
\nonumber \\
\tilde{W}_\alpha
&=& i\hat{\tau}_\alpha 
   +i\theta_\alpha f' 
   +i(\bar{\sigma}\theta)_\alpha f 
   +i(\sigma^m\theta)_\alpha\tilde{w}_m 
   +(\theta\theta)\tau_\alpha. 
\nonumber 
\end{eqnarray}

Before presenting the component expression of 
the classical action (\ref{classical_super_action_01}), 
we would like to clarify the gauge structure of physical component 
fields in the {\UvUa} superstring model. 
In terms of the component fields, 
the gauge transformations (\ref{u1u1_super_trans}) are written down as 
\begin{eqnarray}
\delta\xi^I 
&=& v'\phi^I, 
\nonumber \\
\delta\lambda^I_\alpha
&=& v'\kappa^I_\alpha 
   +\mu'_\alpha\phi^I, 
\nonumber \\
\delta F^I 
&=& v'G^I
   -i(\mu'\kappa^I)
   +M'\phi^I, 
\nonumber \\
\delta\hat{\psi}_\alpha
&=&(\bar{\sigma}\mu)_\alpha
   +\mu'_\alpha, 
\nonumber \\
\delta X' 
&=& M', 
\nonumber \\
\delta X 
&=& M, 
\nonumber \\
\delta\tilde{A}_m 
&=& \frac{1}{e}g_{mn}\ep^{nl}\Big(\del_lv-\frac{i}{2}(\mu\chi_l)\Big) 
   +\del_mv'
   -\frac{i}{2}(\mu'\chi_m), 
\nonumber \\
\delta\psi_\alpha
&=&-\frac{1}{4}\Big((\del_mv\bar{\sigma}+\del_mv')
                    \sigma^n\sigma^m\chi_n
               \Big)_\alpha
   +\frac{1}{2}(\bar{\sigma}\sigma^m\hat{\nabla}_m\mu)_\alpha
   +\frac{1}{2}(\sigma^m\hat{\nabla}_m\mu')_\alpha
\nonumber \\
&& +\frac{i}{8}\bigg(\Big((\mu\chi_m)\bar{\sigma}+(\mu'\chi_m)\Big)
                    \sigma^n\sigma^m\chi_n
               \bigg)_\alpha
   -\frac{1}{4}\Big((M\bar{\sigma}+M')\sigma^m\chi_m
               \Big)_\alpha, 
\nonumber \\
\delta\bar{\phi}^I
&=& u'^I, 
\nonumber \\ 
\delta\bar{\kappa}^I_\alpha
&=& \nu'^I_\alpha, 
\nonumber \\
\delta\bar{G}^I
&=& N'^I, 
\nonumber \\
\delta\hat{\rho}^I_\alpha
&=&-\Big((v\bar{\sigma}-v')\lambda^I\Big)_\alpha 
   +(\bar{\sigma}\nu^I)_\alpha
   +\nu'^I_\alpha 
   -\hat{\tau}_\alpha\phi^I,  
\nonumber \\
\delta Y'^I 
&=& v'F^I 
   +\frac{i}{2}\Big((\mu\bar{\sigma}-\mu')\lambda^I\Big)
   +N'^I
   +\frac{i}{2}(\hat{\tau}\kappa^I) 
   -f'\phi^I, 
\nonumber \\
\delta Y^I 
&=&-vF^I 
   +\frac{i}{2}\Big((\mu-\mu'\bar{\sigma})\lambda^I\Big)
   +N^I
   -\frac{i}{2}(\hat{\tau}\bar{\sigma}\kappa^I)
   -f\phi^I, 
\nonumber \\
\delta\tilde{B}^I_m
&=&-\frac{v}{e}g_{mn}\ep^{nl}\Big(\del_l\xi^I-\frac{i}{2}(\chi_l\lambda^I)\Big)
   +v'\Big(\del_m\xi^I-\frac{i}{2}(\chi_m\lambda^I)\Big)
   -\frac{i}{2}\Big((\mu\bar{\sigma}+\mu')\sigma_m\lambda^I\Big)
\nonumber \\
&& +\frac{1}{e}g_{mn}\ep^{nl}
    \Big(\del_lu^I-\frac{i}{2}(\nu^I\chi_l)\Big)
   +\del_mu'^I  
   -\frac{i}{2}(\nu'^I\chi_m)
   -\frac{i}{2}(\hat{\tau}\sigma_m\kappa^I) 
   -\tilde{w}_m\phi^I, 
\label{u1u1_super_trans_comp} \\
\delta\rho^I_\alpha 
&=&-\frac{1}{2}\Big((v\bar{\sigma}-v')
    \sigma^m\hat{\nabla}_m\lambda^I\Big)_\alpha 
   +\frac{1}{2}\Big((M\bar{\sigma}-M')\lambda^I\Big)_\alpha 
\nonumber \\
&& +\frac{1}{4}\Big((v\bar{\sigma}-v')\sigma^n\sigma^m\chi_n
                    +2\sigma^m(\bar{\sigma}\mu+\mu')
               \Big)_\alpha 
               \Big(\del_m\xi^I-\frac{i}{2}(\chi_m\lambda^I)
               \Big)
\nonumber \\
&& +\frac{1}{4}\Big((v\bar{\sigma}-v')\sigma^m\chi_m
                    -2(\bar{\sigma}\mu-\mu')
               \Big)_\alpha F^I
\nonumber \\
&& -\frac{1}{4}\Big((\del_mu^I\bar{\sigma}+\del_mu'^I)
                    \sigma^n\sigma^m\chi_n
               \Big)_\alpha
   +\frac{1}{2}(\bar{\sigma}\sigma^m\hat{\nabla}_m\nu^I)_\alpha
   +\frac{1}{2}(\sigma^m\hat{\nabla}_m\nu'^I)_\alpha 
\nonumber \\
&& -\frac{1}{4}\Big((N^I\bar{\sigma}+N'^I)\sigma^m\chi_m
               \Big)_\alpha 
   +\frac{i}{8}\bigg(\Big((\nu^I\chi_m)\bar{\sigma}+(\nu'^I\chi_m)\Big)
                    \sigma^n\sigma^m\chi_n
               \bigg)_\alpha
\nonumber \\  
&& +\frac{1}{2}\hat{\tau}_\alpha G^I
   -\frac{1}{2}\Big((f\bar{\sigma}+f'+\tilde{w}_m\sigma^m)\kappa^I
               \Big)_\alpha 
   -\tau_\alpha\phi^I, 
\nonumber \\
\delta H 
&=&-v'X' 
   -\frac{i}{2}(\mu'\hat{\psi})
   +f', 
\nonumber \\
\delta\pi_\alpha 
&=& -v'\psi_\alpha
   -\frac{1}{2}v'(\sigma^m\hat{\nabla}_m\hat{\psi})_\alpha
   +\frac{1}{4}v'\Big(\sigma^m(X'+X\bar{\sigma}+\tilde{A}_n\sigma^n)\chi_m
                 \Big)_\alpha
\nonumber \\
&& -\frac{1}{2}\Big((3X'-X\bar{\sigma}-\tilde{A}_m\sigma^m)\mu'
               \Big)_\alpha 
   -\frac{1}{2}\Bigg(\bigg(M'-\Big(\del_mv'-\frac{i}{2}(\mu'\chi_m)
                              \Big)\sigma^m
                     \bigg)\hat{\psi}
               \Bigg)_\alpha
\nonumber \\
&& +\frac{1}{2}(\sigma^m\hat{\nabla}_m\hat{\tau})_\alpha
   -\frac{1}{4}\Big(\sigma^m(f\bar{\sigma}+f'+\tilde{w}_n\sigma^n)\chi_m
               \Big)_\alpha 
   +\tau_\alpha, 
\nonumber \\
\delta\tilde{C} 
&=& \Big(\del_mv'-\frac{i}{2}(\mu'\chi_m)\Big)
    \Big(\tilde{A}^m-\frac{i}{4}(\hat{\psi}\sigma^n\sigma^m\chi_n)
    \Big) 
\nonumber \\
&& -v'\Big(e^{am}\hat{\nabla}_m\tilde{A}_a
           -\frac{i}{4}(\chi_m\sigma^m\chi_n)\tilde{A}^n
           -\frac{i}{4}(\chi_m\sigma^n\sigma^m\hat{\nabla}_n\hat{\psi})
           +\frac{i}{2}(\psi\sigma^m\chi_m)
\nonumber \\
&&\hspace*{10mm}
    -\frac{i}{2e}\ep^{mn}(\hat{\psi}\bar{\sigma}\hat{\nabla}_m\chi_n)
    +\frac{i}{8}(\chi_m\sigma^n\sigma^m\chi_n)X'
      \Big)
\nonumber \\
&& +2i(\mu'\psi) 
   +\frac{i}{2}(\hat{\psi}\sigma^m\hat{\nabla}_m\mu') 
   +\frac{i}{2}(\mu'\sigma^m\hat{\nabla}_m\hat{\psi}) 
   -\frac{i}{4}\Big(\mu'\sigma^m(X'+X\bar{\sigma}+\tilde{A}_n\sigma^n
                                )\chi_m
               \Big)
\nonumber \\
&& -2M'\Big(X'+\frac{i}{8}(\hat{\psi}\sigma^m\chi_m)\Big) 
   -\frac{i}{4}(\chi_m\sigma^n\sigma^m\hat{\nabla}_n\hat{\tau})
   -\frac{i}{2e}\ep^{mn}(\hat{\tau}\bar{\sigma}\hat{\nabla}_m\chi_n) 
\nonumber \\
&& +\frac{i}{8}f'(\chi_m\sigma^n\sigma^m\chi_n)
   +e^{am}\hat{\nabla}_m\tilde{w}_a
   -\frac{i}{4}\tilde{w}^m(\chi_n\sigma^n\chi_m)
   +\frac{i}{2}(\tau\sigma^m\chi_m),  
\nonumber \\
\delta\phi^I
&=&\delta\kappa^I_\alpha=\delta G^I=0.
\nonumber 
\end{eqnarray}
By using the trivial gauge degrees of freedom for the parameters 
$\mu_\alpha(x)$, $M'(x)$, $M(x)$, $\nu'^I_\alpha(x)$, $\nu^I_\alpha(x)$, 
$N'^I(x)$, $N^I(x)$, $f'(x)$ and $\tau_\alpha(x)$ 
in the gauge transformations (\ref{u1u1_super_trans_comp}), 
we can impose gauge fixing conditions\footnote{
As the {\UvUa} gauge symmetries are essential in our string models, 
we would like to keep these symmetries without gauging away 
the field $\bar{\phi}^I(x)$. 
 } 
\begin{equation}
\hat{\psi}_\alpha
=X'
=X
=\bar{\kappa}^I_\alpha
=\hat{\rho}^I_\alpha
=Y'^I
=Y^I
=H
=\pi_\alpha
=0.
\label{gauge_condition_00}
\end{equation}
In order to compensate the Wess-Zumino gauge (\ref{wess-zumino_gauge}) 
in two-dimensional supergravity within the gauge (\ref{gauge_condition_00}), 
the {\UvUa} gauge parameters should be field dependent. 
From the gauge transformations (\ref{u1u1_super_trans_comp}) and 
the corresponding super-general coordinate, super local Lorentz and 
super-Weyl transformations 
(\ref{super_trans_scalar}) and (\ref{super_trans_spinor}) 
for the fields 
$\hat{\psi}_\alpha(x)$, $X'(x)$, $X(x)$, $\bar{\kappa}^I_\alpha(x)$, 
$\hat{\rho}^I_\alpha(x)$, $Y'^I(x)$, $Y^I(x)$, $H(x)$ and $\pi_\alpha(x)$, 
the gauge parameters allowed within 
the Wess-Zumino gauge (\ref{wess-zumino_gauge}) and the gauge 
(\ref{gauge_condition_00}) take the following forms,    
%
\renewcommand{\arraystretch}{1.8}
%
\begin{equation}
\begin{array}{rcl}
\mu_\alpha 
&=& -(\bar{\sigma}\mu')_\alpha 
    -(\bar{\sigma}\sigma^m\zeta)_\alpha\tilde{A}_m, 
\\
M' 
&=& \displaystyle
    -i(\zeta\psi) 
    +\frac{i}{4}(\zeta\sigma^m\sigma^n\chi_m)\tilde{A}_n, 
\\
M 
&=& \displaystyle
    -i(\zeta\bar{\sigma}\psi) 
    -\frac{i}{4}(\zeta\bar{\sigma}\sigma^m\sigma^n\chi_m)\tilde{A}_n, 
\\
\nu'^I_\alpha
&=& \displaystyle
    -(\sigma^m\zeta)_\alpha\del_m\bar{\phi}^I
    -\zeta_\alpha\bar{G}^I, 
\\
\nu^I_\alpha 
&=& \displaystyle
    \Big((v-v'\bar{\sigma})\lambda^I
    \Big)_\alpha
    -(\bar{\sigma}\nu'^I)_\alpha 
    +(\bar{\sigma}\hat{\tau})_\alpha\phi^I 
    -(\bar{\sigma}\sigma^m\zeta)_\alpha\tilde{B}^I_m, 
\\
N'^I 
&=& \displaystyle
    -v'F^I 
    -\frac{i}{2}(\hat{\tau}\kappa^I) 
    +\frac{i}{2}(\zeta\sigma^m\lambda^I)\tilde{A}_m 
    -i(\zeta\rho^I) 
    +\frac{i}{4}(\zeta\sigma^m\sigma^n\chi_m)\tilde{B}^I_n, 
\\
N^I 
&=& \displaystyle
   vF^I 
   +\frac{i}{2}(\hat{\tau}\bar{\sigma}\kappa^I) 
   +f\phi^I
   +\frac{i}{2}(\zeta\sigma^m\bar{\sigma}\lambda^I)\tilde{A}_m 
   -i(\zeta\bar{\sigma}\rho^I)
   -\frac{i}{4}(\zeta\bar{\sigma}\sigma^m\sigma^n\chi_m)\tilde{B}_n^I, 
\nonumber \\
f' 
&=& 0, 
\\
\tau_\alpha 
&=& \displaystyle
    v'\Big(\psi
          -\frac{1}{4}\tilde{A}_m\sigma^n\sigma^m\chi_n
      \Big)_\alpha
    -\frac{1}{2}(\sigma^m\mu')_\alpha\tilde{A}_m 
    -\frac{1}{2}(\sigma^m\hat{\nabla}_m\hat{\tau})_\alpha
\\   
&& \displaystyle
    +\frac{1}{4}\Big(\sigma^m(f\bar{\sigma}+\tilde{w}_n\sigma^n)\chi_m
                \Big)_\alpha
    -\zeta_\alpha\tilde{C}. 
\end{array}
 \label{redefined_gauge_parameter}
\end{equation}
%
\renewcommand{\arraystretch}{1.6}
%

\vspace*{-3mm}
\noindent
The classical action for the {\UvUa} superstring model 
(\ref{classical_super_action_01}) is then expressed by  
\begin{eqnarray}
S&=& 
\int \! \dd^2xe
\bigg\{-\frac{1}{2}g^{mn}\del_m\xi^I\del_n\xi_I 
     -\frac{i}{2}(\lambda^I\sigma^m\del_m\lambda_I) 
\nonumber \\ 
&&\hspace*{15mm}
     +\frac{i}{2}(\lambda_I\sigma^m\sigma^n\chi_m)\del_n\xi^I 
     -\frac{1}{16}(\chi_m\sigma^n\sigma^m\chi_n)(\lambda^I\lambda_I) 
     +\frac{1}{2}F^IF_I 
\nonumber \\
&&\hspace*{15mm} 
     -g^{mn}\del_m\bar{\phi}^I\del_n\phi_I 
     +\bar{G}^IG_I 
\nonumber \\
&&\hspace*{15mm} 
     +\Big(\tilde{A}^m\del_m\xi^I+i(\psi\lambda^I)\Big)\phi_I 
     +\tilde{B}^{mI}\del_m\phi_I 
     +i(\rho^I\kappa_I)
     -\frac{1}{2}\tilde{C}\phi^I\phi_I
\bigg\}, 
\label{classical_super_action_03}
\end{eqnarray}
where we redefine some of the fields as follows,   
%
\renewcommand{\arraystretch}{2.0}
%
\begin{equation}
\begin{array}{rcl}
\displaystyle
\psi_\alpha
-\frac{1}{4}\tilde{A}_m(\sigma^n\sigma^m\chi_n)_\alpha
&\rightarrow& 
\psi_\alpha, 
\\
\displaystyle
\rho^I_\alpha 
+\frac{1}{2}\del_m\bar{\phi}^I(\sigma^n\sigma^m\chi_n)_\alpha
-\frac{1}{2}\tilde{A}_m(\sigma^m\lambda^I)_\alpha 
-\frac{1}{4}\tilde{B}^I_m(\sigma^n\sigma^m\chi_n)_\alpha 
&\rightarrow&
\rho^I_\alpha. 
\end{array}
\label{field_redefinition}
\end{equation}
%
\renewcommand{\arraystretch}{1.6}
%

\vspace*{-3mm}
\noindent
Under the field redefinitions (\ref{field_redefinition}),  
the gauge transformations within the Wess-Zumino 
gauge (\ref{wess-zumino_gauge}) and the gauge (\ref{gauge_condition_00}) 
are given by 
\begin{eqnarray}
\delta\xi^I 
&=& v'\phi^I 
   +k^n\del_n\xi^I 
   +i(\zeta\lambda^I), 
\nonumber \\
\delta\lambda^I_\alpha
&=& v'\kappa^I_\alpha 
   +\mu'_\alpha\phi^I 
   +k^n\del_n\lambda^I_\alpha
   +(\sigma^m\zeta)_\alpha
    \Big(\del_m\xi^I-\frac{i}{2}(\chi_m\lambda^I)
    \Big)
   +\zeta_\alpha F^I 
   -\frac{1}{2}l(\bar{\sigma}\lambda^I)_\alpha 
   +\frac{1}{2}s\lambda^I_\alpha,  
\nonumber \\
\delta F^I 
&=& v'G^I 
   -i(\mu'\kappa^I)
   +k^n\del_nF^I
   -i(\zeta\psi)\phi^I 
   -\frac{i}{2}(\zeta\sigma^m\sigma^n\chi_m)\del_n\xi^I
\nonumber \\ 
&& +i(\zeta\sigma^m\nabla_m\lambda^I)
   +\frac{1}{8}(\zeta\lambda^I)(\chi_m\sigma^n\sigma^m\chi_n)
   -\frac{i}{2}(\zeta\sigma^m\chi_m)F^I 
   +sF^I, 
\nonumber \\
\delta\tilde{A}^m 
&=&\frac{1}{e}\ep^{mn}\del_nv
   +g^{mn}\del_nv' 
   -\frac{i}{2}(\mu'\sigma^n\sigma^m\chi_n)
\nonumber \\
&& +k^n\del_n\tilde{A}^m
   -\del_nk^m\tilde{A}^n
   +i(\zeta\sigma^m\psi)
   -i(\zeta\sigma^n\chi_n)\tilde{A}^m
   +2s\tilde{A}^m, 
\nonumber \\
\delta\psi_\alpha
&=&-\frac{1}{2}\del_mv'(\sigma^n\sigma^m\chi_n)_\alpha
   +(\sigma^m\nabla_m\mu')_\alpha
   -\frac{i}{8}\mu'_\alpha(\chi_m\sigma^n\sigma^m\chi_n)
\nonumber \\
&& +k^n\del_n\psi_\alpha
   +\zeta_\alpha\nabla_m\tilde{A}^m
   -i(\zeta\sigma^m\chi_m)\psi_\alpha 
   -\frac{i}{2}(\zeta\sigma^m\psi)\chi_{m\alpha}
   -\frac{1}{2}l(\bar{\sigma}\psi)_\alpha
   +\frac{3}{2}s\psi_\alpha, 
\nonumber \\
\delta\phi^I 
&=& k^n\del_n\phi^I 
   +i(\zeta\kappa^I), 
\nonumber \\
\delta\kappa^I_\alpha 
&=& k^n\del_n\kappa^I_\alpha 
   +(\sigma^m\zeta)_\alpha
    \Big(\del_m\phi^I-\frac{i}{2}(\chi_m\kappa^I)
    \Big)
   +\zeta_\alpha G^I 
   -\frac{1}{2}l(\bar{\sigma}\kappa^I)_\alpha
   +\frac{1}{2}s\kappa^I_\alpha, 
\nonumber \\
\delta G^I 
&=& k^n\del_nG^I 
   -\frac{i}{2}(\zeta\sigma^m\sigma^n\chi_m)\del_n\phi^I 
\nonumber \\
&& +i(\zeta\sigma^m\nabla_m\kappa^I)
   +\frac{1}{8}(\zeta\kappa^I)(\chi_m\sigma^n\sigma^m\chi_n) 
   -\frac{i}{2}(\zeta\sigma^m\chi_m)G^I 
   +sG^I, 
\nonumber \\
\delta\bar{\phi}^I 
&=& u'^I 
   +k^n\del_n\bar{\phi}^I,  
\label{gauge_trans_03}\\
\delta\bar{G}^I
&=& -v'F^I 
    -\frac{i}{2}(\hat{\tau}\kappa^I) 
    +k^n\del_n\bar{G}^I
    -i(\zeta\rho^I)
   -\frac{i}{2}(\zeta\sigma^m\chi_m)\bar{G}^I 
   +s\bar{G}^I, 
\nonumber \\
\delta\tilde{B}^{mI} 
&=&-\frac{v}{e}\ep^{mn}\del_n\xi^I 
   +v'g^{mn}\Big(\del_n\xi^I-\frac{i}{2}(\chi_l\sigma_n\sigma^l\lambda^I)
      \Big)
   -i(\mu'\sigma^m\lambda^I)
\nonumber \\
&& +\frac{1}{e}\ep^{mn}\del_nu^I
   +g^{mn}\del_nu'^I
   -\frac{i}{2}(\hat{\tau}\sigma^m\kappa^I)
   -\tilde{w}^m\phi^I
   +k^n\del_n\tilde{B}^{mI}
   -\del_nk^m\tilde{B}^{nI}
\nonumber \\
&& +i(\zeta\lambda^I)\tilde{A}^m
   -i(\zeta\sigma^m\sigma^n\sigma^l\chi_n)\del_l\bar{\phi}^I
   +\frac{i}{2}(\zeta\sigma^n\sigma^m\chi_n)\bar{G}^I
\nonumber \\
&& -i(\zeta\sigma^n\chi_n)\tilde{B}^{mI}
   +i(\zeta\sigma^m\rho^I)
   +2s\tilde{B}^{mI}, 
\nonumber \\
\delta\rho^I_\alpha
&=&v'\Big(\sigma^m\nabla_m\lambda^I
   -\frac{1}{2}\del_m\xi^I\sigma^n\sigma^m\chi_n   
   -\frac{i}{8}(\chi_m\sigma^n\sigma^m\chi_n)\lambda^I
   -\phi^I\psi
   -\frac{1}{2}\tilde{A}^m\sigma_m\kappa^I
     \Big)_\alpha
   +\mu'_\alpha F^I 
\nonumber \\
&& +\frac{1}{2}\Big((G^I-\del_m\phi^I\sigma^m)\hat{\tau}\Big)_\alpha
   -\frac{i}{4}(\hat{\tau}\sigma^m\sigma^n\chi_m)(\sigma_n\kappa^I)_\alpha
   -\frac{1}{2}f(\bar{\sigma}\kappa^I)_\alpha 
   -\frac{1}{2}\tilde{w}^m(\sigma_m\kappa^I)_\alpha
\nonumber \\
&& +k^n\del_n\rho^I_\alpha
   -\zeta_\alpha\Big(g^{mn}\nabla_m\del_n\phi^I
                    +\tilde{A}^m\del_m\xi^I
                    +i(\psi\lambda^I)
                    -\nabla_m\tilde{B}^{mI} 
                    -\tilde{C}\phi^I
                \Big) 
\nonumber \\
&& -(\sigma^m\nabla_m\zeta)_\alpha\bar{G}^I
   -(\sigma^m\zeta)_\alpha\del_m\bar{G}^I
   +\frac{i}{8}\zeta_\alpha(\chi_m\sigma^n\sigma^m\chi_n)\bar{G}^I
\nonumber \\
&& -i(\zeta\sigma^m\chi_m)\rho^I_\alpha
   -\frac{i}{2}(\zeta\sigma^m\rho^I)\chi_{m\alpha}
   -\frac{1}{2}l(\bar{\sigma}\rho^I)_\alpha
   +\frac{3}{2}s\rho^I_\alpha, 
\nonumber \\
\delta\tilde{C} 
&=& \del_mv'\tilde{A}^m
   -v'\nabla_m\tilde{A}^m
   +2i(\mu'\psi)
   +\nabla_m\tilde{w}^m
   +k^n\del_n\tilde{C}
   -i(\zeta\sigma^m\chi_m)\tilde{C}
   +2s\tilde{C}, 
\nonumber 
\end{eqnarray}
where we use the covariant derivative $\nabla_m$ 
for the torsion free connection $\omega_m(x)$ defined via  
(\ref{covariant_derivative_lorentz}),
(\ref{covariant_derivative_curved}),
(\ref{covariant_derivative_mix}) and   
(\ref{covariant_derivative_spinor}) and 
we also redefine the gauge parameter $\tilde{w}^m(x)$ as 
\begin{equation}
\tilde{w}^m-\frac{i}{2e}\ep^{mn}(\hat{\tau}\bar{\sigma}\chi_n)
\rightarrow
\tilde{w}^m.
\label{redefinition_wm}
\end{equation}

Let us consider the reducible symmetry. 
By introducing the reducible parameter $W'(z)$ as  
\begin{equation}
W'=w'+i(\theta\check{w}')+\frac{i}{2}(\theta\theta)\bar{w}', 
\end{equation}
the reducible transformations (\ref{reducibility_condition_01}) 
are expressed in terms of the component fields
\begin{equation}
\begin{array}{rcl}
\delta'u^I
&=& w'\phi^I, 
\\
\delta'\hat{\tau}_\alpha
&=& (\bar{\sigma}\check{w}')_\alpha, 
\\
\delta'f
&=& \bar{w}',
\\
\delta'\tilde{w}^m
&=& \displaystyle
    \frac{1}{e}\ep^{mn}\del_nw', 
\end{array}
\label{reducibility_condition_02}
\end{equation}
where we use the redefinition of the gauge 
parameter (\ref{redefinition_wm}). 
One can easily check the reducible 
transformation (\ref{reducibility_condition_01}) is also consistent with 
the Wess-Zumino gauge.  

Now we are going to an on-shell formulation of the model 
by eliminating the ``auxiliary'' fields $F^I(x)$, $G^I(x)$, 
$\bar{G}^I(x)$, $\kappa^I_\alpha(x)$ and $\rho^I_\alpha(x)$.
All of the gauge transformations for these auxiliary fields are 
proportional to the equations of motion, so that we can eliminate 
these fields within the on-shell formulation.    
Then, the action (\ref{classical_super_action_03}) becomes the 
following form,  
\begin{eqnarray}
S&=& 
\int \! \dd^2xe
\bigg\{-\frac{1}{2}g^{mn}\del_m\xi^I\del_n\xi_I 
     -\frac{i}{2}(\lambda^I\sigma^m\del_m\lambda_I) 
\nonumber \\ 
&&\hspace*{15mm}
     +\frac{i}{2}(\lambda_I\sigma^m\sigma^n\chi_m)\del_n\xi^I 
     -\frac{1}{16}(\chi_m\sigma^n\sigma^m\chi_n)(\lambda^I\lambda_I) 
\nonumber \\
&&\hspace*{15mm} 
     -g^{mn}\del_m\bar{\phi}^I\del_n\phi_I 
     +\Big(\tilde{A}^m\del_m\xi^I+i(\psi\lambda^I)\Big)\phi_I 
     +\tilde{B}^{mI}\del_m\phi_I 
     -\frac{1}{2}\tilde{C}\phi^I\phi_I
\bigg\}, \quad
\label{classical_super_action_04}
\end{eqnarray}
and the gauge transformations are given by  
\begin{eqnarray}
\delta\xi^I 
&=& v'\phi^I 
   +k^n\del_n\xi^I 
   +i(\zeta\lambda^I), 
\nonumber \\
\delta\lambda^I_\alpha
&=&\mu'_\alpha\phi^I 
   +k^n\del_n\lambda^I_\alpha
   +(\sigma^m\zeta)_\alpha
    \Big(\del_m\xi^I-\frac{i}{2}(\chi_m\lambda^I)
    \Big)
   -\frac{1}{2}l(\bar{\sigma}\lambda^I)_\alpha 
   +\frac{1}{2}s\lambda^I_\alpha,  
\nonumber \\
\delta\tilde{A}^m 
&=&\frac{1}{e}\ep^{mn}\del_nv
   +g^{mn}\del_nv' 
   -\frac{i}{2}(\mu'\sigma^n\sigma^m\chi_n)
\nonumber \\
&& +k^n\del_n\tilde{A}^m
   -\del_nk^m\tilde{A}^n
   +i(\zeta\sigma^m\psi)
   -i(\zeta\sigma^n\chi_n)\tilde{A}^m
   +2s\tilde{A}^m, 
\nonumber \\
\delta\psi_\alpha
&=&-\frac{1}{2}\del_mv'(\sigma^n\sigma^m\chi_n)_\alpha
   +(\sigma^m\nabla_m\mu')_\alpha
   -\frac{i}{8}\mu'_\alpha(\chi_m\sigma^n\sigma^m\chi_n)
\nonumber \\
&& +k^n\del_n\psi_\alpha
   +\zeta_\alpha\nabla_m\tilde{A}^m
   -i(\zeta\sigma^m\chi_m)\psi_\alpha 
   -\frac{i}{2}(\zeta\sigma^m\psi)\chi_{m\alpha}
   -\frac{1}{2}l(\bar{\sigma}\psi)_\alpha
   +\frac{3}{2}s\psi_\alpha, 
\nonumber \\
\delta\phi^I 
&=& k^n\del_n\phi^I,  
\nonumber \\
\delta\bar{\phi}^I 
&=& u'^I 
   +k^n\del_n\bar{\phi}^I,  
\label{gauge_trans_04} \\
\delta\tilde{B}^{mI} 
&=&-\frac{v}{e}\ep^{mn}\del_n\xi^I 
   +v'g^{mn}\Big(\del_n\xi^I-\frac{i}{2}(\chi_l\sigma_n\sigma^l\lambda^I)
      \Big)
   -i(\mu'\sigma^m\lambda^I)
\nonumber \\
&& +\frac{1}{e}\ep^{mn}\del_nu^I
   +g^{mn}\del_nu'^I
   -\tilde{w}^m\phi^I
   +k^n\del_n\tilde{B}^{mI}
   -\del_nk^m\tilde{B}^{nI}
\nonumber \\
&& +i(\zeta\lambda^I)\tilde{A}^m
   -i(\zeta\sigma^m\sigma^n\sigma^l\chi_n)\del_l\bar{\phi}^I
   -i(\zeta\sigma^n\chi_n)\tilde{B}^{mI}
   +2s\tilde{B}^{mI}, 
\nonumber \\
\delta\tilde{C} 
&=& \del_mv'\tilde{A}^m
   -v'\nabla_m\tilde{A}^m
   +2i(\mu'\psi)
   +\nabla_m\tilde{w}^m
   +k^n\del_n\tilde{C}
   -i(\zeta\sigma^m\chi_m)\tilde{C}
   +2s\tilde{C},   
\nonumber \\
\delta e_m{}^a 
&=& k^n\del_ne_m{}^a
   +\del_mk^ne_n{}^a
   +i(\zeta\sigma^a\chi_m) 
   +le_m{}^b\ep_b{}^a 
   -se_m{}^a, 
\nonumber \\ 
\delta\chi_{m\alpha} 
&=& \displaystyle 
   k^n\del_n\chi_{m\alpha}
   +\del_mk^n\chi_{n\alpha}
   +2(\nabla_m\zeta)_\alpha
   -\frac{i}{2}(\chi_m\bar{\sigma}\sigma^l\chi_l)
    (\bar{\sigma}\zeta)_\alpha 
\nonumber \\
&& -\frac{1}{2}l(\bar{\sigma}\chi_m)_\alpha 
   -\frac{1}{2}s\chi_{m\alpha} 
   -(\sigma_m\check{s})_\alpha. 
\nonumber 
\end{eqnarray}
Since the gauge parameters $\hat{\tau}_\alpha(x)$ and $f(x)$  
disappear from the gauge transformation (\ref{gauge_trans_04}),  
the reducible transformations with which we would like to work are 
\begin{equation}
\begin{array}{rcl}
\delta'u^I
&=& w'\phi^I, 
\\
\delta'\tilde{w}^m
&=& \displaystyle
    \frac{1}{e}\ep^{mn}\del_nw'.  
\end{array}
\label{reducibility_condition_03}
\end{equation}
In addition to the gauge symmetry, 
the action (\ref{classical_super_action_04}) is also invariant 
under the following global transformations,  
\begin{eqnarray}
\delta\xi^I  
&=& 
\omega^I{}_J \xi^J + a^I, 
\nonumber
\\
\delta\lambda^I_\alpha
&=&
\omega^I{}_J \lambda^J_\alpha, 
\nonumber
\\
\delta\tilde{A}^m  
&=& 
r\tilde{A}^m + \sum_{i=1}^{2g} \alpha_i h^{(i)m}, 
\nonumber 
\\
\delta\psi_\alpha
&=&
r\psi_\alpha, 
\nonumber 
\\
\delta \phi^I 
&=& - r \phi^I + \omega^I{}_J \phi^J, 
\label{global_trans_04} 
\\
\delta \bar{\phi}^I 
&=& r \bar{\phi}^I + \omega^I{}_J \bar{\phi}^J, 
\nonumber 
\\
\delta \tilde{B}^{mI} 
&=& r \tilde{B}^{mI} + \omega^I{}_J \tilde{B}^{mJ}  
    + \sum_{i=1}^{2g} ( \beta_i^I + \alpha_i \xi^I) h^{(i)m}, 
\nonumber \\
\delta \tilde{C} 
&=& 2 r \tilde{C}, 
\nonumber \\
\delta e_m{}^a
&=&
\delta\chi_{m\alpha}
=0.  
\nonumber 
\end{eqnarray}
We take the classical action (\ref{classical_super_action_04}), 
the gauge transformation (\ref{gauge_trans_04}) and 
the reducibility condition (\ref{reducibility_condition_03}) as 
the starting point for the quantization 
we will discuss in this paper.  
The superpartners of the fields 
$\phi^I(x)$, $\tilde{B}^{mI}(x)$ and $\tilde{C}(x)$ in the 
generalized Chern-Simons action disappear 
in the action (\ref{classical_super_action_04}), 
since the supersymmetry transformations of these fields are trivial.  

Before getting into the quantization of the model, 
it is worth to mention semiclassical aspects of the action 
(\ref{classical_super_action_04}),   
by eliminating gauge fields through their equations of motion. 
Indeed, this manipulation might be helpful to understand the heart
of the model.
The field $\bar{\phi}^I(x)$ can 
be gauged away by using the gauge degree of freedom for the 
parameter $u'^I(x)$. 
Equations of motion for the fields $\tilde{B}^{mI}(x)$ and 
$\tilde{C}(x)$ give gauge constraints
\begin{equation}
\begin{array}{rcl}
\del_m\phi^I
&=& 0,
\\
\phi^I\phi_I
&=& 0. 
\end{array}
\end{equation}
It is possible to find nontrivial solutions for these constraints
if the background spacetime metric includes two 
time-like signatures. 
In the light-cone notation\footnote{
We use a convention of the light-cone coordinates for the
background spacetime as 
$x^I = (x^\mu, x^{\hat{+}}, x^{\hat{-}})$ where
$x^{\hat{\pm}}=\frac{1}{\sqrt{2}}(x^{\hat{0}}\pm x^{\hat{1}})$ and
the index $\mu$ runs through $0, 1, \dots, D-3$.
}, 
one of the interesting solutions 
which is naturally related with the usual superstring action 
is
$\phi^{\hat{-}}(x)=\phi^\mu(x)=0$ and 
$\phi^{\hat{+}}(x)=\mbox{const.}$. 
After substituting this solution into the action
(\ref{classical_super_action_04}), the action becomes
\begin{eqnarray}
S&=& 
\int \! \dd^2xe
\bigg\{-\frac{1}{2}g^{mn}\del_m\xi^I\del_n\xi_I 
     -\frac{i}{2}(\lambda^I\sigma^m\del_m\lambda_I) 
\nonumber \\ 
&&\hspace*{15mm}
     +\frac{i}{2}(\lambda_I\sigma^m\sigma^n\chi_m)\del_n\xi^I 
     -\frac{1}{16}(\chi_m\sigma^n\sigma^m\chi_n)(\lambda^I\lambda_I) 
\nonumber \\
&&\hspace*{15mm} 
     -\Big(\tilde{A}^m\del_m\xi^{\hat{-}}+i(\psi\lambda^{\hat{-}})
      \Big)\phi^{\hat{+}} 
\bigg\}. 
\label{reduced_super_action_01}
\end{eqnarray}
In the action (\ref{reduced_super_action_01}),  
relations $\del_m \xi^{\hat{-}}(x) = 0$ and 
$\lambda^{\hat{-}}_\alpha(x)=0$ are given
by the equation of motion for $\tilde{A}^m(x)$ and $\psi^\alpha(x)$, 
respectively.
Then, the final form of the action becomes 
the usual NSR superstring action 
\begin{eqnarray}
S&=& 
\int \! \dd^2xe
\bigg\{-\frac{1}{2}g^{mn}\del_m\xi^\mu\del_n\xi_\mu 
     -\frac{i}{2}(\lambda^\mu\sigma^m\del_m\lambda_\mu) 
\nonumber \\ 
&&\hspace*{15mm}
     +\frac{i}{2}(\lambda_\mu\sigma^m\sigma^n\chi_m)\del_n\xi^\mu 
     -\frac{1}{16}(\chi_m\sigma^n\sigma^m\chi_n)(\lambda^\mu\lambda_\mu) 
\bigg\}. 
\label{reduced_super_action_02}
\end{eqnarray}
Thus, the superstring action (\ref{reduced_super_action_02}) 
is regarded as a gauge-fixed version of the action 
(\ref{classical_super_action_04}). 
The scalar field $\phi^I(x)$ plays an important role
for the covariant formulation of the {\UvUa} superstring model
in the background spacetime which involves
two time coordinates.
From this manipulation it is suggested
that the critical dimension of the background spacetime
is defined as $D-3=9$, {\it i.e.}\ $D=12$. 
However, the dimension $D$ should be determined in the quantum analysis
as we will investigate in this paper. 
We would also like to emphasize that the quantization will be
carried out with preserving $D$-dimensional covariance.

\section{Covariant quantization in the Lagrangian formulation}
\setcounter{equation}{0}
\setcounter{footnote}{0}
  
In this section we consider the covariant quantization of the 
action. 
As we explained in the previous section, 
the action has first-stage reducible and open gauge symmetries. 
In order to quantize the action we adopt the 
field-antifield formulation {\it \'a la} Batalin-Vilkovisky~\cite{bv}.  

In the construction of Batalin-Vilkovisky formulation~\cite{ht, gps}, 
ghost and ghost for ghost 
fields according to the reducibility condition and corresponding
each antifields are introduced. 
The Grassmann parities of the antifields are opposite to 
those of the corresponding fields. 
If a field has ghost number $n$, its antifield has ghost number $-n-1$. 
We denote a set of fields and their antifields 
\begin{eqnarray*}
\Phi^A(x) &=& 
\Big(\varphi^i(x), {\cal C}_0^{a_0}(x), {\cal C}_1^{a_1}(x), 
\ldots, {\cal C}_N^{a_N}(x)
\Big), \\
\Phi^*_A(x) &=& 
\Big(\varphi^*_i(x), {\cal C}^*_{0, a_0}(x), {\cal C}^*_{1, a_1}(x), 
\ldots, {\cal C}^*_{N, a_N}(x)
\Big),  
\end{eqnarray*}
respectively. 
The fields $\varphi^i(x)$ are classical fields, on the other hand, 
the fields ${\cal C}_n^{a_n}(x)$ [$n=0$, $1$, $\ldots$, $N$] 
are ghost and ghost for ghost fields 
corresponding to $N$-th reducible conditions. 
The classical fields $\varphi^i(x)$ and the ghost fields 
${\cal C}_n^{a_n}(x)$ have the ghost number $0$ and $n+1$, respectively.    
Then a minimal action $S_{\rm min}(\Phi, \Phi^*)$ 
is defined by solving the following master equation, 
\begin{equation}
\Big( S_{\rm min}(\Phi, \Phi^*),S_{\rm min}(\Phi, \Phi^*) \Big) =0, 
\label{bv_master_eq}
\end{equation}
with the boundary conditions
\begin{subequations}
\begin{eqnarray}
S_{\rm min}(\Phi, \Phi^*)\bigg|_{\Phi^*=0}
&=& S_{\rm classical}(\varphi),  \\
\frac{\delta_{\rm L}\delta_{\rm R} S_{\rm min}(\Phi, \Phi^*)}
     {\delta{\cal C}^{a_n}_n\delta{\cal C}^*_{n-1, a_{n-1}}}
\bigg|_{\Phi^*=0}
&=& R^{a_{n-1}}_{n, a_{n}}(\Phi), \qquad (n=0, 1, \dots, N).
\label{boundary_condition}
\end{eqnarray}
\end{subequations}

\vspace*{-5mm}
\noindent
Here the antibracket is defined by
\begin{equation}\label{BRSTantibracket}
(X, Y) \equiv 
\frac{\delta_{\rm R}X}{\delta\Phi_A^*}\frac{\delta_{\rm L}Y}{\delta\Phi^A} -
\frac{\delta_{\rm R}X}{\delta\Phi^A}\frac{\delta_{\rm L}Y}{\delta\Phi_A^*}. 
\end{equation}
In this notation, 
${\cal C}^*_{-1, a_{-1}}(x) \equiv \varphi_i^*(x)$ are 
the antifields of the 
classical fields $\varphi^i(x)$. 
The terms $R^{a_{-1}}_{0, a_0}(\Phi)$ and $R^{a_{n-1}}_{n, a_n}(\Phi)$ 
represent the gauge transformations  
and the $n$-th reducibility transformations, respectively. 
The master equation is solved order by order with respect to 
the ghost number.  
The BRST transformations of fields and antifields are given by 
\begin{equation}
s \Phi^A = \Big(S_{\rm min}, \Phi^A\Big), \qquad
s \Phi_A^* = \Big(S_{\rm min}, \Phi_A^*\Big).
\label{bv_brst}
\end{equation}
Eqs.\ (\ref{bv_master_eq}) and (\ref{bv_brst}) assure that the BRST
transformation is nilpotent and the minimal action is invariant 
under the BRST transformation\footnote{
Our convention for the Leipnitz rule of the BRST operation is given by 
$s(XY)=(sX)Y+(-)^{|X|}X(sY)$, where $|X|$ is a Grassmann parity of 
the field $X$.
}. 

Now we apply the Batalin-Vilkovisky formulation to the quantization 
of our model. 
First of all, we take the action (\ref{classical_super_action_04}) as 
the classical action $S_{\rm classical}$.  
The algebra of the gauge transformations (\ref{gauge_trans_04}) 
is given by 
\begin{eqnarray}
{[}\ \delta(v_1), \ \delta(v'_2)\ {]}
&=&
\delta\Big(\tilde{w}^m\equiv-\frac{\ep^{mn}}{e}v_1\del_nv'_2\Big), 
\nonumber \\
{[}\ \delta(v_1), \ \delta(k_2)\ {]}
&=&
\delta\Big(v\equiv k^n_2\del_nv_1\Big),  
\nonumber \\
{[}\ \delta(v_1), \ \delta(\zeta_2)\ {]}
&=& 
\delta\Big(u^I\equiv iv_1(\zeta_2\lambda^I)\Big), 
\nonumber \\
{[}\ \delta(v'_1), \ \delta(v'_2)\ {]}
&=& 
\delta\Big(\tilde{w}^m\equiv v'_1g^{mn}\del_nv'_2 -v'_2g^{mn}\del_nv'_1
      \Big),  
\nonumber \\
{[}\ \delta(v'_1), \ \delta(\mu'_2)\ {]}
&=& 
\delta\Big(\tilde{w}^m\equiv-\frac{i}{2}v'_1(\mu'_2\sigma^l\sigma^m\chi_l)
      \Big), 
\nonumber \\
{[}\ \delta(v'_1), \ \delta(k_2)\ {]}
&=& 
\delta\Big(v'\equiv k^n_2\del_nv'_1\Big), 
\nonumber \\
{[}\ \delta(v'_1), \ \delta(\zeta_2)\ {]}
&=& 
\delta\Big(\mu'_\alpha\equiv(\sigma^m\zeta_2)_\alpha\del_mv'_1
      \Big)
+\delta\Big(u^I\equiv-iv'_1(\zeta_2\bar{\sigma}\lambda^I)
       \Big)
+\delta\Big(\tilde{w}^m\equiv iv'_1(\zeta_2\sigma^m\psi)\Big),  
\nonumber \\
{[}\ \delta(\mu'_1), \ \delta(\mu'_2)\ {]}
&=&
\delta\Big(\tilde{w}^m\equiv-2i(\mu'_1\sigma^m\mu'_2)\Big), 
\nonumber \\
{[}\ \delta(\mu'_1), \ \delta(k_2)\ {]}
&=&
\delta\Big(\mu'_\alpha\equiv k^n_2\del_n\mu'_{1\alpha}\Big), 
\nonumber \\
{[}\ \delta_(\mu'_1), \ \delta(\zeta_2)\ {]}
&=&
\delta\Big(v\equiv i(\mu'_1\bar{\sigma}\zeta_2)\Big)
+\delta\Big(v'\equiv i(\mu'_1\zeta_2)\Big)
\nonumber \\
&&
+\delta\Big(\mu'_\alpha\equiv
            -\frac{i}{2}(\sigma^m\zeta_2)_\alpha(\mu'_1\chi_m)
       \Big)
+\delta\Big(\tilde{w}^m\equiv-i(\mu'_1\zeta_2)\tilde{A}^m\Big), 
\nonumber \\ 
{[}\ \delta(\mu'_1), \ \delta(l_2)\ {]}
&=&
\delta\Big(\mu'_\alpha\equiv-\frac{1}{2}(\bar{\sigma}\mu'_1)_\alpha l_2
      \Big), 
\nonumber \\
{[}\ \delta(\mu_1), \ \delta(s_2)\ {]}
&=& \delta\Big(\mu'_\alpha\equiv\frac{1}{2}\mu'_{1\alpha}s_2
          \Big), 
\nonumber \\
{[}\ \delta(u^I_1), \ \delta(k_2)\ {]}
&=&
\delta\Big(u^I\equiv k^n_2\del_nu^I_1\Big), 
\nonumber \\
{[}\ \delta(u'^I_1), \ \delta(k_2)\ {]}
&=&
\delta\Big(u'^I\equiv k^n_2\del_nu'^I_1\Big),  
\nonumber \\
{[}\ \delta(\tilde{w}_1), \ \delta(k_2)\ {]}
&=& 
\delta\Big(\tilde{w}^m\equiv k_2^n\del_n\tilde{w}_1^m
           -\del_nk^m_2\tilde{w}_1^n
      \Big), 
\nonumber \\
{[}\ \delta(\tilde{w}_1), \ \delta(\zeta_2)\ {]}
&=&
\delta\Big(\tilde{w}^m\equiv-i\tilde{w}^m_1(\zeta_2\sigma^n\chi_n)
      \Big), 
\nonumber \\
{[}\ \delta(\tilde{w}_1), \ \delta(s_2)\ {]}
&=&
\delta\Big(\tilde{w}^m\equiv 2\tilde{w}^m_1s_2\Big), 
\label{gauge_algebra}\\
{[}\ \delta(k_1), \ \delta(k_2)\ {]}
&=&
\delta\Big(k^n\equiv k_2^l\del_lk_1^n-k_1^l\del_lk_2^n\Big), 
\nonumber \\
{[}\ \delta(k_1), \ \delta(\zeta_2)\ {]}
&=&
\delta\Big(\zeta_\alpha\equiv-k_1^n\del_n\zeta_{2\alpha}\Big), 
\nonumber \\
{[}\ \delta(k_1), \ \delta(l_2)\ {]}
&=& 
\delta\Big(l\equiv-k_1^n\del_nl_2\Big), 
\nonumber \\
{[}\ \delta(k_1), \ \delta(s_2) \ {]}
&=&
\delta\Big(s\equiv-k_1^n\del_ns_2\Big), 
\nonumber \\
{[}\ \delta(k_1), \ \delta(\check{s}_2) \ {]}
&=&
\delta(\check{s}_\alpha\equiv-k_1^n\del_n\check{s}_{2\alpha}\Big), 
\nonumber \\
{[}\ \delta(\zeta_1), \ \delta(\zeta_2) \ {]}
&=&
\delta\Big(v\equiv 2ie\ep_{mn}(\zeta_1\sigma^m\zeta_2)\tilde{A}^n
      \Big)
+\delta\Big(\mu'_\alpha\equiv
           i(\zeta_1\sigma^m\zeta_2)(\sigma_m\psi)_\alpha
          +i(\zeta_1\bar{\sigma}\zeta_2)(\bar{\sigma}\psi)_\alpha
      \Big)
\nonumber \\
&&
+\delta\Big(u^I\equiv 2ie\ep_{mn}(\zeta_1\sigma^m\zeta_2)\tilde{B}^{nI}
                     -2i\frac{\ep^{mn}}{e}
                     (\zeta_1\sigma_m\zeta_2)\del_n\bar{\phi}^I
       \Big)
\nonumber \\
&&
+\delta\Big(u'^I\equiv 2i(\zeta_1\sigma^m\zeta_2)\del_m\bar{\phi}^I
       \Big)
+\delta\Big(\tilde{w}^m\equiv 2i(\zeta_1\sigma^m\zeta_2)\tilde{C}
       \Big)
\nonumber \\
&&
+\delta\Big(k^n\equiv-2i(\zeta_1\sigma^n\zeta_2)\Big)
+\delta\Big(\zeta_\alpha\equiv i(\zeta_1\sigma^m\zeta_2)\chi_{m\alpha}
       \Big)
\nonumber \\
&&
+\delta\bigg(l\equiv 2i(\zeta_1\sigma^m\zeta_2)
             \Big(\omega_m-\frac{i}{2}(\chi_m\bar{\sigma}\sigma^n\chi_n)
             \Big)
       \bigg)
\nonumber \\
&& 
+\delta\Bigg(\check{s}_\alpha\equiv 
            i\bigg(\Big((\zeta_1\sigma^n\zeta_2)\sigma_n
              +(\zeta_1\bar{\sigma}\zeta_2)\bar{\sigma}\Big)
              \bar{\sigma}\frac{\ep^{pq}}{e}
             \Big(\nabla_p\chi_q
                 -\frac{i}{4}(\chi_p\bar{\sigma}\sigma^l\chi_l)
                 \bar{\sigma}\chi_q
             \Big)
             \bigg)_\alpha
       \Bigg),  
\nonumber \\ 
{[}\ \delta(\zeta_1), \ \delta(l_2) \ {]}
&=&
\delta\Big(\zeta_\alpha
           \equiv-\frac{1}{2}(\bar{\sigma}\zeta_1)_\alpha l_2
      \Big), 
\nonumber \\
{[}\ \delta(\zeta_1), \ \delta(s_2) \ {]}
&=&
\delta\Big(\zeta_\alpha\equiv-\frac{1}{2}\zeta_{1\alpha}s_2
      \Big)
+\delta\Big(\check{s}_\alpha\equiv-(\sigma^m\zeta_1)_\alpha\del_ms_2
       \Big), 
\nonumber \\
{[}\ \delta(\zeta_1), \ \delta(\check{s}_2) \ {]}
&=&
\delta\Big(l\equiv i(\zeta_1\bar{\sigma}\check{s}_2)
      \Big)
+\delta\Big(s\equiv-i(\zeta_1\check{s}_2)
       \Big)
+\delta\Big(\check{s}_\alpha\equiv
            \frac{i}{2}(\sigma^m\zeta_1)_\alpha(\chi_m\check{s}_2) 
       \Big),
\nonumber \\
{[}\ \delta(l_1), \ \delta(\check{s}_2) \ {]}
&=&
\delta\Big(\check{s}_\alpha\equiv
           \frac{1}{2}(\bar{\sigma}\check{s}_2)_\alpha l_1
      \Big), 
\nonumber \\
{[}\ \delta(s_1), \ \delta(\check{s}_2) \ {]}
&=&
\delta\Big(\check{s}_\alpha\equiv
           -\frac{1}{2}\check{s}_{2\alpha}s_1
      \Big),  
\nonumber \\
(\mbox{others}) 
&=& 0. 
\nonumber 
\end{eqnarray}
In the above gauge algebra, 
some commutation relations are closed within on-shell, 
\begin{eqnarray}
{[}\ \delta(v_1), \ \delta(v'_2)\ {]}\tilde{B}^{mI} 
&=&
\delta(\tilde{w})\tilde{B}^{mI}
+v_1v'_2\frac{\ep^{mn}}{e^2}
\frac{\delta S}{\delta\tilde{B}^n_I}, 
\nonumber \\
{[}\ \delta(v'_1), \ \delta(\zeta_2)\ {]}\lambda^I_\alpha
&=&
\delta(\mu')\lambda^I_\alpha
+(\sigma^m\zeta_2)_\alpha v'_1
 \frac{1}{e} 
 \frac{\delta S}{\delta\tilde{B}^m_I}, 
\nonumber \\
{[}\ \delta(v'_1), \ \delta(\zeta_2)\ {]}\tilde{B}^{mI}
&=&
\delta(\mu')\tilde{B}^{mI}
+\delta(u^I)\tilde{B}^{mI}
+\delta(\tilde{w})\tilde{B}^{mI}
+v'_1\Big(\zeta_2\sigma^m\frac{1}{e}\frac{\delta_LS}{\delta{\lambda_I}}
     \Big), 
\nonumber \\
{[}\ \delta(\zeta_1), \ \delta(\zeta_2) \ {]}\lambda^I_\alpha
&=&
\delta(\mu')\lambda^I_\alpha
+\delta(k)\lambda^I_\alpha
+\delta(\zeta)\lambda^I_\alpha
+\delta(l)\lambda^I_\alpha
\nonumber \\
&&
-(\zeta_1\sigma^m\zeta_2)
 \Big(\sigma_m\frac{1}{e}\frac{\delta_LS}{\delta\lambda_I}
 \Big)_\alpha
-(\zeta_1\bar{\sigma}\zeta_2)
 \Big(\bar{\sigma}\frac{1}{e}\frac{\delta_LS}{\delta\lambda_I}
 \Big)_\alpha, 
\nonumber \\
{[}\ \delta(\zeta_1), \ \delta(\zeta_2) \ {]}\phi^I
&=&
\delta(k)\phi^I
+2i(\zeta_1\sigma^m\zeta_2)\frac{1}{e}
\frac{\delta S}{\delta\tilde{B}^{mI}}, 
\nonumber \\
{[}\ \delta(\zeta_1), \ \delta(\zeta_2) \ {]}\tilde{B}^{mI}
&=&
\delta(v)\tilde{B}^{mI}
+\delta(\mu')\tilde{B}^{mI}
+\delta(u^I)\tilde{B}^{mI}
+\delta(u'^I)\tilde{B}^{mI}
+\delta(\tilde{w})\tilde{B}^{mI}
\nonumber \\
&&
+\delta(k)\tilde{B}^{mI}
+\delta(\zeta)\tilde{B}^{mI}
-2i(\zeta_1\sigma^m\zeta_2)\frac{1}{e}\frac{\delta S}{\delta\phi_I}.  
\nonumber 
\end{eqnarray}
We are now ready to quantize the model in the Batalin-Vilkovisky 
formulation. 

The classical fields $\varphi^i(x)$ consist of
$\xi^I(x)$, $\lambda^I_\alpha(x)$, 
$\tilde{A}^m(x)$, $\psi_\alpha(x)$, 
$\phi^I(x)$, 
$\bar{\phi}^I(x)$, 
$\tilde{B}^{mI}(x)$, 
$\tilde{C}(x)$, 
$e_m{}^a(x)$  
and $\chi_{m\alpha}(x)$. 
Here we introduce ghost fields 
$a(x)$, $a'(x)$, $\alpha'_\alpha(x)$, 
$b^I(x)$, $b'^I(x)$, 
$\tilde{c}^m(x)$,  
$d^m(x)$, $\gamma_\alpha(x)$, $c_{L}(x)$, $c_W(x)$ 
and $\check{c}_{S\alpha}(x)$  
corresponding to the gauge parameters 
$v(x)$, $v'(x)$, $\mu'_\alpha(x)$, 
$u^I(x)$, $u'^I(x)$, 
$\tilde{w}^m(x)$, 
$k^m(x)$, $\zeta_\alpha(x)$, $l(x)$, $s(x)$ and $\check{s}_\alpha(x)$ 
and a ghost for ghost field 
$f(x)$ 
to the reducible parameter  
$w'(x)$. 
The ghost fields 
$a(x)$, $a'(x)$, $b^I(x)$, $b'^I(x)$, $\tilde{c}^m(x)$, 
$d^m(x)$, $c_L(x)$ and $c_W(x)$ are fermionic, whereas 
the ghost fields $\alpha'_\alpha(x)$, $\gamma_\alpha(x)$ 
and $\check{c}_{S\alpha}(x)$ 
and the ghost for ghost field $f(x)$ are bosonic. 
Since the {\UvUa} superstring model is a first-stage reducible system,
the boundary conditions (\ref{boundary_condition}) with $n = 0$, $1$
correspond to 
the gauge transformations (\ref{gauge_trans_04}) and 
the reducibility conditions (\ref{reducibility_condition_03}),
respectively. 
Then, we can solve the master equation
perturbatively in the order of antifields, 
\begin{eqnarray}
S_{\rm min}
&=& S_{\rm classical} 
\nonumber \\
&& +\int\!\dd^2x
\bigg\{
-\xi^*_I
\Big(a'\phi^I+d^n\del_n\xi^I+i(\gamma\lambda^I)
\Big) 
\nonumber \\
&&\hspace*{17.5mm}
+(\lambda^*_I\alpha')\phi^I
+d^n(\lambda^*_I\del_n\lambda^I)
+(\lambda^*_I\sigma^m\gamma) 
 \Big(\del_m\xi^I-\frac{i}{2}(\chi_m\lambda^I)
 \Big)
\nonumber \\
&&\hspace*{17.5mm}
-\frac{1}{2}c_L(\lambda^*_I\bar{\sigma}\lambda^I)
+\frac{1}{2}c_W(\lambda^*_I\lambda^I)
-\frac{1}{e}(\lambda^*_I\sigma^m\gamma)a'\tilde{B}^{*I}_m
\nonumber \\
&&\hspace*{17.5mm}
+\frac{1}{4e}(\lambda^*_I\sigma^m\lambda^{*I})(\gamma\sigma_m\gamma)
+\frac{1}{4e}(\lambda^*_I\bar{\sigma}\lambda^{*I})(\gamma\bar{\sigma}\gamma) 
\nonumber \\
&&\hspace*{17.5mm}
-\tilde{A}^*_m
\Big(
\frac{\ep^{mn}}{e}\del_na
+g^{mn}\del_na'
   -\frac{i}{2}(\alpha'\sigma^n\sigma^m\chi_n)
   +d^n\del_n\tilde{A}^m
   -\del_nd^m\tilde{A}^n
\nonumber \\
&&\hspace*{30.5mm}
+i(\gamma\sigma^m\psi)
-i(\gamma\sigma^n\chi_n)\tilde{A}^m
+2c_W\tilde{A}^m
\Big)
\nonumber \\
&&\hspace*{17.5mm}
-\frac{1}{2}\del_ma'(\psi^*\sigma^n\sigma^m\chi_n)
+(\psi^*\sigma^m\nabla_m\alpha')
-\frac{i}{8}(\psi^*\alpha')(\chi_m\sigma^n\sigma^m\chi_n)
\nonumber \\
&&\hspace*{17.5mm}
+d^n(\psi^*\del_n\psi)
+(\psi^*\gamma)\nabla_m\tilde{A}^m
-i(\gamma\sigma^m\chi_m)(\psi^*\psi)
-\frac{i}{2}(\gamma\sigma^m\psi)(\psi^*\chi_m)
\nonumber \\
&&\hspace*{17.5mm}
-\frac{1}{2}c_L(\psi^*\bar{\sigma}\psi)
+\frac{3}{2}c_W(\psi^*\psi)
\nonumber \\
&&\hspace*{17.5mm}
-\phi^*_I
\Big(
d^n\del_n\phi^I
-\frac{i}{e}(\gamma\sigma^m\gamma)\tilde{B}^{*I}_m
\Big)  
-\bar{\phi}^*_I
\Big(
b'^I+d^n\del_n\bar{\phi}^I
\Big) 
\nonumber \\
&&\hspace*{17.5mm}
-\tilde{B}^*_{mI}
\bigg(
-\frac{a}{e}\ep^{mn}\del_n\xi^I
+a'g^{mn}\Big(\del_n\xi^I-\frac{i}{2}(\chi_l\sigma_n\sigma^l\lambda^I)
         \Big)
-i(\alpha'\sigma^m\lambda^I)
\nonumber \\
&&\hspace*{30.5mm} 
+\frac{1}{e}\ep^{mn}\del_nb^I
+g^{mn}\del_nb'^I
-\tilde{c}^m\phi^I
+d^n\del_n\tilde{B}^{mI}
-\del_nd^m\tilde{B}^{nI}
\nonumber \\
&&\hspace*{30.5mm} 
+i(\gamma\lambda^I)\tilde{A}^m
-i(\gamma\sigma^m\sigma^n\sigma^l\chi_n)\del_l\bar{\phi}^I
-i(\gamma\sigma^n\chi_n)\tilde{B}^{mI}
+2c_W\tilde{B}^{mI}
\nonumber \\
&&\hspace*{30.5mm}
+\frac{1}{2e^2}aa'\ep^{mn}\tilde{B}^{*I}_n
-\frac{1}{e}a(\gamma\sigma^m\lambda^{*I})
+\frac{i}{e}(\gamma\sigma^m\gamma)\phi^{*I}
+\frac{1}{2e^2}\ep^{mn}\tilde{B}^{*I}_nf
\bigg) 
\nonumber \\
&&\hspace*{17.5mm}
-\tilde{C}^*
\Big(
\del_ma'\tilde{A}^m
-a'\nabla_m\tilde{A}^m
+2i(\alpha'\psi)
+\nabla_m\tilde{c}^m
+d^n\del_n\tilde{C}
\nonumber \\
&&\hspace*{30.5mm}
-i(\gamma\sigma^m\chi_m)\tilde{C}
+2c_W\tilde{C}
\Big) 
\nonumber \\
&&\hspace*{17.5mm}
-e^*_a{}^m
\Big(
d^n\del_ne_m{}^a
+\del_md^ne_n{}^a
+i(\gamma\sigma^a\chi_m)
+c_Le_m{}^b\ep_b{}^a
-c_We_m{}^a
\Big)
\nonumber \\
&&\hspace*{17.5mm}
+d^n(\chi^{*m}\del_n\chi_m)
+\del_md^n(\chi^{*m}\chi_n)
+2(\chi^{*m}\nabla_m\gamma)
-\frac{i}{2}(\chi_m\bar{\sigma}\sigma^l\chi_l)
 (\chi^{*m}\bar{\sigma}\gamma)
\nonumber \\
&&\hspace*{17.5mm}
-\frac{1}{2}c_L(\chi^{*m}\bar{\sigma}\chi_m)
-\frac{1}{2}c_W(\chi^{*m}\chi_m)
-(\chi^{*m}\sigma_m\check{c}_S)
\nonumber \\
&&\hspace*{17.5mm}
+a^*
\Big( 
d^n\del_na
+i(\gamma\bar{\sigma}\alpha')
+ie\ep_{mn}(\gamma\sigma^m\gamma)\tilde{A}^n
\Big) 
+a'^*
\Big(
d^n\del_na'
-i(\gamma\alpha')
\Big)
\nonumber \\
&&\hspace*{17.5mm}
+d^n(\alpha'^*\del_n\alpha')
+(\alpha'^*\sigma^m\gamma)\del_ma'
-\frac{i}{2}(\alpha'^*\sigma^m\gamma)(\alpha'\chi_m)
\nonumber \\
&&\hspace*{17.5mm}
-\frac{i}{2}(\gamma\sigma^m\gamma)(\alpha'^*\sigma_m\psi)
-\frac{i}{2}(\gamma\bar{\sigma}\gamma)(\alpha'^*\bar{\sigma}\psi)
-\frac{1}{2}c_L(\alpha'^*\bar{\sigma}\alpha')
+\frac{1}{2}c_W(\alpha'^*\alpha')
\nonumber \\
&&\hspace*{17.5mm}
+b^*_I
\Big(
d^n\del_nb^I
+i(\gamma\lambda^I)a
-i(\gamma\bar{\sigma}\lambda^I)a'
+ie\ep_{mn}(\gamma\sigma^m\gamma)\tilde{B}^{nI}
\nonumber \\
&&\hspace*{27.5mm}
-i\frac{\ep^{mn}}{e}(\gamma\sigma_m\gamma)\del_n\bar{\phi}^I 
+\phi^If
\Big) 
\nonumber \\
&&\hspace*{17.5mm}
+b'^*_I
\Big(
d^n\del_nb'^I
+i(\gamma\sigma^m\gamma)\del_m\bar{\phi}^I
\Big)
\nonumber \\
&&\hspace*{17.5mm}
+\tilde{c}^*_m
\Big(
\frac{\ep^{mn}}{e}a\del_na'
-g^{mn}a'\del_na'
-\frac{i}{2}(\alpha'\sigma^n\sigma^m\chi_n)a'
-i(\alpha'\sigma^m\alpha')
\nonumber \\
&&\hspace*{27.5mm}
+d^n\del_n\tilde{c}^m
-\del_nd^m\tilde{c}^n
+i(\gamma\sigma^m\psi)a'
+i(\gamma\alpha')\tilde{A}^m
-i(\gamma\sigma^n\chi_n)\tilde{c}^m
\nonumber \\
&&\hspace*{27.5mm}
+i(\gamma\sigma^m\gamma)\tilde{C}
+2c_W\tilde{c}^m
+\frac{\ep^{mn}}{e}\del_nf
\Big) 
\nonumber \\
&&\hspace*{17.5mm}
+d^*_m
\Big( 
d^n\del_nd^m
-i(\gamma\sigma^m\gamma)
\Big) 
\nonumber \\
&&\hspace*{17.5mm}
+d^n(\gamma^*\del_n\gamma)
-\frac{i}{2}(\gamma\sigma^m\gamma)(\gamma^*\chi_m)
-\frac{1}{2}c_L(\gamma^*\bar{\sigma}\gamma)
-\frac{1}{2}c_W(\gamma^*\gamma)
\nonumber \\
&&\hspace*{17.5mm}
+c_L^*
\bigg(
d^n\del_nc_L
+i(\gamma\sigma^m\gamma)
 \Big(\omega_m-\frac{i}{2}(\chi_m\bar{\sigma}\sigma^n\chi_n)
 \Big)
+i(\check{c}_S\bar{\sigma}\gamma)
\bigg)
\nonumber \\
&&\hspace*{17.5mm}
+c_W^*
\Big(
d^n\del_nc_W
+i(\check{c}_S\gamma)
\Big)
\nonumber \\
&&\hspace{17.5mm}
+d^n(\check{c}_S^*\del_n\check{c}_S)
-\frac{i}{2}(\gamma\sigma^m\gamma)
\bigg(\check{c}_S^*\sigma_m\bar{\sigma}\frac{\ep^{pq}}{e}
      \Big(\nabla_p\chi_q-\frac{i}{4}(\chi_p\bar{\sigma}\sigma^l\chi_l)
           \bar{\sigma}\chi_q
      \Big)
\bigg)
\nonumber \\
&&\hspace*{17.5mm}
-\frac{i}{2}(\gamma\bar{\sigma}\gamma)
\bigg(\check{c}_S^*\frac{\ep^{pq}}{e}
      \Big(\nabla_p\chi_q-\frac{i}{4}(\chi_p\bar{\sigma}\sigma^l\chi_l)
           \bar{\sigma}\chi_q\Big)
\bigg)
-\del_mc_W(\check{c}_S^*\sigma^m\gamma)
\nonumber \\
&&\hspace*{17.5mm}
-\frac{i}{2}(\check{c}_S\chi_m)(\check{c}_S^*\sigma^m\gamma)
+\frac{1}{2}(\check{c}_S^*\bar{\sigma}\check{c}_S)c_L
-\frac{1}{2}(\check{c}_S^*\check{c}_S)c_W 
\nonumber \\
&&\hspace*{17.5mm}
-f^*
\Big(
d^n\del_nf
-i(\gamma\alpha')a
+i(\gamma\bar{\sigma}\alpha')a'
+ie\ep_{mn}(\gamma\sigma^m\gamma)\tilde{c}^n
\Big)
\bigg\}.
\end{eqnarray}

The gauge degrees of freedom are fixed by introducing a nonminimal 
action which must be added to the minimal one and choosing a suitable  
gauge-fixing fermion. 
By using the gauge parameters for the general coordinate, 
Weyl, local supersymmetry and super-Weyl transformations, 
we here choose super-orthonormal gauge conditions  
$e_m{}^a(x)=\delta_m{}^a$ for the zweibein field and 
$\chi_{m\alpha}(x)=0$ for the gravitino field. 
In the same way for the bosonic model~\cite{tw1},  
the {\UvUa} gauge parameters $\Big(v(x), v'(x), \alpha_i\Big)$, 
$\Big(u^I(x), u'^I(x), \beta^I_i\Big)$ and $\tilde{w}^m$ 
allow to choose gauges  
$\tilde{A}^m(x)=\tilde{B}^{mI}(x)=0$ and $\tilde{C}(x)=\tilde{C}_0$, 
where $\tilde{C}_0$ is a constant parameter. 
We also impose a gauge condition 
$\del_m\Big(eg^{mn}e\ep_{nk}\tilde{c}^k(x)\Big)=0$ to fix the residual 
gauge degrees of freedom from the reducibility condition. 
In addition to these, we fix a gauge $\psi_\alpha(x)=0$ by using 
the gauge parameter $\mu'_\alpha(x)$ in this supersymmetric model. 
In order to adopt all of these gauge fixing conditions, 
we introduce the nonminimal action $S_{\rm nonmin}$, 
\begin{eqnarray}
S_{\rm nonmin} 
=
\int\!\dd^2x
\bigg\{
&&
\ep^{mn}\hat{a}^*_mZ^{(a)}_n
+\ep^{mn}\hat{b}^{*I}_mZ^{(b)}_{nI}
+c^*Z^{(c)}
-(\bar{\alpha}^*\check{Z}^{(\alpha)})
\nonumber 
\\
&& 
+\bar{d}^*_m{}^aZ^m{}_a
-(\beta^*_m\check{Z}^m)
-\bar{f}^*c'
\bigg\}, 
\end{eqnarray}
and the gauge-fixing fermion $\Psi$, 
\begin{eqnarray}
\Psi
=
\int\!\dd^2x
\bigg\{
&&
e\ep_{mn}\hat{a}^m\tilde{A}^n
+e\ep_{mn}\hat{b}^m_I\tilde{B}^{nI}
+ec(\tilde{C}-\tilde{C}_0)
+ie(\bar{\alpha}\psi)
\nonumber 
\\
&&
-e\bar{d}^m{}_a(e_m{}^a-\delta_m{}^a)
+\frac{e}{2}(\beta^m\chi_m)
+\bar{f}\del_m(eg^{mn}e\ep_{nk}\tilde{c}^k)
\bigg\}.
\label{gauge-fixing_fermion}
\end{eqnarray}
The antighost fields $\hat{a}^m(x)$, $\hat{b}^m_I(x)$, 
$c(x)$, $\bar{d}^m{}_a(x)$ and $c'(x)$ are fermionic, 
while $\bar{\alpha}_\alpha(x)$ and $\beta^m{}_\alpha(x)$ 
are bosonic. 
The auxiliary fields $Z^{(a)}_m(x)$, $Z^{(b)}_{mI}(x)$, 
$Z^{(c)}(x)$, $Z^m{}_a(x)$ and $\bar{f}(x)$ are bosonic, 
whereas $\check{Z}^{(\alpha)}_\alpha(x)$ and 
$\check{Z}^m{}_\alpha(x)$ are 
fermionic\footnote{
The auxiliary fields $Z^{(a)}_m(x)$, $Z^{(b)}_{mI}(x)$ and 
$Z^{(c)}(x)$ are 
equivalent to $Z^a_m(x)$, $Z^b_{mI}(x)$ and $Z^c(x)$ in our previous 
paper~\cite{tw1}, respectively.
In order to avoid confusing the attached indices $a$, $b$ and $c$ 
for $Z^a_m(x)$, $Z^b_{mI}(x)$ and $Z^c(x)$   
with local Lorentz indices, we have changed the notations.
}. 

The BRST transformations of the field $\Phi^A(x)$ and the 
antifields $\Phi_A^*(x)$ are now given by 
\begin{equation}
s\Phi^A=\Big(S_{\rm min}+S_{\rm nonmin}, \Phi^A\Big), \qquad
s\Phi_A^*=\Big(S_{\rm min}+S_{\rm nonmin}, \Phi_A^*\Big). 
\end{equation}
Then, the BRST transformations of the fields $\Phi^A(x)$ are 
\begin{subequations}
\begin{eqnarray}
s\xi^I
&=& a'\phi^I+d^n\del_n\xi^I+i(\gamma\lambda^I), 
\nonumber \\
s\lambda^I_\alpha
&=& \alpha'_\alpha\phi^I
   +d^n\del_n\lambda^I_\alpha
   +(\sigma^m\gamma)_\alpha
    \Big(\del_m\xi^I-\frac{i}{2}(\chi_m\lambda^I)
    \Big)
   -\frac{1}{2}c_L(\bar{\sigma}\lambda^I)_\alpha
   +\frac{1}{2}c_W\lambda^I_\alpha
\nonumber \\
&&
-\frac{1}{e}(\sigma^m\gamma)_\alpha a'\tilde{B}^{*I}_m
+\frac{1}{2e}(\sigma^m\lambda^{*I})_\alpha(\gamma\sigma_m\gamma)
+\frac{1}{2e}(\bar{\sigma}\lambda^{*I})_\alpha(\gamma\bar{\sigma}\gamma), 
\nonumber \\
s\tilde{A}^m
&=& \frac{\ep^{mn}}{e}\del_na
   +g^{mn}\del_na'
   -\frac{i}{2}(\alpha'\sigma^n\sigma^m\chi_n)
\nonumber \\
&& +d^n\del_n\tilde{A}^m
   -\del_nd^m\tilde{A}^n
   +i(\gamma\sigma^m\psi)
   -i(\gamma\sigma^n\chi_n)\tilde{A}^m
   +2c_W\tilde{A}^m, 
\nonumber \\
s\psi_\alpha
&=& -\frac{1}{2}\del_ma'(\sigma^n\sigma^m\chi_n)_\alpha
   +(\sigma^m\nabla_m\alpha')_\alpha
   -\frac{i}{8}\alpha'_\alpha(\chi_m\sigma^n\sigma^m\chi_n)
\nonumber \\
&& +d^n\del_n\psi_\alpha
   +\gamma_\alpha\nabla_m\tilde{A}^m
  -i(\gamma\sigma^m\chi_m)\psi_\alpha
  -\frac{i}{2}(\gamma\sigma^m\psi)\chi_{m\alpha}
  -\frac{1}{2}c_L(\bar{\sigma}\psi)_\alpha
  +\frac{3}{2}c_W\psi_\alpha, 
\nonumber \\
s\phi^I
&=& 
d^n\del_n\phi^I
-\frac{i}{e}(\gamma\sigma^m\gamma)\tilde{B}^{*I}_m,  
\nonumber \\
s\bar{\phi}^I
&=& b'^I+d^n\del_n\bar{\phi}^I, 
\nonumber \\
s\tilde{B}^{mI}
&=& -\frac{a}{e}\ep^{mn}\del_n\xi^I
   +a'g^{mn}\Big(\del_n\xi^I-\frac{i}{2}(\chi_l\sigma_n\sigma^l\lambda^I)
            \Big)
   -i(\alpha'\sigma^m\lambda^I)
\nonumber \\
&& +\frac{1}{e}\ep^{mn}\del_nb^I
  +g^{mn}\del_nb'^I
 -\tilde{c}^m\phi^I
 +d^n\del_n\tilde{B}^{mI}
 -\del_nd^m\tilde{B}^{nI}
\nonumber \\
&& +i(\gamma\lambda^I)\tilde{A}^m
  -i(\gamma\sigma^m\sigma^n\sigma^l\chi_n)\del_l\bar{\phi}^I
  -i(\gamma\sigma^n\chi_n)\tilde{B}^{mI}
  +2c_W\tilde{B}^{mI}
\nonumber \\
&&
+\frac{1}{e^2}aa'\ep^{mn}\tilde{B}^{*I}_n
-\frac{1}{e}a(\gamma\sigma^m\lambda^{*I})
+\frac{i}{e}(\gamma\sigma^m\gamma)\phi^{*I}
+\frac{1}{e^2}\ep^{mn}\tilde{B}^{*I}_nf, 
\nonumber \\
s\tilde{C}
&=& \del_ma'\tilde{A}^m
   -a'\nabla_m\tilde{A}^m
   +2i(\alpha'\psi)
   +\nabla_m\tilde{c}^m
   +d^n\del_n\tilde{C}
   -i(\gamma\sigma^m\chi_m)\tilde{C}
   +2c_W\tilde{C}, 
\nonumber \\
se_m{}^a
&=& d^n\del_ne_m{}^a
   +\del_md^ne_n{}^a
   +i(\gamma\sigma^a\chi_m)
   +c_Le_m{}^b\ep_b{}^a
   -c_We_m{}^a, 
\label{gauge_trans_05_1}\\
s\chi_{m\alpha}
&=& d^n\del_n\chi_{m\alpha}
   +\del_md^n\chi_{n\alpha}
   +2(\nabla_m\gamma)_\alpha
   -\frac{i}{2}(\chi_m\bar{\sigma}\sigma^l\chi_l)(\bar{\sigma}\gamma)_\alpha
\nonumber \\
&& -\frac{1}{2}c_L(\bar{\sigma}\chi_m)_\alpha
   -\frac{1}{2}c_W\chi_{m\alpha}
   -(\sigma_m\check{c}_S)_\alpha, 
\nonumber \\
sa 
&=& d^n\del_na
   +i(\gamma\bar{\sigma}\alpha')
   +ie\ep_{mn}(\gamma\sigma^m\gamma)\tilde{A}^n, 
\nonumber \\
sa'
&=& d^n\del_na'
   -i(\gamma\alpha'), 
\nonumber \\
s\alpha'_\alpha
&=&
d^n\del_n\alpha'_\alpha
-(\sigma^m\gamma)_\alpha\del_ma'
+\frac{i}{2}(\sigma^m\gamma)_\alpha(\alpha'\chi_m)
+\frac{i}{2}(\gamma\sigma^m\gamma)(\sigma_m\psi)_\alpha
+\frac{i}{2}(\gamma\bar{\sigma}\gamma)(\bar{\sigma}\psi)_\alpha
\nonumber \\
&&
-\frac{1}{2}c_L(\bar{\sigma}\alpha')_\alpha
+\frac{1}{2}c_W\alpha'_\alpha, 
\nonumber \\
sb^I
&=&
d^n\del_nb^I
+i(\gamma\lambda^I)a
-i(\gamma\bar{\sigma}\lambda^I)a'
+ie\ep_{mn}(\gamma\sigma^m\gamma)\tilde{B}^{nI}
-i\frac{\ep^{mn}}{e}(\gamma\sigma_m\gamma)\del_n\bar{\phi}^I 
+\phi^If, 
\nonumber \\
sb'^I
&=&
d^n\del_nb'^I
+i(\gamma\sigma^m\gamma)\del_m\bar{\phi}^I, 
\nonumber \\
s\tilde{c}^m
&=&
\frac{\ep^{mn}}{e}a\del_na'
-g^{mn}a'\del_na'
-\frac{i}{2}(\alpha'\sigma^n\sigma^m\chi_n)a'
-i(\alpha'\sigma^m\alpha')
+d^n\del_n\tilde{c}^m
-\del_nd^m\tilde{c}^n
\nonumber \\
&&
+i(\gamma\sigma^m\psi)a'
+i(\gamma\alpha')\tilde{A}^m
-i(\gamma\sigma^n\chi_n)\tilde{c}^m
+i(\gamma\sigma^m\gamma)\tilde{C}
+2c_W\tilde{c}^m
+\frac{\ep^{mn}}{e}\del_nf, 
\nonumber \\
sd^m 
&=& 
d^n\del_nd^m
-i(\gamma\sigma^m\gamma), 
\nonumber \\
s\gamma_\alpha
&=& 
d^n\del_n\gamma_\alpha
+\frac{i}{2}(\gamma\sigma^m\gamma)\chi_{m\alpha}
-\frac{1}{2}c_L(\bar{\sigma}\gamma)_\alpha
-\frac{1}{2}c_W\gamma_\alpha, 
\nonumber \\
sc_L
&=&
d^n\del_nc_L
+i(\gamma\sigma^m\gamma)
 \Big(\omega_m-\frac{i}{2}(\chi_m\bar{\sigma}\sigma^n\chi_n)
 \Big)
+i(\check{c}_S\bar{\sigma}\gamma), 
\nonumber \\
sc_W
&=&
d^n\del_nc_W
+i(\check{c}_S\gamma), 
\nonumber \\
s\check{c}_{S\alpha}
&=&
d^n\del_n\check{c}_{S\alpha}
+\frac{i}{2}\bigg(\Big((\gamma\sigma^m\gamma)\sigma_m
                        +(\gamma\bar{\sigma}\gamma)\bar{\sigma}
                  \Big)\bar{\sigma}\frac{\ep^{pq}}{e}
                  \Big(\nabla_p\chi_q
                      -\frac{i}{4}(\chi_p\bar{\sigma}\sigma^l\chi_l)
                      \bar{\sigma}\chi_q
                  \Big)
            \bigg)_\alpha
\nonumber \\
&&
-\del_mc_W(\sigma^m\gamma)_\alpha
-\frac{i}{2}(\check{c}_S\chi_m)(\sigma^m\gamma)_\alpha
-\frac{1}{2}(\bar{\sigma}\check{c}_S)_\alpha c_L
+\frac{1}{2}\check{c}_{S\alpha}c_W, 
\nonumber \\
sf
&=&
d^n\del_nf
-i(\gamma\alpha')a
+i(\gamma\bar{\sigma}\alpha')a'
+ie\ep_{mn}(\gamma\sigma^m\gamma)\tilde{c}^n, 
\nonumber 
\end{eqnarray}
and 
\begin{eqnarray}
s\hat{a}^m
&=&
\ep^{mn}Z_n^{(a)},
\hspace*{20mm}
sZ_m^{(a)}
=0, 
\nonumber 
\\
s\hat{b}^m_I
&=& 
\ep^{mn}Z_{nI}^{(b)},
\hspace*{20.3mm}
sZ_{mI}^{(b)}
=0, 
\nonumber 
\\
sc
&=& 
Z^{(c)}, 
\hspace*{27mm}
sZ^{(c)}
=0, 
\nonumber 
\\
s\bar{\alpha}_\alpha
&=&
\check{Z}_\alpha^{(\alpha)},
\hspace*{25.5mm}
s\check{Z}_\alpha^{(\alpha)}
=0,
\label{gauge_trans_05_2}
\\
s\bar{f}
&=&
c', 
\hspace*{35mm}
sc'
=0, 
\nonumber 
\\
s\bar{d}^m{}_a
&=&
Z^m{}_a, 
\hspace*{25mm}
sZ^m{}_a
=0,
\nonumber 
\\
s\beta^m{}_\alpha
&=&
\check{Z}^m{}_\alpha, 
\hspace*{24.3mm}
s\check{Z}^m{}_\alpha
=0.
\nonumber 
\end{eqnarray}
\end{subequations}

\vspace*{-6mm}

Now, let us construct a gauge-fixed action. 
The antifields are eliminated by using the gauge-fixing fermion 
(\ref{gauge-fixing_fermion}) via equations 
$\Phi^*_A(x)=\delta_{\rm L}\Psi/\delta\Phi^A(x)$. 
In order to specify the physical degrees of freedom of the
two-dimensional supergravity sector, 
it might be useful to decompose the antighost fields $\bar{d}^m{}_a(x)$ and 
$\beta^m{}_\alpha(x)$ as follows, 
\begin{subequations}
\begin{eqnarray}
\bar{d}^m{}_a
&=&
e^n{}_a\bar{d}^m{}_n
+\frac{\ep^{mn}}{e}e_{na}\bar{c}_L
+e^m{}_a\bar{c}_W, 
\\
\beta^m{}_\alpha
&=&
\bar{\beta}^m{}_\alpha
+(\sigma^m\beta)_\alpha, 
\end{eqnarray}
\end{subequations}
where we define 
\begin{eqnarray*}
\bar{d}^m{}_n
&\equiv&
\frac{1}{2}
\Big(\bar{d}^m{}_ae_n{}^a
+\bar{d}^l{}_ae^{ma}g_{ln}
-\delta^m{}_n\bar{d}^l{}_ae_l{}^a
\Big), 
\\
\bar{c}_L
&\equiv&
\frac{1}{2}\ep^{ab}\bar{d}^m{}_ae_{mb}, 
\\
\bar{c}_W
&\equiv&
\frac{1}{2}\bar{d}^m{}_ae_m{}^a,
\\
\bar{\beta}^m{}_\alpha
&\equiv&
\frac{1}{2}(\sigma_n\sigma^m\beta^n)_\alpha, 
\\
\beta_\alpha
&\equiv&
\frac{1}{2}(\sigma_m\beta^m)_\alpha. 
\end{eqnarray*}
The field $\bar{d}^m{}_n(x)$ is symmetric $\ep^n{}_m\bar{d}^m{}_n(x)=0$
and traceless $\delta^n{}_m\bar{d}^m{}_n(x)=0$, 
and the field $\bar{\beta}^m{}_\alpha(x)$ is $\sigma$-traceless 
$(\sigma_m)_\alpha{}^\beta\bar{\beta}^m{}_\beta(x)=0$.  
Then, the gauge-fixed action is given by
\begin{eqnarray}
S_{\rm gauge-fixed}
&=&
S_{\rm min}
+S_{\rm nonmin}\Big|_{\Phi^*=\frac{\delta\Psi}{\delta\Phi}}
\nonumber \\
&=&
\int \!\dd^2x
\bigg\{
-\frac{1}{2}eg^{mn}\del_m\xi^I\del_n\xi_I
-\frac{i}{2}e(\lambda^I\sigma^m\del_m\lambda_I)
-eg^{mn}\del_m\bar{\phi}^I\del_n\phi_I
\nonumber \\
&&\hspace*{13.5mm} 
- \ \hat{a}^m
\Big(
\del_ma
+e\ep_{mk}g^{kn}\del_na'
\Big)
-\hat{b}^m_I
\Big(
\del_mb^I
+e\ep_{mk}g^{kn}\del_nb'^I
\Big)
\nonumber \\
&&\hspace*{13.5mm}
- \ \hat{c}^m
\Big(\del_mc
+e\ep_{mk}g^{kn}\del_nc'
\Big)
+ie(\bar{\alpha}\sigma^m\del_m\alpha')
-eg^{mn}\del_m\bar{f}\del_nf
\nonumber \\
&&\hspace*{13.5mm}
+ \ e\bar{d}^m{}_n\del_md^n
+e(\bar{\beta}^m\del_m\gamma)
\nonumber \\
&&\hspace*{13.5mm}
- \ 2a\hat{b}^m_I\del_m\xi^I
+\ep_{mn}\hat{b}^m_I\hat{c}^n\phi^I
\nonumber \\
&&\hspace*{13.5mm}
+ \ \frac{1}{2}(f+aa')\ep_{mn}\hat{b}^m_I\hat{b}^{nI}
+ie\ep_{mn}\hat{b}^m_I(\alpha'\sigma^n\lambda^I)
\nonumber \\
&&\hspace*{13.5mm}
- \ e\tilde{A}^mZ^{(a)}_m
-e\tilde{B}^{mI}Z^{(b)}_{mI}
+e\tilde{C}Z^{(c)}
+ie(\psi\check{Z}^{(\alpha)})
\nonumber \\
&&\hspace*{13.5mm}
- \ 2e\bar{c}_Lc_L
-2e\bar{c}_Wc_W
-e(\beta\check{c}_S)
\nonumber \\
&&\hspace*{13.5mm}
- \ e(e_m{}^a-\delta_m{}^a)Z^m{}_a
+\frac{e}{2}(\chi_m\check{Z}^m)
\bigg\},
\label{gauge-fixed_action_0}
\end{eqnarray}
where we redefine some of the fields as follows,
\begin{eqnarray*}
Z^{(a)}_m
-\phi_I(\del_m\xi^I)
-\ep_{mn}d^k\del_k\hat{a}^n
-\del_md^k\ep_{kn}\hat{a}^n
+c\del_ma'
+\del_m(ca')
&&
\\
-ie\ep_{mk}g^{kn}\del_n\bar{f}(\gamma\alpha')
-i\ep_{mn}\hat{b}^n_I(\gamma\lambda^I)
+i\del_m(\bar{\alpha}\gamma)
&\rightarrow&
Z^{(a)}_m, 
\\
Z^{(b)}_{mI}
-\del_m\phi_I
-\ep_{mn}d^k\del_k\hat{b}^n_I
-\del_md^k\ep_{kn}\hat{b}^n_I
&\rightarrow&
Z^{(b)}_{mI}, 
\\
Z^{(c)}
-\frac{1}{2}\phi^I\phi_I
-d^n\del_nc
+ie\ep_{mk}g^{kn}\del_n\bar{f}(\gamma\sigma^m\gamma)
&\rightarrow&
Z^{(c)}, 
\\
\check{Z}^{(\alpha)}_\alpha
+\phi_I\lambda^I_\alpha
-d^n\del_n\bar{\alpha}_\alpha
-2c\alpha'_\alpha
+e\ep_{mk}g^{kn}\del_n\bar{f}a'(\sigma^m\gamma)_\alpha
-\ep_{mn}\hat{a}^n(\sigma^m\gamma)_\alpha
&&
\\
-\frac{1}{4}\del_md^n(\sigma^m\sigma_n\bar{\alpha})_\alpha
-\frac{i}{2}(\bar{\alpha}\chi_m)(\sigma^m\gamma)_\alpha
&&
\\
+\frac{1}{2}\Big(c_L-\frac{1}{2}\frac{\ep^{mk}}{e}g_{kn}\del_md^n\Big)
 (\bar{\sigma}\bar{\alpha})_\alpha
-\frac{1}{2}\Big(c_W-\frac{1}{2}\del_nd^n\Big)\bar{\alpha}_\alpha
&\rightarrow&
\check{Z}^{(\alpha)}_\alpha, 
\\
Z^m{}_a
-\frac{i}{2}\frac{\ep^{mn}}{e}\del_n
\bigg(
e_a{}^k
\Big(\alpha'\sigma_k\bar{\sigma}
(\bar{\alpha}-2\bar{f}\bar{\sigma}\alpha')
\Big)
\bigg)
+\del_n(\bar{d}^m{}_ad^n)
-i\bar{d}^m{}_a(\gamma\sigma^n\chi_n)
&&
\\
+2\bar{d}^m{}_a\Big(c_W-\frac{1}{2}\del_nd^n\Big)
-\frac{1}{2}\frac{\ep^{mn}}{e}\del_n
\Big(e_{la}(\beta^l\bar{\sigma}\gamma)
\Big)
&\rightarrow&
Z^m{}_a, 
\\
\check{Z}^m{}_\alpha
+i(\sigma^n\sigma^m\lambda_I)_\alpha\del_n\xi^I
-\frac{1}{8}(\sigma^n\sigma^m\chi_n)_\alpha(\lambda^I\lambda_I)
+ia'\ep_{nk}\hat{b}^k_I(\sigma^n\sigma^m\lambda^I)_\alpha
&&
\\
+2i\ep_{nk}\hat{b}^k_I\del_l\bar{\phi}^I
(\sigma^l\sigma^m\sigma^n\gamma)_\alpha
+ie\ep_{nk}g^{kl}\del_l\bar{f}a'(\sigma^n\sigma^m\alpha')_\alpha
+2i\ep_{nk}g^{kl}\del_l\bar{f}\hat{c}^n(\sigma^m\gamma)_\alpha
&&
\\
-2i\ep_{nk}g^{km}\del_l\bar{f}\hat{c}^n(\sigma^l\gamma)_\alpha
+2i\ep_{nk}g^{ml}\del_l\bar{f}\hat{c}^k(\sigma^n\gamma)_\alpha
-i\ep_{nk}\hat{a}^k(\sigma^n\sigma^m\alpha')_\alpha
&&
\\
-i\del_na'(\sigma^n\sigma^m\bar{\alpha})_\alpha
+\frac{1}{4}(\bar{\alpha}\alpha')(\sigma^n\sigma^m\chi_n)_\alpha
-2i(\sigma^a\gamma)_\alpha\bar{d}^m{}_a
-d^n\del_n\beta^m{}_\alpha
&&
\\
+\del_nd^m\beta^n{}_\alpha
+i(\gamma\sigma^n\chi_n)\beta^m{}_\alpha
-\frac{i}{2}(\bar{\sigma}\sigma^l\chi_l)_\alpha(\beta^m\bar{\sigma}\gamma)
+\frac{1}{2}(\bar{\sigma}\beta^m)_\alpha c_L
-\frac{5}{2}\beta^m{}_\alpha c_W
&\rightarrow&
\check{Z}^m{}_\alpha, 
\\
\hat{a}^m
-\ep^{mn}\del_n(\bar{f}a)
+e\bar{f}g^{mn}\del_na'
-\hat{b}^m_I\xi^I
&\rightarrow&
\hat{a}^m, 
\\
b^I
+a\xi^I
&\rightarrow&
b^I, 
\\
b'^I
+a'\xi^I
&\rightarrow&
b'^I, 
\\
c'
-d^n\del_n\bar{f}
&\rightarrow&
c', 
\\
\bar{\alpha}_\alpha
-2\bar{f}(\bar{\sigma}\alpha')_\alpha
&\rightarrow&
\bar{\alpha}_\alpha, 
\\
c_L-\frac{1}{2}\frac{\ep^{mn}}{e}g_{nl}\del_md^l
&\rightarrow&
c_L, 
\\
c_W-\frac{1}{2}\del_md^m
&\rightarrow&
c_W, 
\\
\check{c}_{S\alpha}
-(\sigma^m\del_m\gamma)_\alpha
&\rightarrow&
\check{c}_{S\alpha}, 
\end{eqnarray*}
and we denote $\hat{c}^m(x)\equiv e\tilde{c}^m(x)$. 
It should be noted that 
we remove a BRST exact term 
$$
-e\tilde{C}_0
\Bigg(
Z^{(c)}
-c\bigg(d^n\del_ne+i(\gamma\sigma^n\chi_n)
-2\Big(c_W-\frac{1}{2}\del_nd^n\Big)
\bigg)
\Bigg)
=
-s(e\tilde{C}_0c), 
$$ 
from the above action. 

Using equations of motion of the gauge-fixed action 
(\ref{gauge-fixed_action_0}), 
thus imposing the gauge fixing conditions, 
we consistently fix the fields as   
\begin{eqnarray}
\tilde{A}^m
&=&\tilde{B}^{mI}=\tilde{C}=\psi_\alpha 
=c_L=c_W=\check{c}_{S\alpha}=\chi_{m\alpha}=0, 
\nonumber 
\\ 
Z^{(a)}_m
&=&Z^{(b)}_{mI}=Z^{(c)}=\check{Z}^{(\alpha)}_\alpha
=\bar{c}_L=\bar{c}_W=\beta_\alpha=\check{Z}^m{}_\alpha=0,
\nonumber 
\\ 
\quad e_m{}^a
&=&\delta_m{}^a,  
\nonumber 
\\
Z^m{}_a
&=& 
-\frac{1}{2}\delta^m{}_a\eta^{kl}\del_k\xi^I\del_l\xi_I
+\delta^n{}_a\eta^{mk}\del_k\xi^I\del_n\xi_I
\nonumber 
\\
&&
-\frac{i}{2}\delta^m{}_a(\lambda^I\sigma^n\del_n\lambda_I)
+\frac{i}{2}\delta^n{}_a(\lambda^I\sigma^m\del_n\lambda_I)
\nonumber 
\\
&& 
-\delta^m{}_a\eta^{kl}\del_k\bar{\phi}^I\del_l\phi_I
+\delta^n{}_a\eta^{mk}\del_k\bar{\phi}^I\del_n\phi_I
+\delta^n{}_a\eta^{mk}\del_n\bar{\phi}^I\del_k\phi_I
\nonumber 
\\
&&
-\delta^m{}_a\hat{a}^n\ep_n{}^l\del_la'
+\delta^n{}_a\hat{a}^k\ep_k{}^m\del_na'
+\delta^n{}_a\hat{a}^k\ep_{kn}\eta^{ml}\del_la'
\\
&&
-\delta^m{}_a\hat{b}^n_I\ep_n{}^l\del_lb'^I
+\delta^n{}_a\hat{b}^k_I\ep_k{}^m\del_nb'^I
+\delta^n{}_a\hat{b}^k_I\ep_{kn}\eta^{ml}\del_lb'^I
\nonumber 
\\
&&
-\delta^m{}_a\hat{c}^n\ep_n{}^l\del_lc'
+\delta^n{}_a\hat{c}^k\ep_k{}^m\del_nc'
+\delta^n{}_a\hat{c}^k\ep_{kn}\eta^{ml}\del_lc'
\nonumber 
\\
&&
+i\delta^m{}_a(\bar{\alpha}\sigma^n\del_n\alpha')
-i\delta^n{}_a(\bar\alpha\sigma^m\del_n\alpha')
\nonumber 
\\
&&
-\delta^m{}_a\eta^{kl}\del_k\bar{f}\del_lf
+\delta^n{}_a\eta^{mk}\del_k\bar{f}\del_nf
+\delta^n{}_a\eta^{mk}\del_n\bar{f}\del_kf
\nonumber 
\\
&&
+\delta^m{}_a\bar{d}^k{}_n\del_kd^n
+\delta^n{}_a\bar{d}^k{}_n\del_kd^m
-\delta^n{}_a\bar{d}^m{}_k\del_nd^k
\nonumber 
\\
&&
+\delta^m{}_a(\bar{\beta}^n\del_n\gamma)
-\delta^n{}_a(\bar{\beta}^m\del_n\gamma)
\nonumber 
\\
&&
+i\delta^m{}_a\ep_{kl}\hat{b}^k_I(\alpha'\sigma^l\lambda^I)
+i\delta^n{}_a\ep_{nk}\hat{b}^k_I(\alpha'\sigma^m\lambda^I). 
\nonumber   
\end{eqnarray}
Then, we finally obtain the gauge-fixed action  
\begin{eqnarray}
S_{\rm gauge-fixed}
=
\int \!\dd^2x
\bigg\{
&&
-\frac{1}{2}\eta^{mn}\del_m\xi^I\del_n\xi_I
-\frac{i}{2}(\lambda^I\sigma^m\del_m\lambda_I)
-\eta^{mn}\del_m\bar{\phi}^I\del_n\phi_I
\nonumber \\
&&
-\hat{a}^m
\Big(\del_ma
+\ep_m{}^n\del_na'
\Big)
-\hat{b}^m_I
\Big(
\del_mb^I
+\ep_m{}^n\del_nb'^I
\Big)
\nonumber \\
&&
-\hat{c}^m
\Big(\del_mc
+\ep_m{}^n\del_nc'
\Big)
+i(\bar{\alpha}\sigma^m\del_m\alpha')
-\eta^{mn}\del_m\bar{f}\del_nf
\nonumber \\
&&
+\eta^{mn}\bar{d}_{mk}\del_nd^k
+(\bar{\beta}^m\del_m\gamma)
\nonumber \\
&&
-2a\hat{b}^m_I\del_m\xi^I
+\ep_{mn}\hat{b}^m_I\hat{c}^n\phi^I
\nonumber \\
&&
+\frac{1}{2}(f+aa')\ep_{mn}\hat{b}^m_I\hat{b}^{nI}
+i\ep_{mn}\hat{b}^m_I(\alpha'\sigma^n\lambda^I)
\bigg\}. 
\label{gauge-fixed_action}
\end{eqnarray}
The action (\ref{gauge-fixed_action}) is invariant under the following 
on-shell nilpotent BRST transformations 
which are obtained from (\ref{gauge_trans_05_1}) 
and (\ref{gauge_trans_05_2}) by eliminating the antifields 
and the auxiliary fields,  
\begin{eqnarray}
s\xi^I
&=&
a'\phi^I
+d^n\del_n\xi^I
+i(\gamma\lambda^I), 
\nonumber 
\\
s\lambda^I_\alpha
&=&
\alpha'_\alpha\phi^I
+d^n\del_n\lambda^I_\alpha
+(\sigma^m\gamma)_\alpha\del_m\xi^I
+\frac{1}{4}\del_md_n(\sigma^m\sigma^n\lambda^I)_\alpha
-\ep_{mn}a'\hat{b}^{mI}(\sigma^n\gamma)_\alpha, 
\nonumber 
\\
s\phi^I
&=&
d^n\del_n\phi^I
+i(\gamma\sigma^m\gamma)\ep_{mn}\hat{b}^{nI}, 
\nonumber 
\\
s\bar{\phi}^I
&=&
b'^I
-a'\xi^I
+d^n\del_n\bar{\phi}^I, 
\nonumber 
\\
sa
&=&
d^n\del_na
+i(\gamma\bar{\sigma}\alpha'), 
\nonumber 
\\
sa'
&=&
d^n\del_na'
-i(\gamma\alpha'), 
\nonumber 
\\
s\alpha'_\alpha
&=&
d^n\del_n\alpha'_\alpha
-(\sigma^m\gamma)_\alpha\del_ma'
+\frac{1}{4}(\sigma^m\sigma^n\alpha')_\alpha\del_md_n, 
\nonumber 
\\
sb^I
&=&
(f-aa')\phi^I
+d^n\del_nb^I
+2i(\gamma\lambda^I)a
-i(\gamma\bar{\sigma}\lambda^I)a'
-i\ep^{mn}(\gamma\sigma_m\gamma)\del_n\bar{\phi}^I
+i(\gamma\bar{\sigma}\alpha')\xi^I, 
\nonumber 
\\
sb'^I
&=&
d^n\del_nb'^I
+i(\gamma\lambda^I)a'
+i(\gamma\sigma^m\gamma)\del_m\bar{\phi}^I
-i(\gamma\alpha')\xi^I, 
\nonumber 
\\
s\hat{c}^m
&=&
(\ep^{mn}a-\eta^{mn}a')\del_na'
-i(\alpha'\sigma^m\alpha')
+\ep^{mn}\del_nf
+\del_n(d^n\hat{c}^m)
-\del_nd^m\hat{c}^n, 
\nonumber 
\\
sd^m
&=&
d^n\del_nd^m
-i(\gamma\sigma^m\gamma), 
\nonumber 
\\
s\gamma_\alpha
&=&
d^n\del_n\gamma_\alpha
-\frac{1}{4}(\sigma^n\sigma^m\gamma)_\alpha\del_md_n, 
\label{gauge_trans}\\
sf
&=&
d^n\del_nf
-i(\gamma\alpha')a
+i(\gamma\bar{\sigma}\alpha')a'
+i\ep_{mn}(\gamma\sigma^m\gamma)\hat{c}^n, 
\nonumber 
\\
s\hat{a}^m
&=&
\ep^{mn}(\phi_I\del_n\xi^I-\del_n\phi_I\xi^I)
-a'\hat{b}^m_I\phi^I
-(\ep^{mn}c-\eta^{mn}c')\del_na'
-\ep^{mn}\del_n(ca'+c'a)
\nonumber 
\\
&&
+\del_n(d^n\hat{a}^m)
-\del_nd^m\hat{a}^n
+i\ep^{mn}\del_n(\gamma\bar{\alpha})
+2i\hat{b}^m_I(\gamma\lambda^I)
+i\del_n\bar{f}(\gamma\sigma^n\sigma^m\alpha'), 
\nonumber 
\\
s\hat{b}^m_I
&=&
\ep^{mn}\del_n\phi_I
+\del_n(d^n\hat{b}^m_I)
-\del_nd^m\hat{b}^n_I, 
\nonumber 
\\
sc
&=&
\frac{1}{2}\phi^I\phi_I
+d^n\del_nc
+i\ep^{mn}\del_m\bar{f}(\gamma\sigma_n\gamma), 
\nonumber 
\\
s\bar{\alpha}_\alpha
&=&
2\Big((c-c'\bar{\sigma})\alpha'
  \Big)_\alpha
-\phi_I\lambda^I_\alpha
+d^n\del_n\bar{\alpha}_\alpha
+\frac{1}{4}\del_md_n(\sigma^m\sigma^n\bar{\alpha})_\alpha
-\ep_{mn}(\hat{a}^m+\hat{b}^m_I\xi^I)(\sigma^n\gamma)_\alpha
\nonumber 
\\
&&
+(\ep^{mn}a'+\eta^{mn}a)\del_m\bar{f}(\sigma_n\gamma)_\alpha, 
\nonumber 
\\
s\bar{f}
&=&
c'
+d^n\del_n\bar{f}, 
\nonumber 
\\
sc'
&=&
d^n\del_nc'
+i(\gamma\sigma^m\gamma)\del_m\bar{f}, 
\nonumber 
\\
s\bar{d}_{mn}
&=&
V_{mn}-\frac{1}{2}\eta_{mn}(\eta^{kl}V_{kl}), 
\nonumber 
\\
s\bar{\beta}_{m\alpha}
&=&
J_{m\alpha}, 
\nonumber 
\end{eqnarray}
where we denote 
\begin{subequations}
\begin{eqnarray}
V_{mn}
&\equiv&
\frac{1}{2}\del_m\xi^I\del_n\xi_I
+\frac{i}{4}(\lambda^I\sigma_m\del_n\lambda_I)
+\del_m\bar{\phi}^I\del_n\phi_I
\nonumber 
\\
&&
+\hat{a}^k\ep_{km}\del_na'
+\hat{b}^k_I\ep_{km}\del_nb'^I
+\hat{c}^k\ep_{km}\del_nc'
-\frac{i}{4}(\bar{\alpha}\sigma_m\del_n\alpha')
+\frac{i}{4}(\alpha'\sigma_m\del_n\bar{\alpha})
\nonumber 
\\
&&
+\del_m\bar{f}\del_nf
-\bar{d}_{mk}\del_nd^k
+\frac{1}{2}d^k\del_k\bar{d}_{mn}
-\frac{3}{4}(\bar{\beta}_m\del_n\gamma)
+\frac{1}{4}(\gamma\del_m\bar{\beta}_n)
+\frac{i}{2}\ep_{mk}\hat{b}^k_I(\alpha'\sigma_n\lambda^I)
\nonumber 
\\
&&
+(m\leftrightarrow n), 
\\
J_{m\alpha}
&\equiv&
-i\Big(\del_n\xi^I+\ep_{nk}a'\hat{b}^{kI}\Big)
(\sigma^n\sigma_m\lambda_I)_\alpha
-2i\Big(
\ep_{ln}\hat{b}^n_I\del_k\bar{\phi}^I
+\ep_{ln}\hat{c}^n\del_k\bar{f}
\Big)
(\sigma^k\sigma_m\sigma^l\gamma)_\alpha
\nonumber 
\\
&&
-i\Big(
\ep_k{}^n(\hat{a}^k+\hat{b}^k_I\xi^I)
-(\ep^{kn}a'+\eta^{kn}a)\del_k\bar{f}
\Big)
(\sigma_n\sigma_m\alpha')_\alpha
+i\del_na'(\sigma^n\sigma_m\bar{\alpha})_\alpha
\nonumber 
\\
&&
+2i\bar{d}_{mn}(\sigma^n\gamma)_\alpha
+\frac{3}{2}\del_md^n\bar{\beta}_{n\alpha}
+d^n\del_n\bar{\beta}_{m\alpha}.
\end{eqnarray}
\end{subequations}

\vspace*{-6mm}
\noindent
We list below the statistics and the ghost numbers of each fields 
for the convenience:  
\begin{center}
\begin{tabular}{c|c|c}
{\bf fields} 
& \ {\bf statistics} \ 
& \ {\bf ghost numbers} \
\\ 
\hline
\hline
$\bar{f}$     
& bosonic 
& $-2$ 
\\
\hline
$\hat{a}^m$, \ $\hat{b}^m$, \ $c$, \ $c'$, \ $\bar{d}_{mn}$ 
& fermionic 
& \begin{minipage}{6mm}
  \vspace*{6mm} $-1$ 
  \end{minipage}
\\
$\bar{\alpha}_\alpha$, \ $\bar{\beta}^m{}_\alpha$ 
& bosonic 
& 
\\
\hline
$\lambda^I_\alpha$ 
& fermionic 
& \begin{minipage}{6mm}
  \vspace*{6mm} \hspace*{2mm}$0$ 
  \end{minipage}
\\
$\xi^I$, \ $\phi^I$, \ $\bar{\phi}^I$ 
& bosonic
& 
\\
\hline
$a$, \ $a'$, \ $b^I$, \ $b'^I$, \ $\hat{c}^m$, \ $d^m$ 
& fermionic 
& \begin{minipage}{6mm}
  \vspace*{6mm} \hspace*{2mm}$1$ 
  \end{minipage}
\\
$\alpha'_\alpha$, \ $\gamma_\alpha$           
& bosonic 
&  \\
\hline
$f$ 
& bosonic 
& \hspace*{1mm}2
\\
\hline
\end{tabular}
\end{center}

\vspace*{2mm}
\noindent
The ghost fields 
$\Big(a(x)$, $a'(x)$, $\hat{a}^m(x)$, $\alpha'_\alpha(x)$, 
$\bar{\alpha}_\alpha(x)\Big)$, 
$\Big(b^I(x)$, $b'^I(x)$, $\hat{c}^m(x)$, 
$\hat{b}^m(x)$, $c(x)$, $c'(x)$, $f(x)$, $\bar{f}(x)\Big)$ 
and 
$\Big(d^m(x)$, $\bar{d}_{mn}(x)$, $\gamma_\alpha(x)$, 
$\bar{\beta}^m{}_\alpha(x)\Big)$ 
come from the symmetries of the {\UvUa}, 
of the generalized Chern-Simons action  
and of the supergravity, respectively. 

We now present a perturbative analysis of the gauge-fixed action 
(\ref{gauge-fixed_action}) and 
investigate BRST Ward identities at the quantum level. 
Then, we find out that nonlocal anomalous terms obtained from 
loop calculations vanish by imposing a condition 
which determines the critical dimension for this superstring model. 
For the explicit calculation it might be convenient to introduce 
light-cone notations on the world-sheet\footnote{
Our convention of the light-cone coordinates on the world-sheet 
is $x^{\pm}=\frac{1}{\sqrt{2}}(x^0\pm x^1)$. 
The metric tensor $\eta_{mn}$ and Levi-Civit\'a symbol $\ep_{mn}$ 
are given by 
$\eta_{++}=\eta_{--}=0$, $\eta_{+-}=\eta_{-+}=-1$ 
and $\ep_{+-}=-\ep_{-+}=-1$, respectively.
}. 
The gauge-fixed action (\ref{gauge-fixed_action}) is then 
expressed with these notations
\begin{eqnarray}
S_{\rm gauge-fixed} 
=
\int\!\dd^2x
\bigg\{
&&
\del_+\xi^I\del_-\xi_I
+\frac{i}{\sqrt{2}}\lambda^I_+\del_-\lambda_{I+}
+\frac{i}{\sqrt{2}}\lambda^I_-\del_+\lambda_{I-}
+2\del_+\bar{\phi}^I\del_-\phi_I
\nonumber 
\\
&&
+\hat{a}_+\del_-a_+
+\hat{a}_-\del_+a_-
+\hat{b}_{+I}\del_-b_+^I
+\hat{b}_{-I}\del_+b_-^I
\nonumber 
\\
&&
+\hat{c}_+\del_-c_+
+\hat{c}_-\del_+c_-
-i\sqrt{2}\bar{\alpha}_+\del_-\alpha'_+
-i\sqrt{2}\bar{\alpha}_-\del_+\alpha'_-
+2\del_+\bar{f}\del_-f
\nonumber 
\\
&&
-\bar{d}_{++}\del_-d^+
-\bar{d}_{--}\del_+d^-
-\bar{\beta}_+\del_-\gamma_+
+\bar{\beta}_-\del_+\gamma_-
\nonumber 
\\
&&
+(a_++a_-)(\hat{b}_{+I}\del_-\xi^I
          +\hat{b}_{-I}\del_+\xi^I)
+\hat{b}_{+I}\hat{c}_-\phi^I
-\hat{b}_{-I}\hat{c}_+\phi^I
\nonumber 
\\
&&
+\Big(f+\frac{1}{2}a_+a_-\Big)\hat{b}_{+I}\hat{b}_-^I
+i\sqrt{2}\hat{b}_{+I}\alpha'_-\lambda^I_-
-i\sqrt{2}\hat{b}_{-I}\alpha'_+\lambda^I_+
\bigg\}, 
\label{gauge-fixed_action_light-cone}
\end{eqnarray}
where we denote 
\begin{eqnarray*}
\lambda^I_+
&\equiv&
\lambda^I_1, 
\hspace*{15mm}
\lambda^I_-
\equiv
\lambda^I_2, 
\\
\alpha'_+
&\equiv&
\alpha'_1, 
\hspace*{14.5mm}
\alpha'_-
\equiv\alpha'_2, 
\hspace*{15mm}
\bar{\alpha}_+
\equiv
\bar{\alpha}_1, 
\hspace*{15mm}  
\bar{\alpha}_-
\equiv
\bar{\alpha}_2,
\\
\gamma_+
&\equiv&
\gamma_2, 
\hspace*{15.5mm}
\gamma_-
\equiv
\gamma_1, 
\hspace*{15.7mm}
\bar{\beta}_+
\equiv
\bar{\beta}_{+\alpha=1}, 
\hspace*{9mm}
\bar{\beta}_-
\equiv
\bar{\beta}_{-\alpha=2}, 
\\
a_{\pm}
&\equiv&
a\mp a', 
\\
b^I_{\pm}
&\equiv&
b^I\mp b'^I, 
\\
c_{\pm}
&\equiv&
c\mp c'. 
\end{eqnarray*}
Propagators are derived by taking inverses of bilinear parts 
in the action (\ref{gauge-fixed_action_light-cone}), 
\begin{eqnarray*}
\langle\xi^I(x)\xi^J(y)\rangle_0
&=&
\langle\bar{\phi}^I(x)\phi^J(y)\rangle_0
\\
&=&
\int\!\frac{\dd^2p}{i(2\pi)^2} \ 
\frac{1}{p^2+i\epsilon} \ 
e^{-ip(x-y)}
\eta^{IJ}, 
\\
\langle\lambda^I_{\pm}(x)\lambda^J_{\pm}(y)\rangle_0
&=&
\int\!\frac{\dd^2p}{i(2\pi)^2} \
\frac{-\sqrt{2}p^\mp}{p^2+i\epsilon} \
e^{-ip(x-y)}
\eta^{IJ}, 
\\
\langle\hat{a}_\pm(x)a_\pm(y)\rangle_0
&=&
\langle\hat{c}_\pm(x)c_\pm(y)\rangle_0
=-\langle\bar{d}_{\pm\pm}(x)d^\pm(y)\rangle_0
\\
&=&
\int\!\frac{\dd^2p}{i(2\pi)^2} \
\frac{-2ip^\mp}{p^2+i\epsilon} \
e^{-ip(x-y)}, 
\\
\langle\hat{b}^I_\pm(x)b^I_\pm(y)\rangle_0
&=&
\int\!\frac{\dd^2p}{i(2\pi)^2} \
\frac{-2ip^\mp}{p^2+i\epsilon} \
e^{-ip(x-y)}\eta^{IJ},  
\\
\langle\bar{\alpha}_\pm(x)\alpha'_\pm(y)\rangle_0
&=&
\int\!\frac{\dd^2p}{i(2\pi)^2} \
\frac{\sqrt{2}p^\mp}{p^2+i\epsilon} \
e^{-ip(x-y)}, 
\\
\langle\bar{\beta}_\pm(x)\gamma_\pm(y)\rangle_0
&=&
\int\!\frac{\dd^2p}{i(2\pi)^2} \
\frac{\mp2ip^\mp}{p^2+i\epsilon} \
e^{-ip(x-y)}, 
\\
\langle\bar{f}(x)f(y)\rangle_0
&=&
\int\!\frac{\dd^2p}{i(2\pi)^2} \
\frac{1}{p^2+i\epsilon} \
e^{-ip(x-y)}. 
\end{eqnarray*}

Now let us consider the following two-point functions,  
\begin{subequations}
\begin{eqnarray}
A(p)_{++}
&\equiv&
\int\!\frac{\dd^2x}{i(2\pi)^2} \ 
\langle V_{++}(x)V_{++}(0)\rangle \ 
e^{ipx}, 
\label{two-point_V++V++}\\
B(p)_{++}
&\equiv&
\int\!\frac{\dd^2x}{i(2\pi)^2} \ 
\langle J_+(x)J_+(0)\rangle \
e^{ipx}, 
\label{two-point_J+J+}
\end{eqnarray}
\end{subequations}

\vspace*{-6mm}
\noindent
where we denote $J_+(x)\equiv J_{+\alpha=1}(x)$. 
Here we mention 
that the two-point functions (\ref{two-point_V++V++}) 
and (\ref{two-point_J+J+}) should vanish from the point of view of the 
BRST symmetries 
$V_{++}(x)=s\bar{d}_{++}(x)$ and $J_+(x)=s\bar{\beta}_+(x)$. 
By estimating all of the contributions arising from 
$(\xi^I, \xi_I)$, $(\lambda^I_+, \lambda_{I+})$, 
$(\bar{\phi}^I, \phi_I)$, $(\hat{a}_+, a_+)$, 
$(\hat{b}_{+I}, b^I_+)$, $(\hat{c}_+, c_+)$, 
$(\bar{\alpha}_+, \alpha'_+)$, $(\bar{f}, f)$, 
$(\bar{d}_{++}, d^+)$ and $(\bar{\beta}_+, \gamma_+)$, 
we can obtain the following result for the two-point function 
(\ref{two-point_V++V++}) up to one-loop order, 
\begin{subequations}
\begin{eqnarray}
A(p)_{++}
&=&
\frac{1}{48\pi^3}
\bigg(D+\frac{1}{2}D+2D-2-2D-2-1+2-26+11\bigg)\frac{(p^-)^3}{p^+} 
\nonumber 
\\
&=& 
\frac{D-12}{32\pi^3}\frac{(p^-)^3}{p^+}.
\end{eqnarray}
Furthermore we can obtain the following result 
for (\ref{two-point_J+J+}), 
\begin{eqnarray}
B(p)_{++}
&=&
\frac{1}{4\sqrt{2}\pi^3}\Big(D-2-10\Big)\frac{(p^-)^2}{p^+}
\nonumber 
\\
&=& 
\frac{D-12}{4\sqrt{2}\pi^3}\frac{(p^-)^2}{p^+}. 
\end{eqnarray}
\end{subequations}

\vspace*{-6mm}
\noindent
In a similar way we can find 
\begin{subequations}
\begin{eqnarray}
A(p)_{--}
&\equiv&
\int\!\frac{\dd^2x}{i(2\pi)^2} \
\langle V_{--}(x)V_{--}(0)\rangle \
e^{ipx}
=
\frac{D-12}{32\pi^3}\frac{(p^+)^3}{p^-}, 
\\
B(p)_{--}
&\equiv&
\int\!\frac{\dd^2x}{i(2\pi)^2} \
\langle J_-(x)J_-(0)\rangle \ 
e^{ipx}
=
\frac{D-12}{4\sqrt{2}\pi^3}\frac{(p^+)^2}{p^-}, 
\end{eqnarray}
\end{subequations}

\vspace*{-6mm}
\noindent
where $J_-(x)\equiv J_{-\alpha=2}(x)$. 
Although we need to check the other two-point functions, 
{\it i.e.}\ $A(p)_{+-}$ and $B(p)_{+-}$, 
these two-point functions are actually divergent.      
However, this divergence can be absorbed adding suitable 
local counter terms to the action. 
Therefore, we conclude 
that the BRST anomalies vanish if and only if 
\begin{equation}
D=12.
\end{equation}
%
 
\section{Quantization in the light-cone gauge formulation}
\setcounter{equation}{0}
\setcounter{footnote}{0}

In this section we carry out the quantization of 
the classical action (\ref{classical_super_action_04}) in the 
light-cone gauge and derive the same critical dimension of 
the model in the covariant gauge.  
In addition, we mention a mass-shell relation of the model.  

In the canonical formulation it might be useful to introduce 
the following new variables for the 
inverse zweibein fields $e_a{}^m(x)$,  
\begin{equation}
e_{\pm}{}^m
\equiv
\frac{e}{2}\Big(e_0{}^m\pm e_1{}^m\Big). 
\end{equation}

According to the ordinary Dirac's procedure\footnote{
Our convention of the generalized Poisson bracket is given in Appendix C. 
We take the independent variables $\Phi^A(x)$ as 
$\xi^I(x)$, $\lambda^I_\alpha(x)$, $\phi^I(x)$, $\bar{\phi}^I(x)$, 
$A_m(x)$, $B_m^I(x)$, $C_{01}(x)$, $\psi_\alpha(x)$, 
$e_\pm{}^m(x)$ and $\chi_{m\alpha}(x)$. 
 },
we introduce canonical momenta defined by 
$P_{\Phi^A}(x) \equiv \delta_{\rm L}S/\delta(\del_0\Phi^A(x))$ 
corresponding to fields $\Phi^A(x)$\footnote{
For the spinor fields, we denote their two components as 
$\lambda^I_+(x)\equiv\lambda^I_1(x)$ and 
$\lambda^I_-(x)\equiv\lambda^I_2(x)$. 
The other spinor fields also obey the same conventions. 
 },
%
\begin{eqnarray}
{P_\xi}_I
&=&
-e
\Big(
g^{0m}\del_m\xi_I
-\tilde{A}^0\phi_I
-\frac{i}{2}(\lambda_I\sigma^m\sigma^a\chi_m)e_a{}^0
\Big), 
\nonumber 
\\
{P_\phi}_I
&=&
-e
\Big(
g^{0m}\del_m\bar{\phi}_I
-\tilde{B}^0_I
\Big), 
\\
{P_{\bar{\phi}}}_I
&=&
-eg^{0m}\del_m\phi_I, 
\nonumber  
\end{eqnarray}
and 
\begin{subequations}
\begin{equation}
{P_\lambda}_{I\pm}
=\frac{\delta_{\rm L}S}{\delta(\del_0\lambda^I_\pm)}
=-ie_\mp{}^0\lambda_{I\pm}, 
\quad\Rightarrow\quad
\theta_{I\pm}
\equiv
{P_\lambda}_{I\pm}
+ie_{\mp}{}^0\lambda_{I\pm}
=0, 
\label{primary_constraint_01}
\end{equation}
\begin{equation}
{P_A}^m={P_B}_I^m=P_{C_{01}}={P_\psi}_\pm
={P_{e_\pm}}_m={P_\chi}^m_\pm
=0. 
\label{primary_constraint_02}
\end{equation}
\end{subequations}

\vspace*{-6mm}
\noindent
The Poisson brackets are defined by 
\begin{equation}
\begin{array}{rcl}
\{\xi^I, {P_\xi}_J\} 
&=&
\{\phi^I, {P_\phi}_J\}
=
\{\bar{\phi}^I, {P_{\bar{\phi}}}_J\}
=\delta^I_J, 
\nonumber 
\\
\{\lambda^I_\pm, {P_\lambda}_{J\pm}\}
&=&
\{{P_\lambda}_{J\pm}, \lambda^I_\pm\}
=-\delta^I_J, 
\nonumber 
\\
\{A_m, {P_A}^n\}
&=&
\delta^n_m, 
\nonumber 
\\
\{B_m^I, {P_B}^n_J\}
&=&
\delta_m^n\delta^I_J, 
\nonumber 
\\
\{C_{01}, P_{C_{01}}\}
&=&
1, 
\nonumber 
\\
\{\psi_\pm, {P_\psi}_\pm\}
&=&
\{{P_\psi}_\pm, \psi_\pm\}
=-1, 
\nonumber 
\\
\{e_\pm{}^m, {P_{e_\pm}}_n\}
&=&
\delta^m_n, 
\nonumber 
\\
\{\chi_{m\pm}, {P_\chi}^n_\pm\}
&=&
\{{P_\chi}^n_\pm, \chi_{m\pm}\}
=-\delta^n_m.  
\nonumber 
\end{array}
\end{equation}
The relations (\ref{primary_constraint_01}) and 
(\ref{primary_constraint_02}) give primary constraints. 
By introducing Lagrange multiplier fields $\rho_i(x)$ for 
primary constraints $\varphi^i(x)$, 
canonical Hamiltonian can be written 
in terms of the phase space variables 
as 
\begin{eqnarray}
H
&=&
\int\!\dd x^1
\bigg\{
\del_0\Phi^AP_{\Phi^A}-{\cal L}+\rho_i\varphi^i
\bigg\}
\nonumber 
\\
&=&
\int\!\dd x^1
\Bigg\{
-\frac{e_-{}^1}{e_-{}^0}
\bigg(
\frac{1}{4}
\Big(
{P_\xi}^I+A_1\phi^I+\del_1\xi^I
\Big)
\Big(
P_{\xi I}+A_1\phi_I+\del_1\xi_I
\Big)
-{P_{\lambda}}_{I+}\del_1\lambda^I_+
\nonumber 
\\
&&\hspace*{27mm}
+\Big(
{P_{\bar{\phi}}}^I+\del_1\phi^I
\Big)
\Big(
P_{\phi I}+B_{1I}+\del_1\bar{\phi}_I
\Big)
-\frac{i}{2}
\Big({P_{\xi}}^I+A_1\phi^I+\del_1\xi^I
\Big)\lambda_{I+}\chi_{1-}
\bigg)
\nonumber 
\\
&&\hspace*{13.5mm}
+ \ \frac{e_+{}^1}{e_+{}^0}
\bigg(
\frac{1}{4}
\Big({P_\xi}^I+A_1\phi^I-\del_1\xi^I
\Big)
\Big(
P_{\xi I}+A_1\phi_I-\del_1\xi_I
\Big)
+{P_{\lambda}}_{I-}\del_1\lambda^I_-
\nonumber 
\\
&&\hspace*{27mm}
+
\Big(
{P_{\bar{\phi}}}^I-\del_1\phi^I
\Big)
\Big(
P_{\phi I}+B_{1I}-\del_1\bar{\phi}_I
\Big)
-\frac{i}{2}
\Big(
{P_{\xi}}^I+A_1\phi^I-\del_1\xi^I
\Big)
\lambda_{I-}\chi_{1+}
\bigg)
\nonumber 
\\
&&\hspace*{13.5mm}
- \ A_0\phi_I\del_1\xi^I
-B_0^I\del_1\phi_I
-\frac{1}{2}C_{01}\phi^I\phi_I
-2i
\Big(e_+{}^0e_-{}^1-e_-{}^0e_+{}^1
\Big)\phi_I
\Big(
\psi_-\lambda^I_+-\psi_+\lambda^I_-
\Big)
\nonumber 
\\
&&\hspace*{13.5mm}
+ \ \frac{i}{2}
\Big(
{P_\xi}^I+A_1\phi^I+\del_1\xi^I
\Big)
\lambda_{I+}\chi_{0-}
-\frac{i}{2}
\Big({P_\xi}^I+A_1\phi^I-\del_1\xi^I
\Big)
\lambda_{I-}\chi_{0+}
\nonumber 
\\
&&\hspace*{13.5mm}
- \ \theta_{I+}{\rho_\lambda}^I_+
-\theta_{I-}{\rho_\lambda}^I_-
\nonumber 
\\
&&\hspace*{13.5mm}
+ \ {\rho_A}_m{P_A}^m
+{\rho_B}^I_m{P_B}_I^m
+\rho_{C_{01}}P_{C_{01}}
+\rho_{\psi +}P_{\psi +}
+\rho_{\psi -}P_{\psi -}
\nonumber 
\\
&&\hspace*{13.5mm}
+ \ {\rho_{e_+}}^m{P_{e_+}}_m
+{\rho_{e_-}}^m{P_{e_-}}_m
+{\rho_{\chi_+}}_m{P_\chi}_+^m
+{\rho_{\chi_-}}_m{P_\chi}_-^m
\Bigg\}, 
\label{hamiltonian_00}
\end{eqnarray}
where we redefine the multiplier ${\rho_\lambda}^I_\pm(x)$ for 
the primary constraint (\ref{primary_constraint_01}) as 
$$
{\rho_\lambda}^I_\pm
+\del_0\lambda^I_\pm
+\frac{e_\mp{}^1}{e_\mp{}^0}\del_1\lambda^I_\pm
\rightarrow
{\rho_\lambda}^I_\pm. 
$$
A consistency check of the primary constraints 
(\ref{primary_constraint_01}) and (\ref{primary_constraint_02}) yields 
a set of secondary constraints
\begin{subequations}
\begin{eqnarray}
&&
\frac{1}{4}
\Big({P_\xi}^I\pm\del_1\xi^I
\Big)
\Big(
{P_\xi}_I\pm\del_1\xi_I
\Big)
\mp{P_\lambda}_{I\pm}\del_1\lambda^I_\pm
=0, 
\label{secondary_constraint_01}\\
&&
\Big(
{P_\xi}^I\pm\del_1\xi^I
\Big)
\lambda_{I\pm}
=0, 
\label{secondary_constraint_02}\\
&&
\phi_I\lambda^I_\pm
=0, 
\label{secondary_constraint_03}\\
&&
\phi_I\del_1\xi^I
=0, 
\label{secondary_constraint_04}\\
&&
\phi_I{P_\xi}^I
=0, 
\label{secondary_constraint_05}\\
&&
\del_1\phi^I
=0, 
\label{secondary_constraint_06}\\
&&
{P_{\bar{\phi}}}^I
=0, 
\label{secondary_constraint_07}\\
&&
\frac{1}{2}\phi^I\phi_I
=0, 
\label{secondary_constraint_08}
\end{eqnarray}
\end{subequations}

\vspace*{-6mm}
\noindent
and these secondary constraints 
give no other relations\footnote{
The consistency checks for some of the constraints 
determine the multipliers 
as functionals of the canonical variables. 
However, these explicit forms are not important because 
these contributions to the canonical Hamiltonian might be ignored 
if we introduce the Dirac brackets. 
}. 
 
Let us specify the algebraic structure of the constraints. 
The constraints (\ref{primary_constraint_01}) have non-vanishing 
Poisson brackets with each other
\begin{equation}
\begin{array}{rcl}
\{\theta^I_\pm, \theta^J_\pm\}
&=&
-2ie_\mp{}^0\eta^{IJ},
\\
\{\theta^I_\pm, \theta^J_\mp\}
&=&
0,  
\end{array}
\end{equation}
and are therefore second class. 
Similarly, the constraints ${P_{e_\pm}}_0(x)=0$ and  
(\ref{secondary_constraint_01})-(\ref{secondary_constraint_03}) 
have non-vanishing Poisson brackets 
with the constraints (\ref{primary_constraint_01}) and hence are also 
second class. 
However, we can take the following linear combinations 
for these constraints,   
\begin{subequations}
\begin{eqnarray}
&&
{P_{e_\pm}}_0=0
\nonumber 
\\
&&
\hspace*{5mm} 
\rightarrow \ 
{P_{e_\pm}}_0
+\frac{1}{2e_\pm{}^0}\theta_{I\mp}\lambda^I_\mp=0,  
\label{modified_constraint_01}\\
&&
\frac{1}{4}
\Big({P_\xi}^I\pm\del_1\xi^I
\Big)
\Big(
{P_\xi}_I\pm\del_1\xi_I
\Big)
\mp{P_\lambda}_{I\pm}\del_1\lambda^I_\pm
=0 
\nonumber 
\\
&&
\hspace*{5mm} 
\rightarrow \ 
\frac{1}{4}
\Big(
{P_\xi}^I\pm\del_1\xi^I
\Big)
\Big(
{P_\xi}_I\pm\del_1\xi_I
\Big)
\mp
\Big(
{P_\lambda}_{I\pm}-\frac{1}{2}\theta_{I\pm}
\Big)
\del_1
\Big(
\lambda^I_\pm+\frac{i}{2e_\mp{}^0}\theta^I_\pm
\Big)
=0, 
\label{modified_constraint_02}\\
&&
\Big(
{P_\xi}^I\pm\del_1\xi^I
\Big)
\lambda_{I\pm}
=0 
\nonumber 
\\
&&
\hspace*{5mm}
\rightarrow \
\Big(
{P_\xi}^I\pm\del_1\xi^I
\Big)
\Big(
\lambda_{I\pm}+\frac{i}{2e_\mp{}^0}\theta_{I\pm}
\Big)
=0, 
\label{modified_constraint_03}\\
&&
\phi_I\lambda^I_\pm=0 
\nonumber 
\\
&&
\hspace*{5mm}
\rightarrow \
\phi_I\Big(\lambda^I_\pm+\frac{i}{2e_\mp{}^0}\theta^I_\pm\Big)
=0, 
\label{modified_constraint_04}
\end{eqnarray}
\end{subequations}

\vspace*{-6mm}
\noindent
so that the above constraints 
(\ref{modified_constraint_01})-(\ref{modified_constraint_04}) 
turn out to have vanishing Poisson brackets with 
the constraints (\ref{primary_constraint_01}). 
Then, we can separate all of the constraints into second class 
(\ref{primary_constraint_01}) and 
first class (\ref{primary_constraint_02}), 
(\ref{modified_constraint_01})-(\ref{modified_constraint_04}) 
and (\ref{secondary_constraint_04})-(\ref{secondary_constraint_08}).  
For the second class constraints, 
we introduce the following Dirac brackets instead of the Poisson 
brackets,
%
\renewcommand{\arraystretch}{2.0}
%
\begin{equation}
\begin{array}{rcl}
\{{P_{e_\mp}}_0, \lambda^I_\pm\}
&=&
\displaystyle
\frac{\lambda^I_\pm}{2e_\mp{}^0}, 
\\
\{\lambda^I_\pm, \lambda^J_\pm\}
&=&
\displaystyle
-\frac{i}{2e_\mp{}^0}\eta^{IJ}, 
\end{array}
\end{equation}    
%
\renewcommand{\arraystretch}{1.6}
%

\vspace*{-4mm}
\noindent
and we set the second class constraints (\ref{primary_constraint_01}) 
as identities. 
Then, the first class constraints 
(\ref{modified_constraint_01})-(\ref{modified_constraint_04}) are simply
replaced to the original ones.  

Now we investigate the dynamics of the model defined by 
the canonical Hamiltonian (\ref{hamiltonian_00}) with the first 
class constraints (\ref{primary_constraint_02}) and 
(\ref{secondary_constraint_01})-(\ref{secondary_constraint_08}).  
Imposing noncovariant gauge fixing conditions, 
we explicitly solve the constraints to some of the variables 
from the equations of motion\footnote{
Hereafter we use the conventions of the world-sheet coordinates 
as $x^0\equiv\tau$ and $x^1\equiv\sigma$. 
We also parameterize the spatial coordinate as $0\le\sigma\le 2\pi$.
}.  

We begin by considering conditions for the scalar field 
$\phi^I(\tau, \sigma)$. 
We find it convenient to introduce Fourier mode expansions of the 
canonical pair 
$\Big(\phi^I(\tau, \sigma), {P_\phi}_J(\tau, \sigma)\Big)$, 
\begin{equation}
\begin{array}{rcl}
\phi^I(\tau, \sigma)
&=& \displaystyle 
    \phi^I(\tau)+\frac{1}{\sqrt{2\pi}} \sum_{m \ne 0}
                   \phi^I_m(\tau) e^{im\sigma}, \\
{P_\phi}_I(\tau, \sigma)
&=& \displaystyle
   \frac{{p_\phi}_I(\tau)}{2\pi}
   + \frac{1}{\sqrt{2\pi}} \sum_{m \ne 0}
     {p_\phi}_{Im}(\tau) e^{im\sigma}.
\end{array}
\end{equation}
Then, Poisson brackets are written by
\begin{eqnarray}
\{\phi^I(\tau), {p_\phi}_J(\tau)\}
&=& \delta^I_J, \nonumber \\
\{\phi_m^I(\tau), {p_\phi}_{Jn}(\tau)\}
&=& \delta^I_J\delta_{m+n}, \\
\hbox{otherwise} &=& 0. \nonumber
\end{eqnarray}
In terms of the Fourier modes, 
the constraint (\ref{secondary_constraint_06}) is
equivalent to $\phi^I_m(\tau)=0$.
On the other hand,
the equation of motion for $\phi^I(\tau, \sigma)$
on the constraint surface is
$\del_\tau\phi^I(\tau, \sigma) = 0$.
Together with the constraint $\phi_m^I(\tau)=0$,
we then set the configuration of the scalar field
as $\phi^I(\tau, \sigma)=\phi^I(\tau)=\phi^I(=\mbox{const.})$.

We first impose orthonormal gauge fixing conditions 
$e_a{}^m(\tau, \sigma)=\delta_a{}^m$ for the constraints 
${P_{e_\pm}}_m(\tau, \sigma)=0$, by using the gauge parameters 
$k^n(\tau, \sigma)$, $l(\tau, \sigma)$ and $s(\tau, \sigma)$ for 
the general coordinate, the local Lorentz and the Weyl scaling 
transformations, respectively. 
In addition, we can also adopt $\chi_{m\pm}(\tau, \sigma)=0$ 
as a gauge fixing condition for the constraint  
${P_\chi}^m_\pm(\tau, \sigma)=0$, by using the 
gauge parameters $\zeta_\pm(\tau, \sigma)$ 
and $\check{s}_\pm(\tau, \sigma)$ for the local supersymmetry and 
the super-Weyl scaling transformations, respectively. 
The bosonic {\UvUa} gauge parameters $v(\tau, \sigma)$, $v'(\tau, \sigma)$ 
and the global parameter $\alpha_i$ can fix to be 
$A_m(\tau, \sigma)=0$ for the constraint  
${P_A}^m(\tau, \sigma)=0$, 
while the fermionic gauge parameters $\mu'_\pm(\tau, \sigma)$ 
can fix to be $\psi_\pm(\tau, \sigma)=0$ for 
the constraint ${P_\psi}_\pm(\tau, \sigma)=0$. 
However, this is not the end of the story.  
The system still has residual symmetries concerned with 
these gauge parameters~\cite{tw1}.  
As we will explain below, taking these symmetries into account, 
we can adopt the following gauge fixing conditions
on ``two'' light-cone coordinates\footnote{
From the definition of the metric (\ref{FlatmetricDefinitionGravity}),
we denote the light-cone coordinates of the 
background spacetime as 
$x^I =(x^+, x^-, x^i, x^{\hat{+}}, x^{\hat{-}})$,
where $x^{\pm}\equiv\frac{1}{\sqrt{2}}(x^0\pm x^{D-3})$ and 
$x^{\hat{\pm}}\equiv\frac{1}{\sqrt{2}}(x^{\hat{0}}\pm x^{\hat{1}})$ and
the index $i$ runs through $1$, $2$, \ldots, $D-4$.
} 
of the background spacetime 
within the gauge $e_a{}^m(\tau, \sigma)=\delta_a{}^m$ and 
$\chi_{m\pm}(\tau, \sigma)=A_m(\tau, \sigma)=\psi_\pm(\tau, \sigma)=0$, 
%
\renewcommand{\arraystretch}{2.2}
%
\begin{equation}
\begin{array}{rclcrcl}
\xi^+ (\tau, \sigma)
&=& \displaystyle\frac{p^+}{2\pi} \tau, &\qquad&
P_\xi^+ (\tau, \sigma)
&=& \displaystyle\frac{p^+}{2\pi}, \\
\xi^{\hat{+}} (\tau, \sigma)
&=& \displaystyle\frac{p^{\hat{+}}}{2\pi} \tau, &\qquad&
P_\xi^{\hat{+}} (\tau, \sigma)
&=& \displaystyle\frac{p^{\hat{+}}}{2\pi}, \\
\lambda^+_\pm(\tau, \sigma)
&=& 0, && \\
\lambda^{\hat{+}}_\pm(\tau, \sigma)
&=& 0, && 
\end{array}
\label{light-cone_gauge_fixing_1}
\end{equation}
%
\renewcommand{\arraystretch}{1.6}

\vspace*{-4mm}
\noindent
where $p^+$ and $p^{\hat{+}}$ are light-cone components of
the center of mass momenta. 
Therefore we can eliminate ``two'' unphysical components of
the coordinates of the background spacetime and their superpartners. 
Indeed the gauge fixing conditions (\ref{light-cone_gauge_fixing_1}) 
correspond to ones for the first class constraints
(\ref{secondary_constraint_01})-(\ref{secondary_constraint_05}). 

In order to show how these conditions (\ref{light-cone_gauge_fixing_1}) 
are accomplished, it might be useful to introduce Fourier mode 
expansions. 
In the gauge $e_a{}^m(\tau, \sigma)=\delta_a{}^m$ and 
$\chi_{m\pm}(\tau, \sigma)=A_m(\tau, \sigma)=\psi_\pm(\tau, \sigma)=0$, 
the dynamics of the coordinates $\xi^I(\tau, \sigma)$ and 
$\lambda^I_\pm(\tau, \sigma)$ turns out to be given by free wave
equations and free Dirac equations with some constraints. 
For the bosonic coordinates, the solutions of the equation of motion 
are  
\begin{equation}
\begin{array}{rcl}
\xi^I(\tau, \sigma) 
&=& \displaystyle 
    x^I + \frac{p^I}{2\pi} \tau
    +\frac{i}{2\sqrt{\pi}} \sum_{m \ne 0}
     \frac{1}{m}\Big(\alpha^I_m e^{-im(\tau-\sigma)}
                     + \tilde{\alpha}^I_m e^{-im(\tau+\sigma)}
                \Big), \\
{P_\xi}^I(\tau, \sigma)
&=& \displaystyle 
    \frac{p^I}{2\pi}
    +\frac{1}{2\sqrt{\pi}} \sum_{m \ne 0}
    \Big(\alpha^I_me^{-im(\tau-\sigma)}
         + \tilde{\alpha}^I_m e^{-im(\tau+\sigma)}
    \Big), 
\end{array}
\end{equation}
and 
Poisson brackets are given by 
\begin{eqnarray}
\{x^I, p^J\} 
&=& \eta^{IJ}, \nonumber \\
\{\alpha^I_m, \alpha^J_n\} 
&=& \{\tilde{\alpha}^I_m, \tilde{\alpha}^J_n\}
   = - i m \eta^{IJ} \delta_{m+n}, \\
\hbox{otherwise}
&=& 0. \nonumber
\end{eqnarray}
For the fermionic coordinates, the solutions depend on 
their boundary conditions {\it i.e.}\ periodic (Ramond model) and 
antiperiodic (Neveu-Schwarz model),  
%
\renewcommand{\arraystretch}{2.2}
%
\begin{equation}
\begin{array}{rcl} 
\lambda^I_-(\tau, \sigma)
&=&
\displaystyle
\frac{1}{\sqrt{2\pi}}
\!\sum_{r\in{\bf Z}+a}\!b^I_re^{-ir(\tau-\sigma)}, 
\\
\lambda^I_+(\tau, \sigma)
&=&
\displaystyle
\frac{1}{\sqrt{2\pi}}
\!\sum_{r\in{\bf Z}+a}\!\tilde{b}^I_re^{-ir(\tau+\sigma)}, 
\end{array}
\end{equation}
%
\renewcommand{\arraystretch}{1.6}
%

\vspace*{-3mm}
\noindent
where $a=0$ for Ramond and $a=1/2$ for Neveu-Schwarz model, respectively.  
Their Poisson brackets are given by 
\begin{equation}
\begin{array}{rcl}
\{b^I_r, b^J_s\}
&=&
\{\tilde{b}^I_r, \tilde{b}^J_s\}
=
-i\eta^{IJ}\delta_{r+s}, 
\\
\{b^I_r, \tilde{b}^J_s\}
&=&
0.
\end{array}
\end{equation}
In terms of the Fourier modes,
the constraints 
(\ref{secondary_constraint_01})-(\ref{secondary_constraint_05}) 
are equivalent to
\begin{eqnarray}
&&
L_m 
=
L_m^{(\alpha)}
+L_m^{(b)}
=0, 
\nonumber 
\\
&&
\tilde{L}_m
=
\tilde{L}_m^{(\alpha)}
+\tilde{L}_m^{(b)}
=0,
\nonumber  
\\
&&
G_r
=
\tilde{G}_r
=
0,
\\
&&
\phi_I \alpha_m^I 
=
\phi_I \tilde{\alpha}_m^I = 0,
\nonumber 
\\
&&
\phi_Ib^I_r
=
\phi_I\tilde{b}^I_r
=0. 
\nonumber 
\end{eqnarray}
In the above eqs., we define the super-Virasoro generators as 
$$
\begin{array}{rclcrcl}
L_m^{(\alpha)} 
&\equiv& 
\displaystyle
\frac{1}{2}\sum_n\alpha_{-n}^I{\alpha_I}_{m+n}, 
&\qquad&
\tilde{L}_m^{(\alpha)}
&\equiv&
\displaystyle
\frac{1}{2}\sum_n{\tilde{\alpha}}_{-n}^I\tilde{\alpha}_{Im+n},
\\
L_m^{(b)}
&\equiv&
\displaystyle
\frac{1}{2}
\!\sum_{r\in{\bf Z}+a}\!\!\Big(r+\frac{m}{2}\Big)
b^I_{-r}b_{Im+r}, 
&\qquad&
\tilde{L}_m^{(b)}
&\equiv&
\displaystyle
\frac{1}{2}
\!\sum_{r\in{\bf Z}+a}\!\!\Big(r+\frac{m}{2}\Big)
\tilde{b}^I_{-r}\tilde{b}_{Im+r},
\\
G_r
&\equiv&
\displaystyle
\sum_n\alpha^I_{-n}b_{Ir+n}, 
&\qquad&
\tilde{G}_r
&\equiv&
\displaystyle
\sum_n\tilde{\alpha}^I_{-n}\tilde{b}_{Ir+n}, 
\end{array}
$$
where we denote 
$\alpha_0^I=\tilde{\alpha}^I_0\equiv p^I/(2\sqrt{\pi})$.
The gauge fixing conditions (\ref{light-cone_gauge_fixing_1}) are 
equivalent to 
\begin{subequations}
\begin{equation}
\begin{array}{l}
x^+
=x^{\hat{+}}
=0, 
\\
\alpha^+_m
=\alpha^{\hat{+}}_m
=\tilde{\alpha}^+_m
=\tilde{\alpha}^{\hat{+}}_m
=0, 
\quad (m\ne 0), 
\end{array}
\label{light-cone_gauge_fixing_boson}
\end{equation}
and 
\begin{equation}
b^+_r
=b^{\hat{+}}_r
=\tilde{b}^+_r
=\tilde{b}^{\hat{+}}_r
=0.  
\label{light-cone_gauge_fixing_fermion}
\end{equation}
\end{subequations}

\vspace*{-6mm}

Now let us explain the procedure to get the gauge fixing 
conditions 
(\ref{light-cone_gauge_fixing_boson}) and 
(\ref{light-cone_gauge_fixing_fermion}). 
For the fermionic sectors, within the super-orthonormal gauge,   
we can change the fermionic coordinates 
$\lambda^I_\pm(\tau, \sigma)$  
with the gauge parameters $\zeta_\pm(\tau, \sigma)$ 
provided that conditions
$\del_\tau\zeta_+(\tau, \sigma)$ $=-\del_\sigma\zeta_+(\tau, \sigma)$ 
and $\del_\tau\zeta_-(\tau, \sigma)=\del_\sigma\zeta_-(\tau, \sigma)$
are satisfied. 
Here we take the following forms which realize these conditions, 
\begin{eqnarray*}
\zeta_+(\tau, \sigma)
&=&
\frac{1}{\sqrt{2\pi}}
\!\sum_{r\in {\bf Z}+a}\!\zeta_re^{-ir(\tau-\sigma)}, 
\\
\zeta_-(\tau, \sigma)
&=&
\frac{1}{\sqrt{2\pi}}
\!\sum_{r\in{\bf Z}+a}\!\tilde{\zeta}_re^{-ir(\tau+\sigma)}. 
\end{eqnarray*}
In analogy to the bosonic {\UvUa} string case~\cite{tw1},  
the fermionic {\UvUa} gauge 
parameters $\mu'_\pm(\tau, \sigma)$ can
be also used to change the
fermionic coordinates within the gauge $\psi_\pm(\tau, \sigma)=0$  
provided that conditions 
$\del_\tau\mu'_+(\tau, \sigma)=\del_\sigma\mu'_+(\tau, \sigma)$
and $\del_\tau\mu'_-(\tau, \sigma) =-\del_\sigma\mu'_-(\tau, \sigma)$
are satisfied. 
We take the following forms for $\mu'_\pm(\tau, \sigma)$ 
to realize these conditions, 
\begin{eqnarray*}
\mu'_-(\tau, \sigma)
&=&
\frac{1}{\sqrt{2\pi}}
\!\sum_{r\in{\bf Z}+a}\!\mu'_re^{-ir(\tau-\sigma)}, 
\\
\mu'_+(\tau, \sigma)
&=&
\frac{1}{\sqrt{2\pi}}
\!\sum_{r\in{\bf Z}+a}\!\tilde{\mu}'_re^{-ir(\tau+\sigma)}. 
\end{eqnarray*}
The gauge transformations corresponding to these parameters are 
consistent with the equations of motion for the fermionic 
coordinates $\lambda^I_\pm(\tau, \sigma)$, 
because, in terms of the Fourier modes, the gauge transformations 
are given by 
%
\renewcommand{\arraystretch}{2.2}
%
\begin{equation}
\begin{array}{rcl}
\delta b^I_r
&=&
\displaystyle
\frac{1}{\sqrt{\pi}}\sum_n\zeta_{r-n}\alpha^I_n
+\mu'_r\phi^I, 
\\
\delta\tilde{b}^I_r
&=&
\displaystyle
-\frac{1}{\sqrt{\pi}}\sum_n\tilde{\zeta}_{r-n}\tilde{\alpha}^I_n
+\tilde{\mu}'_r\phi^I.
\end{array}
\label{gauge_trans_fourier_1}
\end{equation}
%
\renewcommand{\arraystretch}{1.6}
%

\vspace*{-3mm}
\noindent
It is worth to mention that these gauge transformations
are the same ones
in usual string theories, 
except for the gauge transformations corresponding to the parameters
$\mu'_r$ and $\tilde{\mu}'_r$. 
However, we would like to emphasize that these gauge transformations
can be disappear on the following combinations,
\begin{eqnarray*}
\delta(\phi^{\hat{+}}b_r^+-\phi^+b^{\hat{+}}_r)
&=&
\frac{1}{\sqrt{\pi}}
\sum_n\zeta_{r-n}
(\phi^{\hat{+}}\alpha^+_n-\phi^+\alpha^{\hat{+}}_n), 
\\
\delta(\phi^{\hat{+}}\tilde{b}_r^+-\phi^+\tilde{b}^{\hat{+}}_r)
&=&
-\frac{1}{\sqrt{\pi}}
\sum_n\tilde{\zeta}_{r-n}
(\phi^{\hat{+}}\tilde{\alpha}^+_n-\phi^+\tilde{\alpha}^{\hat{+}}_n).  
\end{eqnarray*}
In analogy to taking the light-cone gauge in usual string theories, 
by using the gauge degrees of freedom for $\zeta_r$ and 
$\tilde{\zeta}_r$, we can recursively adopt gauge conditions 
\begin{equation}
\begin{array}{rcl}
\phi^{\hat{+}}b^+_r-\phi^+b^{\hat{+}}_r
&=&
0, 
\\
\phi^{\hat{+}}\tilde{b}^+_r-\phi^+\tilde{b}^{\hat{+}}_r
&=&
0, 
\end{array}
\label{light-cone_gauge_fixing_fourier_1-1}
\end{equation}
if the following condition is satisfied, 
\begin{equation}
\phi^{\hat{+}}p^+-\phi^+p^{\hat{+}}
\ne
0. 
\label{light-cone_ass}
\end{equation}
Next we use the gauge degrees of freedom for $\mu'_r$ and
$\tilde{\mu}'_r$ in (\ref{gauge_trans_fourier_1}). 
In order to keep the condition (\ref{light-cone_ass})
one cannot vanish both of the scalar fields $\phi^{\hat{+}}$
and $\phi^+$ simultaneously. 
If $\phi^{\hat{+}}\ne 0$, 
we can adopt the following gauge fixing conditions
of the $\hat{+}$ component,  
\begin{equation}
b^{\hat{+}}_r=\tilde{b}^{\hat{+}}_r
=0, 
\label{light-cone_gauge_fixing_fourier_1-2}
\end{equation}
without spoiling the gauge fixing 
conditions (\ref{light-cone_gauge_fixing_fourier_1-1}).
From (\ref{light-cone_gauge_fixing_fourier_1-1}) and
(\ref{light-cone_gauge_fixing_fourier_1-2}) we can then arrive at 
the gauge fixing conditions (\ref{light-cone_gauge_fixing_fermion}). 
In the similar way, we also conclude the same gauge fixing conditions
(\ref{light-cone_gauge_fixing_fermion}), in the case $\phi^+\ne 0$.
Therefore we may assume the case $\phi^{\hat{+}} \ne 0$
throughout this paper without loss of generality.
For the bosonic sectors, the procedure to obtain the light-cone gauge 
(\ref{light-cone_gauge_fixing_boson}) within the gauge 
(\ref{light-cone_gauge_fixing_fermion}) is essentially as  
same as in the previous work~\cite{tw1}, and therefore 
does not need to be repeated here. 

The gauge fixing procedures for the remaining constraints 
${P_B}^m_I(\tau, \sigma)={P_{C_{01}}}(\tau, \sigma)=0$ and 
(\ref{secondary_constraint_06})-(\ref{secondary_constraint_08})  
are also as same as in the bosonic {\UvUa} string model.  
As we explicitly showed in the paper~\cite{tw1}, 
after imposing suitable gauge fixing conditions, 
the dynamics for the remaining phase space variables  
is completely determined,  
so that it is described by 
the zero-modes of the fields $\phi^I(\tau, \sigma)$ and 
${P_\phi}^I(\tau, \sigma)\Big(=-B_\sigma^I(\tau, \sigma)\Big)$, 
\begin{equation}
\begin{array}{rcl}
\phi^I(\tau, \sigma)
&=&
\phi^I, 
\\
{P_\phi}^I(\tau, \sigma)
&=&
\displaystyle
\frac{{p_\phi}^I}{2\pi}-\hat{C}_0\phi^I\tau, 
\end{array}
\end{equation}
with second class constraints 
\begin{equation}
\frac{1}{2}\phi^I\phi_I=0, 
\quad 
{p_\phi}^{\hat{+}}=0, 
\end{equation}
where the constant $\hat{C}_0$ arises from a gauge fixing condition 
$C_{\tau\sigma}(\tau, \sigma)=-\hat{C}_0$. 
In the two-time physics, 
these zero-modes are regarded as canonical particle 
modes~\cite{bdk, rz}.   

Now let us summarize the correspondence between the constraints 
and the gauge fixing conditions\footnote{
As in usual closed string theories, 
a constraint $L_0-\tilde{L}_0=0$ leads residual gauge invariance 
{\it i.e.}\  
the translation along the world-sheet coordinate $\sigma$. 
This constraint results in the level matching condition 
for physical states in quantum theories. 
}: 
\begin{eqnarray*}
{\hbox{\bf constraints}} &\qquad\qquad& {\hbox{\bf gauge fixing conditions}}
\nonumber \\
L_0 + \tilde{L}_0 = 0, 
&\qquad\qquad& x^+ = 0,
\nonumber \\
L_m =\tilde{L}_m = 0,
&\qquad\qquad& 
\alpha^{+}_m=\tilde{\alpha}^{+}_m=0, \quad (m\ne 0),
\nonumber \\
G_r=\tilde{G}_r=0, 
&\qquad\qquad& b_r^+=\tilde{b}_r^+=0, 
\nonumber \\
\phi_I p^I = 0, 
&\qquad\qquad& x^{\hat{+}} = 0, 
\nonumber \\
\phi_I \alpha_m^I = \phi_I {\tilde{\alpha}}_m^I = 0,
&\qquad\qquad& \alpha^{\hat{+}}_m={\tilde{\alpha}}^{\hat{+}}_m = 0,
\quad (m\ne 0),
\nonumber \\
\phi_Ib^I_r=\phi_I\tilde{b}^I_r=0, 
&\qquad\qquad&
b^{\hat{+}}_r=\tilde{b}^{\hat{+}}_r=0, 
\nonumber \\ 
\frac{1}{2}\phi^I\phi_I = 0,
&\qquad\qquad& {p_\phi}^{\hat{+}}=0.
\nonumber
\end{eqnarray*}
We are now ready to discuss the dynamics of the system.   

As the constraints are quadratic in the Fourier modes, we can solve
these directly and the dependent variables are expressed
in terms of the independent variables.
Here are the non-vanishing Poisson brackets of the  
independent canonical variables
\begin{equation}
\begin{array}{rcl}
\{x^-, p^+\} 
&=& 
\{x^{\hat{-}}, p^{\hat{+}}\} = -1,
\\
\{x^i, p^j\} 
&=& 
\delta^{ij},
\\
\{\alpha^i_m, \alpha^j_n\}
&=& 
\{\tilde{\alpha}^i_m, \tilde{\alpha}^j_n\} 
= -i m \delta^{ij} \delta_{m+n},  
\\
\{b^i_r, b^j_s\}
&=&
\{\tilde{b}^i_r, \tilde{b}^j_s\}
=
-i\delta^{ij}\delta_{r+s}, 
\\
\{\phi^+, {p_\phi}^-\}
&=& 
\{\phi^-, {p_\phi}^+\} = \{\phi^{\hat{+}}, {p_\phi}^{\hat{-}}\} = -1,
\\
\{\phi^i, {p_\phi}^j\} 
&=& \delta^{ij},
\end{array}
\label{independent_variables} 
\end{equation}
and the remaining non-vanishing dependent variables are written down as 
\begin{eqnarray}
p^-
&=& \frac{-1}{\phi^{\hat{+}}p^+-\phi^+p^{\hat{+}}}
    \bigg(\frac{p^{\hat{+}}p^{\hat{+}}}{\phi^{\hat{+}}}
          \Big(\phi^+\phi^--\frac{1}{2}\phi^i\phi_i
          \Big)
          -p^{\hat{+}}
          \Big(\phi^-p^+-\phi^ip_i
          \Big)
          -2\pi\phi^{\hat{+}}
          \Big(L_0^{\rm tr}+\tilde{L}_0^{\rm tr}
          \Big)
    \bigg),
\nonumber \\
\alpha^-_m 
&=& \frac{-1}{\phi^{\hat{+}}p^+-\phi^+p^{\hat{+}}}
    \bigg(p^{\hat{+}}\phi^i\alpha_{im} 
         -2\sqrt{\pi}\phi^{\hat{+}}
         L_{m}^{\rm tr}
    \bigg), \qquad (m \ne 0),
\nonumber \\
\tilde{\alpha}^-_m 
&=& \frac{-1}{\phi^{\hat{+}}p^+-\phi^+p^{\hat{+}}}
    \bigg(p^{\hat{+}}\phi^i\tilde{\alpha}_{im}
         -2\sqrt{\pi}\phi^{\hat{+}}
          \tilde{L}_{m}^{\rm tr}
    \bigg), \qquad (m \ne 0), 
\nonumber \\
p^{\hat{-}} 
&=& \frac{1}{\phi^{\hat{+}}p^+-\phi^+p^{\hat{+}}}
    \bigg(\frac{p^{\hat{+}}p^+}{\phi^{\hat{+}}}
         \Big(\phi^+\phi^--\frac{1}{2}\phi^i\phi_i
         \Big) 
    -p^+\Big(\phi^-p^+-\phi^ip_i
         \Big) 
     -2\pi\phi^+\Big(L_0^{\rm tr}+\tilde{L}_0^{\rm tr}
                \Big)
    \bigg), 
\nonumber 
\\
\alpha^{\hat{-}}_m
&=& \frac{1}{\phi^{\hat{+}}p^+-\phi^+p^{\hat{+}}}
    \bigg(p^+\phi^i\alpha_{im}
         -2\sqrt{\pi}\phi^+
         L_{m}^{\rm tr}
    \bigg), \qquad (m\ne 0), \qquad 
\nonumber 
\\
\tilde{\alpha}^{\hat{-}}_m 
&=& \frac{1}{\phi^{\hat{+}}p^+-\phi^+p^{\hat{+}}}
    \bigg(p^+\phi^i\tilde{\alpha}_{im}
         -2\sqrt{\pi}\phi^+
          \tilde{L}_{m}^{\rm tr}
    \bigg), \qquad (m\ne 0), 
\label{dependent_variables} 
\\
b^-_r
&=&
\frac{-1}{\phi^{\hat{+}}p^+-\phi^+p^{\hat{+}}}
\Big(p^{\hat{+}}\phi^ib_{ir}-2\sqrt{\pi}\phi^{\hat{+}}
    G^{\rm tr}_r
\Big), 
\nonumber 
\\
\tilde{b}^-_r
&=&
\frac{-1}{\phi^{\hat{+}}p^+-\phi^+p^{\hat{+}}}
\Big(p^{\hat{+}}\phi^i\tilde{b}_{ir}-2\sqrt{\pi}\phi^{\hat{+}}
    \tilde{G}^{\rm tr}_r
\Big), 
\nonumber 
\\
b^{\hat{-}}_r
&=&
\frac{1}{\phi^{\hat{+}}p^+-\phi^+p^{\hat{+}}}
\Big(p^+\phi^ib_{ir}-2\sqrt{\pi}\phi^+G^{\rm tr}_r\Big), 
\nonumber 
\\
\tilde{b}^{\hat{-}}_r
&=&
\frac{1}{\phi^{\hat{+}}p^+-\phi^+p^{\hat{+}}}
\Big(p^+\phi^i\tilde{b}_{ir}-2\sqrt{\pi}\phi^+\tilde{G}^{\rm tr}_r\Big), 
\nonumber 
\\
\phi^{\hat{-}} 
&=& - \, \frac{1}{\phi^{\hat{+}}}
    \Big(\phi^+\phi^--\frac{1}{2}\phi^i\phi_i
    \Big), 
\nonumber 
\end{eqnarray}
where transverse parts of the super-Virasoro generators 
$L^{\rm tr}_m(=L^{(\alpha)\rm tr}_m+L^{(b)\rm tr}_m)$,  
$\tilde{L}^{\rm tr}_m(=\tilde{L}^{(\alpha)\rm tr}_m
+\tilde{L}^{(b)\rm tr}_m)$, 
$G_r^{\rm tr}$ 
and 
$\tilde{G}_r^{\rm tr}$ 
are defined by
$$
\begin{array}{rclcrcl}
L_m^{(\alpha)\rm tr} 
&\equiv& 
\displaystyle
\frac{1}{2}\sum_n\alpha_{-n}^i{\alpha_i}_{m+n}, 
&\qquad&
\tilde{L}_m^{(\alpha)\rm tr}
&\equiv&
\displaystyle
\frac{1}{2}\sum_n{\tilde{\alpha}}_{-n}^i\tilde{\alpha}_{i m+n},
\\
L_m^{(b)\rm tr}
&\equiv&
\displaystyle
\frac{1}{2}
\!\sum_{r\in{\bf Z}+a}\!\!\Big(r+\frac{m}{2}\Big)
b^i_{-r}b_{im+r}, 
&\qquad&
\tilde{L}_m^{(b)\rm tr}
&\equiv&
\displaystyle
\frac{1}{2}
\!\sum_{r\in{\bf Z}+a}\!\!\Big(r+\frac{m}{2}\Big)
\tilde{b}^i_{-r}\tilde{b}_{im+r},
\\
G_r^{\rm tr}
&\equiv&
\displaystyle
\sum_n\alpha^i_{-n}b_{ir+n}, 
&\qquad&
\tilde{G}_r^{\rm tr}
&\equiv&
\displaystyle
\sum_n\tilde{\alpha}^i_{-n}\tilde{b}_{ir+n}.    
\end{array}
$$

Now let us investigate the symmetry of the $D$-dimensional
background spacetime. 
The generators for the translation and the Lorentz transformation 
are derived from the classical action
(\ref{classical_super_action_04}). 
In terms of the Fourier modes, these are 
\begin{subequations}
\begin{eqnarray}
P^I 
&=& 
p^I, 
\\
M^{IJ}
&=& 
x^Ip^J
-\frac{i}{2}\sum_{m\ne0}\frac{1}{m}
\Big(\alpha^I_{-m} \alpha^J_m
    +\tilde{\alpha}^I_{-m}\tilde{\alpha}^J_m
\Big) 
-\frac{i}{2}\!\sum_{r\in{\bf Z}+a}\!\!
\Big(b^I_{-r}b^J_r
    +\tilde{b}^I_{-r}\tilde{b}^J_r
\Big)
+\phi^I {p_\phi}^J
\nonumber 
\\
&&
-(I \leftrightarrow J).
\end{eqnarray}
\end{subequations}

\vspace*{-6mm} 
\noindent
Using the independent canonical variables,  
the Poincar\'e algebra ISO($D-2, 2$) is satisfied, 
\begin{eqnarray}
\{P^I, P^J\} 
&=& 0, 
\nonumber \\
\{M^{IJ}, P^K\} 
&=& \eta^{IK} P^J - \eta^{JK} P^I, 
\label{poincare} \\
\{M^{IJ}, M^{KL}\} 
&=& \eta^{IK} M^{JL} - \eta^{JK} M^{IL}
  - \eta^{IL} M^{JK} + \eta^{JL} M^{IK}  , 
\nonumber
\end{eqnarray}
if a constraint $L_0^{\rm tr}=\tilde{L}_0^{\rm tr}$ 
is imposed. 
The gauge fixing procedure we considered is the way 
to preserve the full $D$-dimensional Poincar\'e symmetry. 

According to the ordinary string theories in the light-cone gauge, 
we have to examine Poincar\'e algebra (\ref{poincare}) 
in the quantum theory~\cite{ggrt}. 
The checking of the Poincar\'e algebra is again straightforward, 
except for commutation relations 
$[M^{i-}, M^{j-}]$, $[M^{i\hat{-}}, M^{j\hat{-}}]$, 
$[M^{i-}, M^{j\hat{-}}]$, $[M^{i-}, M^{-\hat{-}}]$ 
and $[M^{i\hat{-}}, M^{-\hat{-}}]$.  
The explicit forms of these Lorentz generators are given in Appendix D. 
After lengthy computation, 
we can obtain the following results,  
\begin{eqnarray}
{[}M^{i-}, M^{j-}{]} 
&=& 
\frac{4\pi{\phi^{\hat{+}}}^2}{(\phi^{\hat{+}}p^+-\phi^+p^{\hat{+}})^2} 
A^{ij}, 
\nonumber 
\\
{[}M^{i\hat{-}}, M^{j\hat{-}}{]} 
&=& 
\frac{4\pi{\phi^+}^2}{(\phi^{\hat{+}}p^+-\phi^+p^{\hat{+}})^2}
A^{ij}, 
\nonumber 
\\
{[}M^{i-}, M^{j\hat{-}}{]} 
&=& 
i\delta^{ij}M^{-\hat{-}} 
-\frac{4\pi\phi^{\hat{+}}\phi^+}{(\phi^{\hat{+}}p^+-\phi^+p^{\hat{+}})^2} 
A^{ij}, 
\\
{[}M^{i-}, M^{-\hat{-}}{]}
&=&
\frac{4\pi\phi^{\hat{+}}\phi_j}{(\phi^{\hat{+}}p^+-\phi^+p^{\hat{+}})^2} 
A^{ij}, 
\nonumber 
\\
{[}M^{i\hat{-}}, M^{-\hat{-}}{]}
&=&
-\frac{4\pi\phi^+\phi_j}{(\phi^{\hat{+}}p^+-\phi^+p^{\hat{+}})^2} 
A^{ij}. 
\nonumber
\end{eqnarray}
Anomalous terms $A^{ij}$ are 
\begin{subequations}
\begin{eqnarray}
A^{ij} 
&=&
-
\Bigg( 
\sum_{m=1}^\infty m 
\Big(  
\alpha^i_{-m}\alpha^j_{m}
+\tilde{\alpha}^i_{-m}\tilde{\alpha}^j_m
\Big)
+4\sum_{r=\frac{1}{2}}^\infty r^2
\Big(
b^i_{-r}b^j_r
+\tilde{b}^i_{-r}\tilde{b}^j_r
\Big)
\Bigg)
\bigg(1-\frac{D-4}{8}
\bigg)
\nonumber 
\\
&&   
+ \,
\Bigg(
\sum_{m=1}^\infty \frac{1}{m}
\alpha^i_{-m}\alpha^j_m 
+\sum_{r=\frac{1}{2}}^\infty
b^i_{-r}b^j_r
\Bigg)
\bigg(
\hspace*{-1mm}:\hspace*{-1mm}L_0^{\rm tr}\hspace*{-1mm}: 
-:\hspace*{-1mm}\tilde{L}_0^{\rm tr}\hspace*{-1mm}: 
+a_0+\tilde{a}_0-\frac{D-4}{8}
\bigg)
\nonumber 
\\
&&   
+ \,
\Bigg(
\sum_{m=1}^\infty \frac{1}{m}
\tilde{\alpha}^i_{-m}\tilde{\alpha}^j_m
+\sum_{r=\frac{1}{2}}^\infty
\tilde{b}^i_{-r}\tilde{b}^j_r
\Bigg)
\bigg(
\hspace*{-1mm}:\hspace*{-1mm}\tilde{L}_0^{\rm tr}\hspace*{-1mm}: 
-:\hspace*{-1mm}L_0^{\rm tr}\hspace*{-1mm}:  
+\tilde{a}_0+a_0-\frac{D-4}{8}
\bigg)
\nonumber 
\\
&&
-(i \leftrightarrow j), 
\end{eqnarray}
for (NS, NS) sector, 
which means both right and left movers 
satisfy the Neveu-Schwarz boundary conditions, 
\begin{eqnarray}
A^{ij} 
&=&
-
\Bigg( 
\sum_{m=1}^\infty m 
\Big(  
\alpha^i_{-m}\alpha^j_{m}
+\tilde{\alpha}^i_{-m}\tilde{\alpha}^j_m
\Big)
+4\sum_{r=1}^\infty r^2
\Big(
b^i_{-r}b^j_r
+\tilde{b}^i_{-r}\tilde{b}^j_r
\Big)
\Bigg)
\bigg(1-\frac{D-4}{8}
\bigg)
\nonumber 
\\
&&   
+ \,
\Bigg(
\sum_{m=1}^\infty\frac{1}{m}
\alpha^i_{-m}\alpha^j_m 
+\sum_{r=1}^\infty
b^i_{-r}b^j_r
+\frac{1}{2}
b_0^ib_0^j
\Bigg)
\bigg(
\hspace*{-1mm}:\hspace*{-1mm}L^{\rm tr}_0\hspace*{-1mm}:
-:\hspace*{-1mm}\tilde{L}^{\rm tr}_0\hspace*{-1mm}:
+a_0+\tilde{a}_0
\bigg)
\nonumber 
\\
&&   
+ \,
\Bigg(
\sum_{m=1}^\infty\frac{1}{m}
\tilde{\alpha}^i_{-m}\tilde{\alpha}^j_m
+\sum_{r=1}^\infty
\tilde{b}^i_{-r}\tilde{b}^j_r
+\frac{1}{2}
\tilde{b}_0^i\tilde{b}_0^j
\Bigg)
\bigg(
\hspace*{-1mm}:\hspace*{-1mm}\tilde{L}^{\rm tr}_0\hspace*{-1mm}:
-:\hspace*{-1mm}L^{\rm tr}_0\hspace*{-1mm}:
+\tilde{a}_0+a_0
\bigg)
\nonumber 
\\
&&
-(i \leftrightarrow j), 
\end{eqnarray}
for (R, R) sector and 
\begin{eqnarray}
A^{ij} 
&=&
-
\Bigg( 
\sum_{m=1}^\infty m 
\Big(  
\alpha^i_{-m}\alpha^j_{m}
+\tilde{\alpha}^i_{-m}\tilde{\alpha}^j_m
\Big)
+4\sum_{r=\frac{1}{2}}^\infty r^2
b^i_{-r}b^j_r
+4\sum_{r=1}^\infty r^2
\tilde{b}^i_{-r}\tilde{b}^j_r
\Bigg)
\bigg(1-\frac{D-4}{8}
\bigg)
\nonumber 
\\
&&   
+ \,
\Bigg(
\sum_{m=1}^\infty \frac{1}{m}
\alpha^i_{-m}\alpha^j_m 
+\sum_{r=\frac{1}{2}}^\infty
b^i_{-r}b^j_r
\Bigg)
\bigg(
\hspace*{-1mm}:\hspace*{-1mm}L_0^{\rm tr}\hspace*{-1mm}: 
-:\hspace*{-1mm}\tilde{L}_0^{\rm tr}\hspace*{-1mm}: 
+a_0+\tilde{a}_0-\frac{D-4}{8}
\bigg)
\nonumber 
\\
&&   
+ \,
\Bigg(
\sum_{m=1}^\infty \frac{1}{m}
\tilde{\alpha}^i_{-m}\tilde{\alpha}^j_m
+\sum_{r=1}^\infty
\tilde{b}^i_{-r}\tilde{b}^j_r
+\frac{1}{2}\tilde{b}_0^i\tilde{b}_0^j
\Bigg)
\bigg(
\hspace*{-1mm}:\hspace*{-1mm}\tilde{L}_0^{\rm tr}\hspace*{-1mm}: 
-:\hspace*{-1mm}L_0^{\rm tr}\hspace*{-1mm}:  
+\tilde{a}_0+a_0
\bigg)
\nonumber 
\\
&&
-(i \leftrightarrow j), 
\end{eqnarray}
\end{subequations}

\vspace*{-6mm}

\noindent
for (NS, R) sector. 
The constants $a_0$ and $\tilde{a}_0$ denote 
the ordering ambiguity of the operators 
$L_0^{\rm tr}$ and $\tilde{L}_0^{\rm tr}$ 
by adopting the normal-ordering prescription, 
$$
L_0^{\rm tr}
\ \rightarrow \ \ 
:\hspace*{-1mm}L_0^{\rm tr}\hspace*{-1mm}: - \ a_0, 
\qquad
\tilde{L}_0^{\rm tr} 
\ \rightarrow \ \ 
:\hspace*{-1mm}\tilde{L}_0^{\rm tr}\hspace*{-1mm}: - \ \tilde{a}_0.  
$$
The level matching condition for physical states is now expressed as 
\begin{equation}
:\hspace*{-1mm}L^{\rm tr}_0\hspace*{-1mm}:- \ a_0
= \ :\hspace*{-1mm}\tilde{L}^{\rm tr}_0\hspace*{-1mm}:- \ \tilde{a}_0, 
\quad\leftrightarrow\quad 
N-a_0=\tilde{N}-\tilde{a}_0, 
\label{level_matching_condition}
\end{equation}
where number operators $N$ and $\tilde{N}$ are defined by 
$$
N\equiv
\sum_{m=1}^\infty\alpha^i_{-m}\alpha_{i m}
+\!\!\!\!\sum_{r\in{\bf Z}+a>0}^\infty\!\!\!\!\!
rb^i_{-r}b_{ir}, \qquad 
\tilde{N}\equiv
\sum_{m=1}^\infty\tilde{\alpha}^i_{-m}\tilde{\alpha}_{i m}
+\!\!\!\!\sum_{r\in{\bf Z}+a>0}^\infty\!\!\!\!\!
r\tilde{b}^{i}_{-r}\tilde{b}_{ir}.
$$
Imposing the level matching condition (\ref{level_matching_condition}), 
the anomalous terms $A^{ij}$ vanish on physical states if and only if 
\begin{equation}
 D=12, \qquad\quad 
\left. 
\begin{array}{ll}
\displaystyle
a_0=\tilde{a}_0=\frac{1}{2},  
& \quad 
\Big(\mbox{(NS, NS) sector}
\Big), 
\\
a_0=\tilde{a}_0=0,  
& \quad
\Big(\mbox{(R, R) sector}
\Big), 
\\
\displaystyle
a_0=\frac{1}{2}, \quad \tilde{a}_0=0,   
& \quad
\Big(\mbox{(NS, R) sector}
\Big).  
\end{array}
\right.   
\label{critical_dimensions}
\end{equation}
Then, the Poincar\'e algebra ISO(10, 2) is satisfied 
in the quantum theory. 

A mass-shell relation of this superstring model is given by 
\begin{eqnarray}
m^2 
&=& -P^IP_I \nonumber \\
&=& 4\pi\Big(N+\tilde{N}-a_0-\tilde{a}_0\Big). 
\end{eqnarray}
As a common future of our string models~\cite{tw1} 
and the two-time physics~\cite{b},  
on-shell degrees of freedom are equivalent to usual string theories, 
because our extra spacetime coordinates are introduced by the ``gauge'' 
symmetries.    

\section{Conclusions and discussions}
\setcounter{equation}{0}
\setcounter{footnote}{0}

We have explicitly constructed the {\UvUa} NSR superstring model 
by using the superfield formulation and discussed the quantization 
of the model.  
Even though the system had reducible and open gauge symmetries,
we have shown that the covariant quantization has been successfully
carried out in the Lagrangian formulation {\it \'a la} Batalin
and Vilkovisky. 
Furthermore we have presented the noncovariant light-cone gauge 
formulation and investigated the symmetry of the background spacetime.
With careful considerations of the residual {\UvUa} gauge symmetries,  
we have specified the gauge fixing conditions corresponding to the
first-class constraints.  
Under these suitable conditions, 
we have been able to clarify dynamical independent variables
and solve the first-class constraints explicitly. 
Although manifest covariance has been lost,
we have confirmed the full $D$-dimensional Poincar\'e algebra of
the background spacetime by direct computation.

Since the quantizations of the model have been successfully
carried out, we could argue the critical dimension of the superstring 
model.
In our case, it turns out to be 10+2.
This means the background spacetime involves two time coordinates.
Conversely, the requirement of two negative signatures in 
the background metric is natural one due to the gauge invariance
of our model. 
The critical dimension has been obtained from both
the BRST Ward identity in the BRST formulation
and the $D$-dimensional quantum Poincar\'e algebra
in the noncovariant light-cone gauge formulation.
Therefore, we have concluded a consistent quantum theory of
our {\UvUa} superstring model has been formulated in 10+2-dimensional
background spacetime. 
We have also discussed the quantum states. 
Contributions toward the mass-shell relation 
from zero-modes of the scalar field $\phi^I(x)$ are completely canceled,
so that our superstring model possesses
the same spectra as usual string theories. 

We propose our string models as devices to formulate
the physics involving two time coordinates and to search 
for a fundamental theory with an underlying 
complex nature of spacetime which would be linked via dualities 
to M-theory, type II string theories and F-theory 
from higher-dimensional points of view. 
Our explicit Lagrangian formulation might be a clue to understand 
spacetime itself. 

The classical actions and the gauge symmetries of our 
string models strongly suggest that these model should be more 
naturally defined in higher-dimensional field theories, namely, 
that membranes or $p$-branes are more fundamental
than strings in our formulation. 
From the point of view of constraint algebras, 
the action might be derived from
a membrane action by adopting a compactification prescription. 
In this case, Kaluza-Klein fields which arise from the compactification  
might correspond to the gauge fields in our present formulation.  

One of the remarkable features of our higher-dimensional formulation 
is that models are allowed to have pairs of the extra time and 
space coordinates. 
Therefore, by applying our mechanism, 
one can construct a succession of supersymmetric models 
formulated in background spacetimes, 
involving some time coordinates, 
where Majorana-Weyl fermions can live consistently. 

\vspace*{1cm}

\noindent
{\Large{\bf Acknowledgments}}
\vspace{2mm}

We would like to thank Dublin Institute for Advanced Studies 
for warm hospitality. 
We would also like to thank Niels Bohr Institute for 
warm hospitality during the final stage of this work.   
T.T. wishes to thank W. Nahm for interesting comments and 
discussions. 
Y.W. wishes to thank J. Ambj{\o}rn for interesting comments and 
discussions. 
    
\vspace*{1cm}

\noindent
{\large{\bf Appendix A. Two-dimensional world-sheet}}

\renewcommand{\theequation}{A.\arabic{equation}}

\setcounter{equation}{0}
\setcounter{footnote}{0}

\vspace*{5mm}

We denote the characters $a, b, c, \cdots$ and $m, n, l, \cdots$ as 
flat local Lorentz and curved spacetime indices, respectively. 

The two-dimensional flat metric $\eta_{ab}$ and 
Levi-Civit\`a symbol $\varepsilon_{ab}$ are given by  
\vspace*{1mm}
%
\renewcommand{\arraystretch}{1.0}
%
\begin{equation}
\eta_{ab} =  
\eta^{ab} =  
\left( \begin{array}{cc}
         \ -1 \ & \ 0 \ \\
         \ 0  \ & \ 1 \
       \end{array}
\right),  
\qquad
\varepsilon_{ab} =  
- \varepsilon^{ab} =  
\left( \begin{array}{cc}
         \ 0 \ & \ 1 \ \\
         \ -1\ & \ 0 \
       \end{array}
\right). 
\end{equation}
As the two-dimensional Clifford algebra, 
the $\sigma$-matrices satisfy 
\begin{equation}
\{\sigma^a, \sigma^b\}=2\eta^{ab}.  
\end{equation}
Their explicit representations are 
$$
(\sigma^0)_\alpha{}^\beta=
\left( \begin{array}{cc}
        \ 0 \ & \ 1 \ \\
        \ -1\ & \ 0 \ 
       \end{array}
\right), \qquad 
(\sigma^1)_\alpha{}^\beta= 
\left( \begin{array}{cc}
        \ 0 \ & \ 1 \ \\
        \ 1 \ & \ 0 \ 
       \end{array}
\right), 
$$
and  
$$
(\bar{\sigma})_\alpha{}^\beta 
\equiv(\sigma^0\sigma^1)_\alpha{}^\beta=
\left( \begin{array}{cc} 
        \ 1 \ & \ 0 \ \\
        \ 0 \ & \ -1\ 
       \end{array}
\right). 
$$
The spinor metric is given by  
\begin{equation}
\eta_{\alpha\beta}=\eta^{\alpha\beta}= 
\left( \begin{array}{cc}
         \ 0 \ & \ 1 \ \\
         \ -1\ & \ 0 \ 
       \end{array}
\right), 
\end{equation}
and the spinor indices are raised or lowered by using the metric 
$\eta^{\alpha\beta}$ and $\eta_{\alpha\beta}$\footnote{
One might prefer to regard the equation 
$\theta^\alpha=\theta_\beta\eta^{\beta\alpha}$ 
as the Dirac conjugate relation 
$\bar{\psi}=\psi^\dagger\sigma^0=\psi^TC$, 
where the spinor $\psi$ is a real (Majorana) spinor and 
the charge conjugation matrix is defined as $C\equiv\sigma^0$.   
},  
\begin{equation}
\theta^\alpha=\theta_\beta\eta^{\beta\alpha}, \qquad
\theta_\alpha=\eta_{\alpha\beta}\theta^\beta. 
\end{equation}
%
\renewcommand{\arraystretch}{1.6}
%

\vspace*{-8mm}

\noindent
A bilinear form of spinors $\theta$ and $\chi$ is defined by 
$$
(\theta{\cal M}_\sigma\chi) 
\equiv\theta^\alpha({\cal M}_\sigma)_\alpha{}^\beta\chi_\beta, 
$$
where ${\cal M}_\sigma$ denotes any products of the $\sigma$-matrices. 
An integration of spinor coordinates is given by 
\begin{equation}
\int\!\dd^2\theta\equiv\frac{1}{2}\int\!\dd\theta^2\dd\theta^1, 
\end{equation}
and its normalization is defined by 
$$
\int\!\dd^2\theta(\theta\theta)=1. 
$$  

One can introduce an orthonormal basis ``zweibein'' one-form field 
of the local Lorentz flame 
for each cotangent space of two-dimensional spacetime 
\begin{equation}
e^a=\dd x^me_m{}^a, 
\end{equation}
where the indexes $a(=0, 1)$ and $m(=0, 1)$ label the local Lorentz 
flame and curved spacetime, respectively.  
Orthonormality for the zweibein $e_m{}^a(x)$ implies 
$$
g^{mn}e_m{}^ae_n{}^b=\eta^{ab},  
$$
by using inverse metric $g^{mn}(x)$. 
One may assume the invertibility of the zweibein, 
$$
e_m{}^ae_a{}^n=\delta_m{}^n, \qquad 
e_a{}^me_m{}^b=\delta_a{}^b.
$$
In the curved spacetime, the metric $g_{mn}(x)$ and 
the Levi-Civit\`a symbol $\ep^{mn}$ are given by 
\begin{equation}
g_{mn}=e_m{}^ae_n{}^b\eta_{ab}, \qquad 
\ep^{mn}=ee_a{}^me_b{}^n\ep^{ab}, \qquad
\ep_{mn}=\frac{1}{e}e_m{}^ae_n{}^b\ep_{ab},  
\end{equation}
where $e(x)\equiv\det e_m{}^a(x)=\sqrt{-g(x)}$. 
One might prefer to use the Levi-Civit\`a tensors, 
$$
{\cal E}^{mn}\equiv\frac{1}{e}\ep^{mn}, \qquad  
{\cal E}_{mn}\equiv e\ep_{mn}.  
$$
The zweibein $e_m{}^a(x)$ allows to covert the local Lorentz  
indices to the spacetime indices and back, 
$$
A^a=e_m{}^aA^m, \qquad 
A^m=A^ae_a{}^m. 
$$
One can also define 
$\sigma$-matrices in curved spacetime $\sigma^m(x)$ as 
\begin{equation}
\sigma^m\equiv\sigma^ae_a{}^m. 
\end{equation}
The zweibein obeys the following SO$(1,1)$ local Lorentz transformation 
\begin{equation}
\delta e_m{}^a=le_m{}^b\ep_b{}^a, 
\label{local_Lorentz_zweibein}
\end{equation}
where the symbol $\ep_b{}^a$ is a generator of the vectorial representation 
for the local Lorentz group SO$(1, 1)$ 
and the function $l(x)$ is a corresponding gauge parameter. 
In order to define a covariant derivative on the local Lorentz group, 
a spin connection one-form $\omega_a{}^b(x)$ is introduced as 
\begin{equation}
\omega_a{}^b=\dd x^m\omega_m\ep_a{}^b=\omega\ep_a{}^b, 
\end{equation}
and its local Lorentz transformation is defined by 
\begin{equation}
\delta\omega_m=\del_ml. 
\end{equation}
The connections allow us to define covariant derivatives  
acting on the local Lorentz indices as 
\begin{eqnarray}
\nabla_m\phi 
&=& \del_m\phi, 
\nonumber \\
\nabla_mA_a 
&=& \del_mA_a+\omega_m\ep_a{}^bA_b, 
\label{covariant_derivative_lorentz}\\
\nabla_mA^a
&=& \del_mA^a-\omega_mA^b\ep_b{}^a, 
\nonumber
\end{eqnarray}
whereas covariant derivatives acting on the 
ordinary curved spacetime indices are defined by 
\begin{eqnarray}
\nabla_m\phi 
&=& \del_m\phi, 
\nonumber \\
\nabla_mA_n 
&=& \del_mA_n
   -\Gamma^l{}_{mn}A_l, 
\label{covariant_derivative_curved}\\
\nabla_mA^n
&=& \del_mA^n 
   +\Gamma^n{}_{ml}A^l, 
\nonumber 
\end{eqnarray}
where the coefficient $\Gamma^l{}_{mn}(x)$ is usual Christoffel
connection.  
From the equivalence for the covariant derivatives between 
local Lorentz flame and curved spacetime flame, 
one can obtain the following relation between spin connections 
and Christoffel connections, 
\begin{equation}
\Gamma^l{}_{mn}=e_a{}^l(\del_me_n{}^a-\omega_me_n{}^b\ep_b{}^a), 
\end{equation}
or equivalently 
\begin{equation}
\nabla_me_n{}^a=0. 
\label{vielbein_postulate} 
\end{equation}
In (\ref{vielbein_postulate}), we use the following definition 
for the covariant derivative for the field $A_m{}^a(x)$ involving 
``mixed'' indices, 
\begin{equation}
\nabla_mA_n{}^a
\equiv\del_mA_n{}^a-\Gamma^l{}_{mn}A_l{}^a-\omega_mA_n{}^b\ep_b{}^a.
\label{covariant_derivative_mix}
\end{equation}
In two-dimensional case, 
the metric compatibility which corresponds to $\nabla_kg_{mn}(x)=0$ 
is automatically satisfied by the following way 
\begin{eqnarray*}
\nabla_m\eta_{ab}
&=& \del_m\eta_{ab}
   +\omega_m\ep_a{}^c\eta_{cb}
   +\omega_m\ep_b{}^c\eta_{ac}
\\
&=& \omega_m\ep_a{}^c\eta_{cb} 
   +\omega_m\ep_b{}^c\eta_{ac}
\\
&=& 0. 
\end{eqnarray*}

The torsion two-form $T^a(x)$ and the curvature two-form $R_a{}^b(x)$ 
are then defined as 
\begin{subequations}
\begin{eqnarray}
T^a
&=& \dd e^a -\omega e^b\ep_b{}^a 
   \equiv
   -\frac{1}{2}e^ce^bT_{bc}{}^a, 
\\
R_a{}^b
&=& \dd \omega_a{}^b+\omega_a{}^c\omega_c{}^b 
   \equiv 
   -\frac{1}{2}e^de^cR_{cda}{}^b. 
\end{eqnarray}
\end{subequations}

\vspace*{-6mm}
\noindent
In the curved spacetime, the torsion components are given as 
\begin{eqnarray*}
T_{mn}{}^l
&\equiv& e_a{}^lT_{mn}{}^a 
\nonumber \\
&=& e_a{}^l\Big(\del_me_n{}^a-\omega_me_n{}^b\ep_b{}^a-(m\leftrightarrow n)
           \Big)
\nonumber \\
&=& \Gamma^l{}_{mn}-\Gamma^l{}_{nm}.
\end{eqnarray*}
If one impose the usual torsion free condition $T_{mn}{}^l(x)=0$, 
the spin connections $\omega_m(x)$ are expressed in terms of the
zweibein fields   
\begin{equation}
\omega_m=\frac{1}{e}e_{am}\ep^{nl}\del_ne_l{}^a.
\label{torsion_free_connection}
\end{equation}

In addition to the above definition for the covariant derivative on 
the bosonic fields, 
one can also define covariant derivatives acting on spinor 
fields. 
The generator of the spinorial representation of the local 
Lorentz group SO$(1, 1)$ is given by 
$\frac{1}{2}(\bar{\sigma})_\alpha{}^\beta$. 
The local Lorentz transformations of spinor fields are then given by 
%
\renewcommand{\arraystretch}{2.0}
%
\begin{equation}
\begin{array}{rcl}
\delta\psi_\alpha
&=& \displaystyle
   -\frac{1}{2}l(\bar{\sigma}\psi)_\alpha, 
\\
\delta\psi^\alpha
&=& \displaystyle
   -\frac{1}{2}l(\bar{\sigma}\psi)^\alpha.  
\end{array}
\end{equation}
From these transformations, one can define the following 
covariant derivative acting on the spinor fields $\psi_\alpha(x)$ and 
$\psi^\alpha(x)$,  
\begin{equation}
\begin{array}{rcl}
(\nabla_m\psi)_\alpha
&=& \displaystyle 
    \del_m\psi_\alpha+\frac{1}{2}\omega_m(\bar{\sigma}\psi)_\alpha, 
\\
(\nabla_m\psi)^\alpha
&=& \displaystyle 
    \del_m\psi^\alpha+\frac{1}{2}\omega_m(\bar{\sigma}\psi)^\alpha. 
\end{array}
\label{covariant_derivative_spinor}
\end{equation}
One can easily check the metric compatibility for the spinor metric 
$\eta^{\alpha\beta}$ and $\eta_{\alpha\beta}$, 
$$  
\nabla_m\eta_{\alpha\beta}=\nabla_m\eta^{\alpha\beta}=0, 
$$
and the following covariant constant relations 
$$
\nabla_m\delta_\alpha{}^\beta
=\nabla_m(\sigma^a)_\alpha{}^\beta
=\nabla_m(\bar{\sigma})_\alpha{}^\beta
=0. 
$$

\vspace*{10mm}

\noindent
{\large{\bf Appendix B. Geometry of superspace}}

\renewcommand{\theequation}{B.\arabic{equation}}

\setcounter{equation}{0}
\setcounter{footnote}{0}

\vspace*{5mm}

In order to construct supersymmetric theory on the 
two-dimensional world-sheet,  
we use the $(1,1)$ type superspace with coordinates 
$z^M=(x^m, \ \theta^\mu) \ (m=0, 1; \ \mu=1, 2)$. 
The coordinates $x^m$ and $\theta^\mu$ are bosonic and fermionic, 
respectively, 
\begin{equation}
z^Mz^N=(-)^{|M||N|}z^Nz^M, 
\end{equation}
where $|M|$ is a Grassmann parity of the coordinate $z^M$. 

One can define a differential $p$-form in $(1, 1)$ superspace as 
\begin{equation}
\Phi(z)
\equiv
\frac{(-)^{p(p-1)/2}}{p!}
\dd z^{M_p}\cdots \dd z^{M_1}\Phi_{M_1\cdots M_p}(z), 
\end{equation}
where the coefficient function $\Phi_{M_1\cdots M_p}(z)$ is 
``graded'' antisymmetric in its indices, due to the ``graded'' 
anticommutativity of the differential forms, 
$$
\dd z^M\dd z^N=-(-)^{|M||N|}\dd z^N\dd z^M.
$$  
The exterior derivative for the $p$-form is defined by 
\begin{eqnarray*}
\dd \Phi(z) 
&\equiv&
\dd z^L\del_L 
\Big(\frac{(-)^{p(p-1)/2}}{p!}
\dd z^{M_p}\cdots\dd z^{M_1}\Phi_{M_1\cdots M_p}(z)
\Big) 
\\
&=&\frac{(-)^{(p+1)p/2}}{p!}
\dd z^{M_p}\cdots\dd z^{M_1}\dd z^L\del_L\Phi(z).
\end{eqnarray*}
Under the ``usual'' wedge product of $p$-form $\Phi(z)$ and $q$-form 
$\Psi(z)$, 
$$
\Phi(z)\Psi(z)
\equiv
\frac{(-)^{p(p-1)/2+q(q-1)/2}}{p!q!}
\dd z^{M_p}\cdots\dd z^{M_1}\Phi_{M_1\cdots M_p}(z) 
\dd z^{N_q}\cdots\dd z^{N_1}\Psi_{N_1\cdots N_q}(z),
$$
the Leipnitz rule for the 
product of $p$-form $\Phi(z)$ and $q$-form $\Psi(z)$ is given by 
$$
\dd (\Phi(z)\Psi(z)) 
=\dd\Phi(z)\Psi(z)+(-)^p\Phi(z)\dd\Psi(z). 
$$

At each points on superspace, a local Lorentz frame is defined by 
introducing a vielbein one-form 
\begin{equation}
E^A(z)=\dd z^ME_M{}^A(z), 
\end{equation}
where the indices $A=(a, \alpha)$ $(a=0, 1; \ \alpha=1, 2)$ denote 
the local Lorentz flame.  
The vielbeins $E_M{}^A(z)$ are invertible superfields 
\begin{equation}
E_M{}^A(z)E_A{}^N(z)=\delta_M{}^N, \qquad 
E_A{}^M(z)E_M{}^B(z)=\delta_A{}^B.
\end{equation}
In the local Lorentz frame, the exterior derivative is written by 
\begin{equation}
\dd=\dd z^M\del_M=E^A\del_A, 
\end{equation}
where we denote the derivative $\del_A$ as 
$$
\del_A\equiv E_A{}^M\del_M. 
$$
The ``graded'' Lie bracket is given by 
\begin{equation}
[\del_A, \del_B\} 
= C_{AB}{}^C\del_C,   
\end{equation}
where 
$$
C_{AB}{}^C \equiv (\del_AE_{B}{}^N)E_N{}^C 
             -(-)^{|A||B|}(\del_BE_A{}^N)E_N{}^C.
$$
The vielbeins obey the following 
local Lorentz transformations, 
\begin{equation}
\delta E^A(z) = E^B(z)L(z)\varepsilon_B{}^A, 
\end{equation}
where the generator of SO$(1, 1)$ local Lorentz group is defined by 
$$
\ep_A{}^B  \ = \ 
    \left(\begin{array}{cc}
           \  \ep_a{}^b \ & \ 0                                        \ \\
           \   0        \ & \ \frac{1}{2} (\bar{\sigma})_\alpha{}^\beta \
    \end{array}\right), 
$$
and the superfield $L(z)$ is a gauge parameter. 
In order to define supercovariant derivative on SO(1,1) local Lorentz  
group, a connection one-form is introduced as 
\begin{equation}
\Omega_A{}^B(z)=\dd z^M\Omega_M(z)\ep_A{}^B=\Omega(z)\ep_A{}^B, 
\end{equation}
and its local Lorentz transformation is defined by 
\begin{equation}
\delta\Omega_M(z) = \del_ML(z). 
\end{equation}
The connections allow us to define supercovariant derivatives on 
scalar field $\Phi(z)$ and vector fields $\Psi^A(z)$ and $\Psi_A(z)$ 
as 
\begin{eqnarray}
{\cal D}\Phi 
&=& \dd\Phi 
= E^C\del_C\Phi, 
\nonumber \\
{\cal D}\Psi_A 
&=& \dd\Psi_A+\Omega\ep_A{}^B\Psi_B
=E^C\Big(\del_C\Psi_A+\Omega_C\ep_A{}^B\Psi_B\Big) 
\equiv E^C{\cal D}_C\Psi_A, 
\\
{\cal D}\Psi^A &=& \dd\Psi^A-\Omega\Psi^B\ep_B{}^A 
= E^C\Big(\del_C\Psi^A-\Omega_C\Psi^B\ep_B{}^A\Big)  
\equiv E^C{\cal D}_C\Psi^A,  
\nonumber 
\end{eqnarray}
where we  denote $\Omega_A(z)\equiv E_A{}^N\Omega_N(z)$. 

The torsion two-form $T^A(z)$ and the curvature two-form $R_A{}^B(z)$
are then defined as 
\begin{subequations}
\begin{eqnarray}
T^A&=&{\cal D}E^A
\equiv-\frac{1}{2}E^CE^BT_{BC}{}^A, \\
R_A{}^B&=&\dd\Omega_A{}^B+\Omega_A{}^C\Omega_C{}^B
\equiv-\frac{1}{2}E^DE^CR_{CD A}{}^B. 
\end{eqnarray}
\end{subequations}
In the local Lorentz flame, 
the torsion and the curvature are explicitly given by 
\begin{eqnarray*}
T_{BC}{}^A 
&=& -C_{BC}{}^A -\Omega_B\ep_C{}^A+(-)^{|B||C|}\Omega_C\ep_B{}^A,  
\\
R_{CD A}{}^B 
&=& \Big(\del_C\Omega_D-(-)^{|C||D|}\del_D\Omega_C 
        -C_{CD}{}^E\Omega_E
    \Big)\ep_{A}{}^B 
\\
&\equiv& F_{CD}\ep_A{}^B.
\end{eqnarray*}
The torsion and the curvature satisfy the Bianchi identities 
\begin{subequations}
\begin{eqnarray}
{\cal D}T^A 
&=& -E^BR_B{}^A, 
\label{bianchi_1} \\
{\cal D}R_A{}^B 
&=&0.
\label{bianchi_2}
\end{eqnarray}
\end{subequations}

\vspace*{-8mm}
\noindent
Because the tangent group SO(1,1) is Abelian,   
the identity (\ref{bianchi_2}) becomes trivial relation. 
On the other hand, the identity (\ref{bianchi_1}) becomes 
the following nontrivial one in the local Lorentz flame, 
\begin{eqnarray*}
&&
{\cal D}_AT_{BC}{}^D 
+(-)^{|A|(|B|+|C|)}{\cal D}_BT_{CA}{}^D 
+(-)^{|C|(|A|+|B|)}{\cal D}_CT_{AB}{}^D 
\\
&&
+T_{AB}{}^ET_{EC}{}^D 
+(-)^{|A|(|B|+|C|)}T_{BC}{}^ET_{EA}{}^D 
+(-)^{|C|(|A|+|B|)}T_{CA}{}^ET_{EB}{}^D 
\\
&=& 
-R_{ABC}{}^D
-(-)^{|A|(|B|+|C|)}R_{BCA}{}^D
-(-)^{|C|(|A|+|B|)}R_{CAB}{}^D. 
\end{eqnarray*}

The dynamical variables of two-dimensional supergravity are 
the vielbein $E_M{}^A(z)$ and the connection $\Omega_M(z)$. 
The degrees of freedom of these superfields are 
$80=(4\times 4\times 4)+ (4\times 4)$, 
because two bosonic fields and one Majorana spinor field are 
contained in one single superfield. 
In order to clarify true physical degrees of freedom,   
it may be useful to impose the following kinematic constraints 
on some of the torsion components $T_{AB}{}^C(z)$,   
\begin{eqnarray}
T_{\beta\gamma}{}^a=T_{\gamma\beta}{}^a
&=& 2i(\sigma^a)_{\beta\gamma}, 
\nonumber \\ 
T_{bc}{}^a=-T_{cb}{}^a
&=& 0, 
\label{kinematic_constraint} \\
T_{\beta\gamma}{}^\alpha=T_{\gamma\beta}{}^\alpha
&=& 0. 
\nonumber 
\end{eqnarray} 
These constraints reduce the degrees of freedom to 
$24=80-\Big((2\times 3\times 4)+(2\times 1\times 4)+(2\times 3\times 4)
\Big)$. 
Other torsion and curvature components are determined from 
the above constraints by introducing one single scalar superfield 
$\hat{S}(z)$~\cite{howe}:    
\begin{eqnarray}
T_{\beta c}{}^a=-T_{c\beta}{}^a
&=&0, 
\nonumber \\
T_{b\gamma}{}^\alpha=-T_{\gamma b}{}^\alpha 
&=&\frac{1}{4}(\sigma_b)_\gamma{}^\alpha\hat{S}, 
\\ 
T_{bc}{}^\alpha=-T_{cb}{}^\alpha
&=& -\frac{i}{4}\ep_{bc}(\bar{\sigma})^{\alpha\beta}{\cal D}_\beta\hat{S}, 
\nonumber 
\end{eqnarray}
and 
\begin{eqnarray}
F_{\alpha\beta}=F_{\beta\alpha}
&=& i(\bar{\sigma})_{\alpha\beta}\hat{S}, 
\nonumber \\
F_{\alpha b}=-F_{b\alpha}
&=&\frac{1}{2}(\bar{\sigma}\sigma_b)_\alpha{}^\beta{\cal D}_\beta\hat{S}, 
\\
F_{ab}=-F_{ba}
&=& \frac{i}{4}\ep_{ab}{\cal D}^\alpha{\cal D}_\alpha\hat{S} 
+\frac{1}{4}\ep_{ab}\hat{S}^2. 
\nonumber 
\end{eqnarray}

Under the super-general coordinate and the super local Lorentz 
transformations, the vielbeins $E_M{}^A(z)$ and the connections  
$\Omega_M(z)$ transform as 
\begin{subequations}
\begin{eqnarray}
\delta E_M{}^A
&=& K^N\del_NE_M{}^A+\del_MK^NE_N{}^A+E_M{}^BL\ep_B{}^A, 
\\
\delta \Omega_M
&=& K^N\del_N\Omega_M+\del_MK^N\Omega_N+\del_ML,  
\end{eqnarray}
\end{subequations}

\vspace*{-8mm}
\noindent
where $K^N(z)$ and $L(z)$ are local parameters 
for the super-general coordinate and super local Lorentz transformations, 
respectively. 
If one denotes these gauge parameters as  
\begin{eqnarray*}
K^n
&=& k_{(0)}^n+i\theta^\rho k_{(1)\rho}^n 
   +\frac{i}{2}(\theta\theta)k_{(2)}^n, 
\\
K^\nu 
&=& k_{(0)}^\nu+i\theta^\rho k_{(1)\rho}^\nu 
   +\frac{i}{2}(\theta\theta)k_{(2)}^\nu, 
\\
L
&=& l_{(0)}+i\theta^\rho l_{(1)\rho} 
   +\frac{i}{2}(\theta\theta)l_{(2)},  
\end{eqnarray*}
and use the degrees of freedom for the parameters 
$k_{(1)\rho}^n(x)$, $k_{(2)}^n(x)$, $k_{(1)\rho}^\nu(x)$  
$k_{(2)}^\nu(x)$, $l_{(1)\rho}(x)$ and $l_{(2)}(x)$,    
one can impose the following Wess-Zumino gauge,  
\begin{eqnarray}
E_\mu{}^a 
&=& i\theta^\rho E_{\rho\mu}{}^a + {\cal O}(\theta^2), 
\nonumber 
\\
E_\mu{}^\alpha 
&=& \delta_\mu{}^\alpha + i\theta^\rho E_{\rho\mu}{}^\alpha 
   +{\cal O}(\theta^2), 
\label{wess-zumino_gauge} \\
\Omega_\mu 
&=& i\theta^\rho\omega_{\rho\mu} + {\cal O}(\theta^2), 
\nonumber 
\end{eqnarray}
where $E_{\rho\mu}{}^a(x)=E_{\mu\rho}{}^a(x)$, 
$E_{\rho\mu}{}^\alpha(x)=E_{\mu\rho}{}^\alpha(x)$ and 
$\omega_{\rho\mu}(x)=\omega_{\mu\rho}(x)$. 
Since the degrees of freedom of the Wess-Zumino gauge are  
$(2 \times 2) + 2 + (2 \times 2) + 2 + 2 + 1 = 15$, 
the remaining degrees of freedom are 
$24 - 15 = 9$ and 
the remaining gauge degrees of freedom are 
$20 - 15 = 5$ which 
we should specify later.    
We identify these 9 degrees of freedom to the following field 
contents, 
\begin{eqnarray}
E_m{}^a|_{\theta=0} 
&\equiv& e_m{}^a, \hspace*{12mm} (\mbox{zweibein}), 
\nonumber \\
E_m{}^\alpha|_{\theta=0} 
&\equiv& \frac{1}{2}\chi_m{}^\alpha, \qquad 
(\mbox{Rarita-Schwinger field}), 
\\
\hat{S}|_{\theta=0}
&\equiv& A, \hspace*{16mm} (\mbox{auxiliary field}). 
\nonumber
\end{eqnarray}
By using these independent variables $e_m{}^a(x)$, $\chi_m{}^\alpha(x)$ 
and $A(x)$, we can then determine all of the field variables: 
%
\renewcommand{\arraystretch}{2.0}
%
\begin{itemize}
\item vielbeins: 
\begin{equation}
\begin{array}{rcl}
E_m{}^a 
&=& \displaystyle 
   e_m{}^a + i(\theta\sigma^a\chi_m) 
   +\frac{i}{4}(\theta\theta)e_m{}^aA, 
\\
E_m{}^\alpha 
&=& \displaystyle 
    \frac{1}{2}\chi_m{}^\alpha 
   -\frac{1}{4}(\theta\sigma_m)^\alpha A 
   -\frac{1}{2}(\theta\bar{\sigma})^\alpha\hat{\omega}_m 
   -\frac{3}{16}i(\theta\theta)\chi_m{}^\alpha A 
   +\frac{i}{8}(\theta\theta)(\check{\upsilon}\sigma_m)^\alpha, 
\\
E_\mu{}^a 
&=& \displaystyle 
   i\theta^\lambda(\sigma^a)_{\lambda\mu}, 
\\
E_\mu{}^\alpha
&=& \displaystyle 
   \delta_\mu{}^\alpha 
   -\frac{i}{8}(\theta\theta)\delta_\mu{}^\alpha A.  
\end{array}
\end{equation}
\item inverse vielbeins: 
\begin{equation}
\begin{array}{rcl}
E_a{}^m 
&=& \displaystyle 
    e_a{}^m 
   -\frac{i}{2}(\theta\sigma^m\chi_a)
   -\frac{1}{8}(\theta\theta)(\chi_a\sigma^n\sigma^m\chi_n), 
\\
E_\alpha{}^m 
&=& \displaystyle 
    -i\theta^\lambda(\sigma^m)_{\lambda\alpha} 
    -\frac{1}{4}(\theta\theta)(\sigma^n\sigma^m\chi_n)_\alpha, 
\\
E_a{}^\mu 
&=& \displaystyle
   -\frac{1}{2}\chi_a{}^\mu 
   +\frac{i}{4}(\theta\sigma^n\chi_a)\chi_n{}^\mu 
   +\frac{1}{4}(\theta\sigma_a)^\mu A
   +\frac{1}{2}(\theta\bar{\sigma})^\mu\hat{\omega}_a 
\\
&& \displaystyle
  +\frac{1}{16}(\theta\theta)(\chi_a\sigma^l\sigma^n\chi_l)\chi_n{}^\mu 
  -\frac{i}{8}(\theta\theta)(\chi_a\sigma^n\bar{\sigma})^\mu\hat{\omega}_n 
  -\frac{i}{8}(\theta\theta)(\check{\upsilon}\sigma_a)^\mu, 
\\
E_\alpha{}^\mu 
&=& \displaystyle 
   \delta_\alpha{}^\mu 
   +\frac{i}{2}\theta^\lambda(\sigma^n)_{\lambda\alpha}\chi_n{}^\mu 
\\
&& \displaystyle 
   +\frac{1}{8}(\theta\theta)(\sigma^l\sigma^n\chi_l)_\alpha\chi_n{}^\mu 
   -\frac{i}{4}(\theta\theta)(\sigma^n\bar{\sigma})_\alpha{}^\mu
    \hat{\omega}_n 
   -\frac{i}{8}(\theta\theta)\delta_\alpha{}^\mu A.  
\end{array}
\end{equation}
\item connections: 
\begin{equation}
\begin{array}{rcl}
\Omega_m 
&=& \displaystyle 
    \hat{\omega}_m 
   +\frac{i}{2}(\theta\bar{\sigma}\sigma_m\check{\upsilon}) 
   +\frac{i}{2}(\theta\bar{\sigma}\chi_m)A 
\\
&& \displaystyle
   -\frac{i}{4e}(\theta\theta)g_{mn}\ep^{nl}\del_lA 
   +\frac{i}{4}(\theta\theta)\hat{\omega}_mA 
   -\frac{1}{8}(\theta\theta)
    (\check{\upsilon}\bar{\sigma}\sigma_n\sigma_m\chi^n), 
\\
\Omega_\mu 
&=& \displaystyle
   \frac{i}{2}\theta^\lambda(\bar{\sigma})_{\lambda\mu}A. 
\end{array}
\end{equation}
\item auxiliary field: 
\begin{equation}
\hat{S} = A + i(\theta\check{\upsilon}) + \frac{i}{2}(\theta\theta)B. 
\end{equation}
\end{itemize}
In the above relations, we denote 
\begin{eqnarray*}
\hat{\omega}_m 
&=& \omega_m 
   -\frac{i}{2}(\chi_m\bar{\sigma}\sigma^n\chi_n), 
\\
\check{\upsilon}_\mu 
&=& -\frac{2}{e}
   \ep^{mn}(\bar{\sigma}\hat{\nabla}_m\chi_n)_\mu 
   -\frac{1}{2}(\sigma^m\chi_m)_\mu A,  
\\
B
&=& \frac{2}{e}\ep^{mn}\del_m\hat{\omega}_n 
   -\frac{i}{2}(\check{\upsilon}\sigma^m\chi_m) 
   -\frac{i}{4e}\ep^{mn}(\chi_m\bar{\sigma}\chi_n)A 
   +\frac{1}{2}A^2, 
\end{eqnarray*}
where 
$\omega_m(x)\Big(=\frac{1}{e}e_{am}\ep^{nl}\del_ne_l{}^a(x)\Big)$ is the  
usual torsion free spin connection defined via
(\ref{torsion_free_connection}). 
The covariant derivative on spinor fields $\hat{\nabla}_m$ is defined by 
(\ref{covariant_derivative_spinor}), 
except for using the connection $\hat{\omega}_m(x)$ instead of the 
torsion free connection $\omega_m(x)$. 
It is worth to mention that 
since $E_\mu{}^\alpha(z)$ is essentially Kronecker delta 
between $\mu$ and $\alpha$ indices, spinor indices might be written 
either $\mu$ or $\alpha$ in the Wess-Zumino gauge. 

In addition to the above super-general coordinate and 
super local Lorentz transformations,  
one can define the super-Weyl scaling transformation~\cite{howe},  
which is consistent with the kinematic constraints 
(\ref{kinematic_constraint}), 
\begin{equation}
\begin{array}{rcl}
\delta E_M{}^a 
&=& -SE_M{}^a, 
\\
\delta E_M{}^\alpha 
&=& \displaystyle 
    -\frac{1}{2}SE_M{}^\alpha 
    +\frac{i}{2}E_M{}^a(\sigma_a)^{\alpha\beta}{\cal D}_\beta S, 
\\
\delta\Omega_M 
&=& E_M{}^a\ep_a{}^b{\cal D}_bS 
   +E_M{}^\alpha(\bar{\sigma})_\alpha{}^\beta{\cal D}_\beta S, 
\\
\delta\hat{S} 
&=& S\hat{S} 
   -i{\cal D}^\alpha{\cal D}_\alpha S,  
\end{array}
\end{equation}
where $S(z)$ is a scalar superfield parameter for the super-Weyl 
scaling.  
If one expands the scalar superfield $S(z)$ by the following way,  
\begin{equation}
S=s+i(\theta\check{s})+\frac{i}{2}(\theta\theta)\bar{s}, 
\end{equation}
the auxiliary field $A(x)$ can be gauged away by using the gauge 
degree of freedom $\bar{s}(x)$. 
At this stage, we have $8(=9-1)$ degrees of freedom and 
$8(=5+3)$ gauge degrees of freedom.  

Now let us specify residual gauge symmetries. 
As we have explained,  
we still have 8 gauge degrees of freedom and then identify these 
to the following gauge parameters,    
\begin{eqnarray}
K^n|_{\theta=0} 
&\equiv& k^n, 
\qquad\qquad{\mbox{(general coordinate transformation)}}, 
\nonumber \\
K^\nu|_{\theta=0}
&\equiv& \zeta^\nu, 
\qquad\qquad(\mbox{local supersymmetry transformation}), 
\nonumber \\
L|_{\theta=0} 
&\equiv& l, 
\hspace*{19mm}(\mbox{local Lorentz transformation}), 
\\
S|_{\theta=0} 
&\equiv& s, 
\hspace*{19mm}(\mbox{Weyl scaling transformation}), 
\nonumber \\
-i\del_\mu S|_{\theta=0} 
&\equiv& \check{s}_\mu, 
\hspace*{17mm}(\mbox{super-Weyl scaling transformation}).  
\nonumber 
\end{eqnarray}
In order to preserve the Wess-Zumino gauge, 
the other components of the gauge parameters should be determined as 
\begin{eqnarray}
K^n 
&=& k^n
   +i(\theta\sigma^n\zeta)
   +\frac{1}{4}(\theta\theta)(\chi_l\sigma^n\sigma^l\zeta), 
\nonumber \\
K^\nu 
&=& \zeta^\nu 
   -\frac{i}{2}(\theta\sigma^l\zeta)\chi_l{}^\nu
   -\frac{1}{2}l(\theta\bar{\sigma})^\nu
   +\frac{1}{2}\theta^\nu s 
   +\frac{i}{4}(\theta\theta)(\bar{\sigma}\sigma^l\zeta)^\nu\hat{\omega}_l 
\nonumber \\
&& -\frac{1}{8}(\theta\theta)(\chi_k\sigma^l\sigma^k\zeta)\chi_l{}^\nu, 
\\
L 
&=& l 
   -i(\theta\sigma^l\zeta)\hat{\omega}_l
   -i(\theta\bar{\sigma}\check{s})
   -\frac{i}{4}(\theta\theta)(\check{\upsilon}\bar{\sigma}\zeta)
   -\frac{1}{4}(\theta\theta)(\chi_k\sigma^l\sigma^k\zeta)\hat{\omega}_l,  
\nonumber \\
S 
&=& s
   +i(\theta\check{s})
   +\frac{1}{2e}(\theta\theta)\ep^{mn}
    (\zeta\bar{\sigma}\hat{\nabla}_m\chi_n). 
\nonumber 
\end{eqnarray}
In particular, the residual symmetries of the supergravity multiplet 
$e_m{}^a(x)$ and $\chi_m{}^\alpha(x)$  
are given by 
\begin{subequations}
\begin{eqnarray}
\delta e_m{}^a 
&=& k^n\del_ne_m{}^a
   +\del_mk^ne_n{}^a
   +i(\zeta\sigma^a\chi_m) 
   +le_m{}^b\ep_b{}^a 
   -se_m{}^a, 
\\ 
\delta\chi_m{}^\alpha 
&=& \displaystyle 
   k^n\del_n\chi_m{}^\alpha
   +\del_mk^n\chi_n{}^\alpha
   +2(\hat{\nabla}_m\zeta)^\alpha 
  -\frac{1}{2}l(\bar{\sigma}\chi_m)^\alpha 
   -\frac{1}{2}s\chi_m{}^\alpha 
   -(\sigma_m\check{s})^\alpha. 
\label{super_trans_gravitino}
\end{eqnarray}
\end{subequations}

\vspace*{-6mm}

One may introduce a scalar superfield $\Phi(z)$ and a spinor
superfield $\tilde{\Psi}_\alpha(z)$ as    
\begin{subequations}
\begin{eqnarray}
\Phi 
&=& \phi+i(\theta\kappa)+\frac{i}{2}(\theta\theta)G, 
\label{scalar_superfield} 
\\
\tilde{\Psi}_\alpha
&=& i\hat{\psi}_\alpha+i\theta_\alpha X'+i(\bar{\sigma}\theta)_\alpha X
   +i(\sigma^m\theta)_\alpha\tilde{A}_m 
   +(\theta\theta)\psi_\alpha. 
\label{spinor_superfield}
\end{eqnarray}
\end{subequations}

\vspace*{-8mm}
\noindent
The super-general coordinate, super-Lorentz and super-Weyl 
transformations are given by  

\vspace*{-6mm}
\begin{subequations}
\begin{eqnarray}
\delta\Phi 
&=& K^N\del_N\Phi, 
\\
\delta\tilde{\Psi}_\alpha
&=& K^N\del_N\tilde{\Psi}_\alpha
   -\frac{1}{2}L(\bar{\sigma})_\alpha{}^\beta\tilde{\Psi}_\beta
   +\frac{1}{2}S\tilde{\Psi}_\alpha, 
\end{eqnarray}
\end{subequations}

\vspace*{-6mm}
\noindent
In the Wess-Zumino gauge, one can express these transformations  
for the component fields, 
\begin{subequations}
\begin{eqnarray}
\delta\phi 
&=& k^n\del_n\phi+i(\zeta\kappa), 
\nonumber \\
\delta\kappa_\alpha 
&=& k^n\del_n\kappa_\alpha 
   +(\sigma^m\zeta)_\alpha\Big(\del_m\phi-\frac{i}{2}(\chi_m\kappa)\Big)
   +\zeta_\alpha G
   -\frac{1}{2}l(\bar{\sigma}\kappa)_\alpha 
   +\frac{1}{2}s\kappa_\alpha, 
\label{super_trans_scalar} \\
\delta G
&=& k^n\del_nG 
   -\frac{i}{2}(\zeta\sigma^m\sigma^n\chi_m)
               \Big(\del_n\phi-\frac{i}{2}(\chi_n\kappa)\Big)
   +i(\zeta\sigma^m\hat{\nabla}_m\kappa) 
   -\frac{i}{2}(\zeta\sigma^m\chi_m)G
   +sG, 
\nonumber 
\end{eqnarray}
and 
\begin{eqnarray}
\delta\hat{\psi}_\alpha 
&=& k^n\del_n\hat{\psi}_\alpha 
   +\Big((X'+X\bar{\sigma}+\tilde{A}_m\sigma^m)\zeta\Big)_\alpha
   -\frac{1}{2}l(\bar{\sigma}\hat{\psi})_\alpha 
   +\frac{1}{2}s\hat{\psi}_\alpha, 
\nonumber \\
\delta X' 
&=& k^n\del_nX'
   +i(\zeta\psi) 
   +\frac{i}{2}(\zeta\sigma^m\hat{\nabla}_m\hat{\psi}) 
   -\frac{i}{4}\Big(\zeta\sigma^m(X'+X\bar{\sigma}+\tilde{A}_n\sigma^n
                                  )\chi_m
               \Big) 
   +sX', 
\nonumber \\
\delta X
&=& k^n\del_nX 
   +i(\zeta\bar{\sigma}\psi)
   -\frac{i}{2}(\zeta\bar{\sigma}\sigma^m\hat{\nabla}_m\hat{\psi})
   +\frac{i}{4}\Big(\zeta\bar{\sigma}\sigma^m
                    (X'+X\bar{\sigma}+\tilde{A}_n\sigma^n
                    )\chi_m
               \Big)
   +sX, 
\nonumber \\
\delta\tilde{A}_m 
&=& \del_mk^n\tilde{A}_n
   +k^n\del_n\tilde{A}_m
   +i(\zeta\sigma_m\psi)
   +\frac{i}{2}(\zeta\sigma^n\sigma_m\hat{\nabla}_n\hat{\psi})
   +i(\zeta\sigma^n\chi_m)\tilde{A}_n
\nonumber \\
&& -\frac{i}{4}\Big(\zeta\sigma^n\sigma_m
                    (X'+X\bar{\sigma}+\tilde{A}_l\sigma^l
                    )\chi_n
               \Big)
   -\frac{i}{2}(\check{s}\sigma_m\hat{\psi}),  
\label{super_trans_spinor} \\
\delta\psi_\alpha 
&=& k^n\del_n\psi_\alpha 
   +\frac{i}{4}(\zeta\sigma^m\sigma^n\chi_m)
    (\hat{\nabla}_n\hat{\psi})_\alpha 
   -\frac{i}{2}(\zeta\sigma^m\chi_m)\psi_\alpha
\nonumber \\
&& +\frac{1}{2}(\sigma^m\zeta)_\alpha\del_mX'
   +\frac{1}{2}(\bar{\sigma}\sigma^m\zeta)_\alpha\del_mX 
   +\frac{1}{2}(\sigma^m\sigma^n\zeta)_\alpha e_m{}^a\hat{\nabla}_n
     \tilde{A}_a 
\nonumber \\
&& -\frac{i}{8}(\zeta\sigma^m\sigma^n\chi_m)
               \Big((X'+X\bar{\sigma}+\tilde{A}_l\sigma^l)\chi_n
               \Big)_\alpha
\nonumber \\
&& -\frac{i}{4e}\ep^{mn}
    (\zeta\hat{\nabla}_m\chi_n)(\bar{\sigma}\hat{\psi})_\alpha 
   +\frac{i}{4e}\ep^{mn}
    (\zeta\bar{\sigma}\hat{\nabla}_m\chi_n)\hat{\psi}_\alpha
\nonumber \\
&& -\frac{1}{2}l(\bar{\sigma}\psi)_\alpha 
   +\frac{3}{2}s\psi_\alpha 
   +\frac{1}{2}\Big((X'+X\bar{\sigma})\check{s}\Big)_\alpha, 
\nonumber 
\end{eqnarray}
\end{subequations}
   
\vspace*{-6mm}
\noindent
where we denote the covariant derivative 
for the vector field 
$\tilde{A}_a(x)\equiv e_a{}^m\tilde{A}_m(x)$ as 
\begin{eqnarray*}
\hat{\nabla}_m\tilde{A}_a 
&=& \del_m\tilde{A}_a
   +\hat{\omega}_m\ep_a{}^b\tilde{A}_b. 
\end{eqnarray*}

The superdeterminant $E(z)$ is defined by 
\begin{eqnarray}
E&\equiv&
\mbox{sdet}E_M{}^A
=\det E_m{}^a\det E_\alpha{}^\mu
\nonumber \\
&=& e
   +\frac{i}{2}e(\theta\sigma^m\chi_m) 
   +\frac{i}{4}e(\theta\theta)A 
   +\frac{1}{8}(\theta\theta)\ep^{mn}(\chi_m\bar{\sigma}\chi_n). 
\end{eqnarray}

It might be useful to give the following relation for the checking of 
the gauge invariance of the action in the superfield formulation,   
\begin{equation}
\int\!\dd^2\theta E{\cal D}^\alpha\tilde{\Psi}_\alpha 
= \del_m\Big(eg^{mn}\tilde{A}_n
 -\frac{i}{2}\ep^{mn}(\hat{\psi}\bar{\sigma}\chi_n)\Big),
\end{equation}
for an arbitrary spinor superfield $\tilde{\Psi}_\alpha(z)$ 
defined by (\ref{spinor_superfield}).   
We also give the following relation, 
\begin{equation}
(\bar{\sigma})^{\alpha\beta}{\cal D}_\alpha{\cal D}_\beta\Phi
=0,  
\end{equation}
for an arbitrary scalar superfield $\Phi(z)$. 

\vspace*{10mm}

\noindent
{\large{\bf Appendix C. Generalized Poisson bracket}}

\renewcommand{\theequation}{C.\arabic{equation}}

\setcounter{equation}{0}
\setcounter{footnote}{0}

\vspace*{5mm}

A generalized Poisson bracket~\cite{ht} is defined by 
\begin{equation}
\{F, G\} \equiv
\bigg(\frac{\delta_{\rm L} F}{\delta\varphi^i}
      \frac{\delta_{\rm L} G}{\delta P_{\varphi^i}}
     -\frac{\delta_{\rm L} F}{\delta P_{\varphi^i}}
      \frac{\delta_{\rm L} G}{\delta\varphi^i}
\bigg) 
+(-)^{|F|}
\bigg(\frac{\delta_{\rm L} F}{\delta\theta^\alpha}
      \frac{\delta_{\rm L} G}{\delta P_{\theta^\alpha}}
     +\frac{\delta_{\rm L} F}{\delta P_{\theta^\alpha}}
      \frac{\delta_{\rm L} G}{\delta\theta^\alpha}
\bigg), 
\end{equation}
where canonical variables $\varphi^i$ and $P_{\varphi^i}$ are bosonic, 
and $\theta^\alpha$ and $P_{\theta^\alpha}$ are fermionic.  
In the above definition the contraction of the indices contains 
the integration of space or spacetime and $|F|$ is the Grassmann parity 
of $F$. 
This generalized Poisson bracket will be replaced 
by the graded commutation relation multiplied by $-i$ upon 
quantization, as usual, 
\begin{equation}
\{\quad, \quad\} 
\rightarrow
\frac{1}{i}[\quad, \quad\}.
\end{equation}
The explicit forms of the basic Poisson brackets are given by  
\begin{eqnarray*}
\{\varphi^i, P_{\varphi^j}\} 
&=& -\{P_{\varphi^j}, \varphi^i\} = \delta^i_j, \\
\{\theta^\alpha, P_{\theta^\beta}\} 
&=& \{P_{\theta^\beta}, \theta^\alpha\} = -\delta^\alpha_\beta. 
\end{eqnarray*}
The algebraic properties of the Poisson bracket are as follows: 
\begin{eqnarray*}
\{F, G\}
&=&-(-)^{|F||G|}\{G, F\}, 
\\
\{F, G_1G_2\}
&=& \{F, G_1\}G_2 + (-)^{|F||G_1|}G_1\{F, G_2\}, 
\\
\{F_1F_2, G\}
&=&
F_1\{F_2, G\}
+(-)^{|G||F_2|}\{F_1, G\}F_2.  
\end{eqnarray*} 

\vspace*{10mm}

\noindent
{\large{\bf Appendix D. Lorentz generators in the 
light-cone gauge formulation}}

\renewcommand{\theequation}{D.\arabic{equation}}

\setcounter{equation}{0}
\setcounter{footnote}{0}

\vspace*{5mm}

\noindent
We give bellow some of the explicit operator forms of the Lorentz generators 
in the light-cone gauge formulation: 

$\bullet$ (NS, NS) sector: 

\vspace*{-5mm}

\begin{subequations}
\begin{eqnarray}
M^{i-}
&=&
x^i
\bigg(
\frac{(p^{\hat{+}})^2}{\phi^{\hat{+}}(\phi^{\hat{+}}p^+-\phi^+p^{\hat{+}})}
\Big(-\phi^+\phi^-+\frac{1}{2}\phi^j\phi_j
\Big)
-\frac{p^{\hat{+}}}{\phi^{\hat{+}}p^+-\phi^+p^{\hat{+}}}
\Big(-\phi^-p^++\phi^jp_j
\Big) 
\nonumber 
\\
&&
\hspace*{6.5mm}
+\frac{2\pi\phi^{\hat{+}}}{\phi^{\hat{+}}p^+-\phi^+p^{\hat{+}}}
\Big(
\hspace*{-1mm}
:\hspace*{-1mm}
L^{\rm tr}_0
\hspace*{-1mm}:
+
:\hspace*{-1mm}
\tilde{L}^{\rm tr}_0
\hspace*{-1mm}:
-a_0-\tilde{a}_0 
\Big)
\bigg)
\nonumber 
\\
&&
-x^{-}p^i
\nonumber 
\\
&&
+i\frac{p^{\hat{+}}\phi_k}{\phi^{\hat{+}}p^+-\phi^+p^{\hat{+}}}
\bigg(
\sum_{m=1}^\infty\frac{1}{m}
\Big(
\alpha^i_{-m}\alpha^k_m
-\alpha^k_{-m}\alpha^i_m
\Big)
+\sum_{r=\frac{1}{2}}^\infty 
\Big(
b^i_{-r}b^k_r
-b^k_{-r}b^i_r
\Big)
\bigg)
\nonumber 
\\
&&
-i\frac{2\sqrt{\pi}\phi^{\hat{+}}}{\phi^{\hat{+}}p^+-\phi^+p^{\hat{+}}}
\bigg(
\sum_{m=1}^\infty
\frac{1}{m}
\Big(\alpha^i_{-m}L_m^{\rm tr}-L_{-m}^{\rm tr}\alpha^i_m
\Big)
+\sum_{r=\frac{1}{2}}^\infty
\Big(b^i_{-r}G_r^{\rm tr}-G_{-r}^{\rm tr}b^i_r
\Big)
\bigg)
\nonumber 
\\
&&
+i\frac{p^{\hat{+}}\phi_k}{\phi^{\hat{+}}p^+-\phi^+p^{\hat{+}}}
\bigg(
\sum_{m=1}^\infty\frac{1}{m}
\Big(
\tilde{\alpha}^i_{-m}\tilde{\alpha}^k_m
-\tilde{\alpha}^k_{-m}\tilde{\alpha}^i_m
\Big)
+\sum_{r=\frac{1}{2}}^\infty
\Big(
\tilde{b}^i_{-r}\tilde{b}^k_r
-\tilde{b}^k_{-r}\tilde{b}^i_r
\Big)
\bigg)
\nonumber 
\\
&&
-i\frac{2\sqrt{\pi}\phi^{\hat{+}}}{\phi^{\hat{+}}p^+-\phi^+p^{\hat{+}}}
\bigg(
\sum_{m=1}^\infty
\frac{1}{m}
\Big(
\tilde{\alpha}^i_{-m}\tilde{L}_m^{\rm tr}
-\tilde{L}_{-m}^{\rm tr}\tilde{\alpha}^i_m
\Big)
+\sum_{r=\frac{1}{2}}^\infty
\Big(
\tilde{b}^i_{-r}\tilde{G}_r^{\rm tr}
-\tilde{G}_{-r}^{\rm tr}\tilde{b}^i_r
\Big)
\bigg)
\nonumber 
\\
&&
+\phi^i{p_\phi}^-
-\phi^-{p_\phi}^i, 
\\
M^{i\hat{-}}
&=&
x^i
\bigg(
\frac{p^+p^{\hat{+}}}{\phi^{\hat{+}}(\phi^{\hat{+}}p^+-\phi^+p^{\hat{+}})}
\Big(\phi^+\phi^--\frac{1}{2}\phi^j\phi_j
\Big)
-\frac{p^+}{\phi^{\hat{+}}p^+-\phi^+p^{\hat{+}}}
\Big(\phi^-p^+-\phi^jp_j
\Big)
\nonumber 
\\
&&
\hspace*{6.5mm}
-\frac{2\pi\phi^+}{\phi^{\hat{+}}p^+-\phi^+p^{\hat{+}}}
\Big(
\hspace*{-1mm}
:\hspace*{-1mm}
L^{\rm tr}_0
\hspace*{-1mm}:
+
:\hspace*{-1mm}
\tilde{L}^{\rm tr}_0
\hspace*{-1mm}:
-a_0-\tilde{a}_0
\Big)
\bigg)
\nonumber 
\\
&&
-x^{\hat{-}}p^i
\nonumber 
\\
&&
-i\frac{p^+\phi_k}{\phi^{\hat{+}}p^+-\phi^+p^{\hat{+}}}
\bigg(
\sum_{m=1}^\infty\frac{1}{m}
\Big(
\alpha^i_{-m}\alpha^k_m
-\alpha^k_{-m}\alpha^i_m
\Big)
+\sum_{r=\frac{1}{2}}^\infty
\Big(
b^i_{-r}b^k_r
-b^k_{-r}b^i_r
\Big)
\bigg)
\nonumber 
\\
&&
+i\frac{2\sqrt{\pi}\phi^+}{\phi^{\hat{+}}p^+-\phi^+p^{\hat{+}}}
\bigg(
\sum_{m=1}^\infty
\frac{1}{m}
\Big(
\alpha^i_{-m}L_m^{\rm tr}-L_{-m}^{\rm tr}\alpha^i_m
\Big)
+\sum_{r=\frac{1}{2}}^\infty
\Big(b^i_{-r}G_r^{\rm tr}-G_{-r}^{\rm tr}b^i_r
\Big)
\bigg)
\nonumber 
\\
&&
-i\frac{p^+\phi_k}{\phi^{\hat{+}}p^+-\phi^+p^{\hat{+}}}
\bigg(
\sum_{m=1}^\infty\frac{1}{m}
\Big(
\tilde{\alpha}^i_{-m}\tilde{\alpha}^k_m
-\tilde{\alpha}^k_{-m}\tilde{\alpha}^i_m
\Big)
+\sum_{r=\frac{1}{2}}^\infty
\Big(
\tilde{b}^i_{-r}\tilde{b}^k_r
-\tilde{b}^k_{-r}\tilde{b}^i_r
\Big)
\bigg)
\nonumber 
\\
&&
+i\frac{2\sqrt{\pi}\phi^+}{\phi^{\hat{+}}p^+-\phi^+p^{\hat{+}}}
\bigg(
\sum_{m=1}^\infty
\frac{1}{m}
\Big(
\tilde{\alpha}^i_{-m}\tilde{L}_m^{\rm tr}
-\tilde{L}_{-m}^{\rm tr}\tilde{\alpha}^i_m
\Big)
+\sum_{r=\frac{1}{2}}^\infty
\Big(
\tilde{b}^i_{-r}\tilde{G}_r^{\rm tr}
-\tilde{G}_{-r}^{\rm tr}\tilde{b}^i_r
\Big)
\bigg)
\nonumber 
\\
&&
+\phi^i{p_\phi}^{\hat{-}}
+\frac{1}{\phi^{\hat{+}}}
\Big(
\phi^+\phi^--\frac{1}{2}\phi^j\phi_j
\Big)
{p_\phi}^i, 
\\
M^{-\hat{-}}
&=&
x^-
\bigg(
\frac{p^+p^{\hat{+}}}{\phi^{\hat{+}}(\phi^{\hat{+}}p^+-\phi^+p^{\hat{+}})}
\Big(
\phi^+\phi^--\frac{1}{2}\phi^j\phi_j
\Big)
-\frac{p^+}{\phi^{\hat{+}}p^+-\phi^+p^{\hat{+}}}
\Big(
\phi^-p^+-\phi^jp_j
\Big)
\nonumber 
\\
&&
\hspace*{8mm}
-\frac{2\pi\phi^+}{\phi^{\hat{+}}p^+-\phi^+p^{\hat{+}}}
\Big(
\hspace*{-1mm}
:\hspace*{-1mm}
L^{\rm tr}_0
\hspace*{-1mm}:
+
:\hspace*{-1mm}
\tilde{L}^{\rm tr}_0
\hspace*{-1mm}:
-a_0-\tilde{a}_0
\Big)
\bigg)
\nonumber 
\\
&&
+x^{\hat{-}}
\bigg(
\frac{(p^{\hat{+}})^2}{\phi^{\hat{+}}(\phi^{\hat{+}}p^+-\phi^+p^{\hat{+}})}
\Big(
\phi^+\phi^--\frac{1}{2}\phi^j\phi_j
\Big)
-\frac{p^{\hat{+}}}{\phi^{\hat{+}}p^+-\phi^+p^{\hat{+}}}
\Big(
\phi^-p^+-\phi^jp_j
\Big)
\nonumber 
\\
&&
\hspace*{11.5mm}
-\frac{2\pi\phi^{\hat{+}}}{\phi^{\hat{+}}p^+-\phi^+p^{\hat{+}}}
\Big(
\hspace*{-1mm}
:\hspace*{-1mm}
L^{\rm tr}_0
\hspace*{-1mm}:
+
:\hspace*{-1mm}
\tilde{L}^{\rm tr}_0
\hspace*{-1mm}:
-a_0-\tilde{a}_0
\Big)
\bigg)
\nonumber 
\\
&&
+i\frac{2\sqrt{\pi}\phi_j}{\phi^{\hat{+}}p^+-\phi^+p^{\hat{+}}}
\bigg(
\sum_{m=1}^\infty\frac{1}{m}
\Big(
\alpha^j_{-m}L^{\rm tr}_m
-L^{\rm tr}_{-m}\alpha_m^j
\Big)
+\sum_{r=\frac{1}{2}}^\infty
\Big(b_{-r}^jG_r^{\rm tr}
-G_{-r}^{\rm tr}b_r^j
\Big)
\bigg)
\nonumber 
\\
&&
+i\frac{2\sqrt{\pi}\phi_j}{\phi^{\hat{+}}p^+-\phi^+p^{\hat{+}}}
\bigg(
\sum_{m=1}^\infty\frac{1}{m}
\Big(
\tilde{\alpha}^j_{-m}\tilde{L}^{\rm tr}_m
-\tilde{L}^{\rm tr}_{-m}\tilde{\alpha}_m^j
\Big)
+\sum_{r=\frac{1}{2}}^\infty
\Big(
\tilde{b}_{-r}^j\tilde{G}_r^{\rm tr}
-\tilde{G}_{-r}^{\rm tr}\tilde{b}_r^j
\Big)
\bigg)
\nonumber 
\\
&&
+\phi^-{p_\phi}^{\hat{-}}
+\frac{1}{\phi^{\hat{+}}}
\Big(\phi^+\phi^--\frac{1}{2}\phi^j\phi_j
\Big){p_\phi}^-. 
\end{eqnarray}
\end{subequations}

\vspace*{-6mm}

$\bullet$ (R, R) sector: 
\begin{subequations}
\begin{eqnarray}
M^{i-}
&=&
x^i
\bigg(
\frac{(p^{\hat{+}})^2}{\phi^{\hat{+}}(\phi^{\hat{+}}p^+-\phi^+p^{\hat{+}})}
\Big(
-\phi^+\phi^-+\frac{1}{2}\phi^j\phi_j
\Big)
-\frac{p^{\hat{+}}}{\phi^{\hat{+}}p^+-\phi^+p^{\hat{+}}}
\Big(-\phi^-p^++\phi^jp_j
\Big)
\nonumber 
\\
&&
\hspace*{6.5mm}
+\frac{2\pi\phi^{\hat{+}}}{\phi^{\hat{+}}p^+-\phi^+p^{\hat{+}}}
\Big(
\hspace*{-1mm}
:\hspace*{-1mm}
L^{\rm tr}_0
\hspace*{-1mm}:
+
:\hspace*{-1mm}
\tilde{L}^{\rm tr}_0
\hspace*{-1mm}:
-a_0-\tilde{a}_0
\Big)
\bigg)
\nonumber 
\\
&&
-x^{-}p^i
\nonumber 
\\
&&
+i\frac{p^{\hat{+}}\phi_k}{\phi^{\hat{+}}p^+-\phi^+p^{\hat{+}}}
\bigg(
\sum_{m=1}^\infty\frac{1}{m}
\Big(
\alpha^i_{-m}\alpha^k_m
-\alpha^k_{-m}\alpha^i_m
\Big)
\nonumber 
\\
&&
\hspace*{29.5mm}
+\sum_{r=1}^\infty 
\Big(
b^i_{-r}b^k_r
-b^k_{-r}b^i_r
\Big)
+\frac{1}{2}
\Big(
b^i_0b^k_0
-b^k_0b^i_0
\Big)
\bigg)
\nonumber 
\\
&&
-i\frac{2\sqrt{\pi}\phi^{\hat{+}}}{\phi^{\hat{+}}p^+-\phi^+p^{\hat{+}}}
\bigg(
\sum_{m=1}^\infty
\frac{1}{m}
\Big(\alpha^i_{-m}L_m^{\rm tr}-L_{-m}^{\rm tr}\alpha^i_m
\Big)
\nonumber 
\\
&&
\hspace*{29.5mm}
+\sum_{r=1}^\infty
\Big(
b^i_{-r}G_r^{\rm tr}
-G_{-r}^{\rm tr}b^i_r
\Big)
+\frac{1}{2}
\Big(
b^i_0G_0^{\rm tr}-G_0^{\rm tr}b^i_0
\Big)
\bigg)
\nonumber 
\\
&&
+i\frac{p^{\hat{+}}\phi_k}{\phi^{\hat{+}}p^+-\phi^+p^{\hat{+}}}
\bigg(
\sum_{m=1}^\infty\frac{1}{m}
\Big(
\tilde{\alpha}^i_{-m}\tilde{\alpha}^k_m
-\tilde{\alpha}^k_{-m}\tilde{\alpha}^i_m
\Big)
\nonumber 
\\
&&
\hspace*{29.5mm}
+\sum_{r=1}^\infty
\Big(
\tilde{b}^i_{-r}\tilde{b}^k_r
-\tilde{b}^k_{-r}\tilde{b}^i_r
\Big)
+\frac{1}{2}
\Big(
\tilde{b}_0^i\tilde{b}_0^k
-\tilde{b}_0^k\tilde{b}_0^i
\Big)
\bigg)
\nonumber 
\\
&&
-i\frac{2\sqrt{\pi}\phi^{\hat{+}}}{\phi^{\hat{+}}p^+-\phi^+p^{\hat{+}}}
\bigg(
\sum_{m=1}^\infty
\frac{1}{m}
\Big(
\tilde{\alpha}^i_{-m}\tilde{L}_m^{\rm tr}
-\tilde{L}_{-m}^{\rm tr}\tilde{\alpha}^i_m
\Big)
\nonumber 
\\
&&
\hspace*{29.5mm}
+\sum_{r=1}^\infty
\Big(
\tilde{b}^i_{-r}\tilde{G}_r^{\rm tr}
-\tilde{G}_{-r}^{\rm tr}\tilde{b}^i_r
\Big)
+\frac{1}{2}
\Big(
\tilde{b}_0^i\tilde{G}_0^{\rm tr}
-\tilde{G}_0^{\rm tr}\tilde{b}_0^i
\Big)
\bigg)
\nonumber 
\\
&&
+\phi^i{p_\phi}^-
-\phi^-{p_\phi}^i, 
\\
M^{i\hat{-}}
&=&
x^i
\bigg(
\frac{p^+p^{\hat{+}}}{\phi^{\hat{+}}(\phi^{\hat{+}}p^+-\phi^+p^{\hat{+}})}
\Big(
\phi^+\phi^--\frac{1}{2}\phi^j\phi_j
\Big)
-\frac{p^+}{\phi^{\hat{+}}p^+-\phi^+p^{\hat{+}}}
\Big(
\phi^-p^+-\phi^jp_j
\Big)
\nonumber 
\\
&&
\hspace*{6.5mm}
-\frac{2\pi\phi^+}{\phi^{\hat{+}}p^+-\phi^+p^{\hat{+}}}
\Big(
\hspace*{-1mm}
:\hspace*{-1mm}
L^{\rm tr}_0
\hspace*{-1mm}:
+
:\hspace*{-1mm}
\tilde{L}^{\rm tr}_0
\hspace*{-1mm}:
-a_0-\tilde{a}_0
\Big)
\bigg)
\nonumber 
\\
&&
-x^{\hat{-}}p^i
\nonumber 
\\
&&
-i\frac{p^+\phi_k}{\phi^{\hat{+}}p^+-\phi^+p^{\hat{+}}}
\bigg(
\sum_{m=1}^\infty\frac{1}{m}
\Big(
\alpha^i_{-m}\alpha^k_m
-\alpha^k_{-m}\alpha^i_m
\Big)
\nonumber 
\\
&&
\hspace*{30mm}
+\sum_{r=1}^\infty
\Big(
b^i_{-r}b^k_r
-b^k_{-r}b^i_r
\Big)
+\frac{1}{2}
\Big(
b^i_0b^k_0
-b^k_0b^i_0
\Big)
\bigg)
\nonumber 
\\
&&
+i\frac{2\sqrt{\pi}\phi^+}{\phi^{\hat{+}}p^+-\phi^+p^{\hat{+}}}
\bigg(
\sum_{m=1}^\infty
\frac{1}{m}
\Big(
\alpha^i_{-m}L_m^{\rm tr}
-L_{-m}^{\rm tr}\alpha^i_m
\Big)
\nonumber 
\\
&&
\hspace*{30mm}
+\sum_{r=1}^\infty
\Big(b^i_{-r}G_r^{\rm tr}-G_{-r}^{\rm tr}b^i_r
\Big)
+\frac{1}{2}
\Big(
b^i_0G_0^{\rm tr}
-G_0^{\rm tr}b^i_0
\Big)
\bigg)
\nonumber 
\\
&&
-i\frac{p^+\phi_k}{\phi^{\hat{+}}p^+-\phi^+p^{\hat{+}}}
\bigg(
\sum_{m=1}^\infty\frac{1}{m}
\Big(
\tilde{\alpha}^i_{-m}\tilde{\alpha}^k_m
-\tilde{\alpha}^k_{-m}\tilde{\alpha}^i_m
\Big)
\nonumber 
\\
&&
\hspace*{30mm}
+\sum_{r=1}^\infty
\Big(
\tilde{b}^i_{-r}\tilde{b}^k_r
-\tilde{b}^k_{-r}\tilde{b}^i_r
\Big)
+\frac{1}{2}
\Big(
\tilde{b}_0^i\tilde{b}_0^k
-\tilde{b}_0^k\tilde{b}_0^i
\Big)
\bigg)
\nonumber 
\\
&&
+i\frac{2\sqrt{\pi}\phi^+}{\phi^{\hat{+}}p^+-\phi^+p^{\hat{+}}}
\bigg(
\sum_{m=1}^\infty
\frac{1}{m}
\Big(
\tilde{\alpha}^i_{-m}\tilde{L}_m^{\rm tr}
-\tilde{L}_{-m}^{\rm tr}\tilde{\alpha}^i_m
\Big)
\nonumber 
\\
&&
\hspace*{30mm}
+\sum_{r=1}^\infty
\Big(
\tilde{b}^i_{-r}\tilde{G}_r^{\rm tr}
-\tilde{G}_{-r}^{\rm tr}\tilde{b}^i_r
\Big)
+\frac{1}{2}
\Big(
\tilde{b}^i_0\tilde{G}_0^{\rm tr}
-\tilde{G}_0^{\rm tr}\tilde{b}_0^i
\Big)
\bigg)
\nonumber 
\\
&&
+\phi^i{p_\phi}^{\hat{-}}
+\frac{1}{\phi^{\hat{+}}}
\Big(
\phi^+\phi^--\frac{1}{2}\phi^j\phi_j
\Big)
{p_\phi}^i, 
\\
M^{-\hat{-}}
&=&
x^-
\bigg(
\frac{p^+p^{\hat{+}}}{\phi^{\hat{+}}(\phi^{\hat{+}}p^+-\phi^+p^{\hat{+}})}
\Big(
\phi^+\phi^--\frac{1}{2}\phi^j\phi_j
\Big)
-\frac{p^+}{\phi^{\hat{+}}p^+-\phi^+p^{\hat{+}}}
\Big(
\phi^-p^+-\phi^jp_j
\Big)
\nonumber 
\\
&&
\hspace*{8mm}
-\frac{2\pi\phi^+}{\phi^{\hat{+}}p^+-\phi^+p^{\hat{+}}}
\Big(
\hspace*{-1mm}
:\hspace*{-1mm}
L^{\rm tr}_0
\hspace*{-1mm}:
+
:\hspace*{-1mm}
\tilde{L}^{\rm tr}_0
\hspace*{-1mm}:
-a_0-\tilde{a}_0
\Big)
\bigg)
\nonumber 
\\
&&
+x^{\hat{-}}
\bigg(
\frac{(p^{\hat{+}})^2}{\phi^{\hat{+}}(\phi^{\hat{+}}p^+-\phi^+p^{\hat{+}})}
\Big(
\phi^+\phi^--\frac{1}{2}\phi^j\phi_j
\Big)
-\frac{p^{\hat{+}}}{\phi^{\hat{+}}p^+-\phi^+p^{\hat{+}}}
\Big(
\phi^-p^+-\phi^jp_j
\Big)
\nonumber 
\\
&&
\hspace*{11.5mm}
-\frac{2\pi\phi^{\hat{+}}}{\phi^{\hat{+}}p^+-\phi^+p^{\hat{+}}}
\Big(
\hspace*{-1mm}
:\hspace*{-1mm}
L^{\rm tr}_0
\hspace*{-1mm}:
+
:\hspace*{-1mm}
\tilde{L}^{\rm tr}_0
\hspace*{-1mm}:
-a_0-\tilde{a}_0
\Big)
\bigg)
\nonumber 
\\
&&
+i\frac{2\sqrt{\pi}\phi_j}{\phi^{\hat{+}}p^+-\phi^+p^{\hat{+}}}
\bigg(
\sum_{m=1}^\infty\frac{1}{m}
\Big(
\alpha^j_{-m}L^{\rm tr}_m
-L^{\rm tr}_{-m}\alpha_m^j
\Big)
\nonumber 
\\
&&
\hspace*{29.5mm}
+\sum_{r=1}^\infty
\Big(
b_{-r}^jG_r^{\rm tr}
-G_{-r}^{\rm tr}b_r^j
\Big)
+\frac{1}{2}
\Big(
b^j_0G_0^{\rm tr}
-G_0^{\rm tr}b^j_0
\Big)
\bigg)
\nonumber 
\\
&&
+i\frac{2\sqrt{\pi}\phi_j}{\phi^{\hat{+}}p^+-\phi^+p^{\hat{+}}}
\bigg(
\sum_{m=1}^\infty\frac{1}{m}
\Big(
\tilde{\alpha}^j_{-m}\tilde{L}^{\rm tr}_m
-\tilde{L}^{\rm tr}_{-m}\tilde{\alpha}_m^j
\Big)
\nonumber 
\\
&&
\hspace*{29.5mm}
+\sum_{r=1}^\infty
\Big(
\tilde{b}_{-r}^j\tilde{G}_r^{\rm tr}
-\tilde{G}_{-r}^{\rm tr}\tilde{b}_r^j
\Big)
+\frac{1}{2}
\Big(
\tilde{b}^j_0\tilde{G}_0^{\rm tr}
-\tilde{G}_0^{\rm tr}\tilde{b}^j_0
\Big)
\bigg)
\nonumber 
\\
&&
+\phi^-{p_\phi}^{\hat{-}}
+\frac{1}{\phi^{\hat{+}}}
\Big(\phi^+\phi^--\frac{1}{2}\phi^j\phi_j
\Big){p_\phi}^-. 
\end{eqnarray}
\end{subequations}

\vspace*{-6mm}

$\bullet$ (NS, R) sector: 
\begin{subequations}
\begin{eqnarray}
M^{i-}
&=&
x^i
\bigg(
\frac{(p^{\hat{+}})^2}{\phi^{\hat{+}}(\phi^{\hat{+}}p^+-\phi^+p^{\hat{+}})}
\Big(-\phi^+\phi^-+\frac{1}{2}\phi^j\phi_j
\Big)
-\frac{p^{\hat{+}}}{\phi^{\hat{+}}p^+-\phi^+p^{\hat{+}}}
\Big(-\phi^-p^++\phi^jp_j
\Big) 
\nonumber 
\\
&&
\hspace*{6.5mm}
+\frac{2\pi\phi^{\hat{+}}}{\phi^{\hat{+}}p^+-\phi^+p^{\hat{+}}}
\Big(
\hspace*{-1mm}
:\hspace*{-1mm}
L^{\rm tr}_0
\hspace*{-1mm}:
+
:\hspace*{-1mm}
\tilde{L}^{\rm tr}_0
\hspace*{-1mm}:
-a_0-\tilde{a}_0 
\Big)
\bigg)
\nonumber 
\\
&&
-x^{-}p^i
\nonumber 
\\
&&
+i\frac{p^{\hat{+}}\phi_k}{\phi^{\hat{+}}p^+-\phi^+p^{\hat{+}}}
\bigg(
\sum_{m=1}^\infty\frac{1}{m}
\Big(
\alpha^i_{-m}\alpha^k_m
-\alpha^k_{-m}\alpha^i_m
\Big)
+\sum_{r=\frac{1}{2}}^\infty 
\Big(
b^i_{-r}b^k_r
-b^k_{-r}b^i_r
\Big)
\bigg)
\nonumber 
\\
&&
-i\frac{2\sqrt{\pi}\phi^{\hat{+}}}{\phi^{\hat{+}}p^+-\phi^+p^{\hat{+}}}
\bigg(
\sum_{m=1}^\infty
\frac{1}{m}
\Big(\alpha^i_{-m}L_m^{\rm tr}-L_{-m}^{\rm tr}\alpha^i_m
\Big)
+\sum_{r=\frac{1}{2}}^\infty
\Big(b^i_{-r}G_r^{\rm tr}-G_{-r}^{\rm tr}b^i_r
\Big)
\bigg)
\nonumber 
\\
&&
+i\frac{p^{\hat{+}}\phi_k}{\phi^{\hat{+}}p^+-\phi^+p^{\hat{+}}}
\bigg(
\sum_{m=1}^\infty\frac{1}{m}
\Big(
\tilde{\alpha}^i_{-m}\tilde{\alpha}^k_m
-\tilde{\alpha}^k_{-m}\tilde{\alpha}^i_m
\Big)
\nonumber 
\\
&&
\hspace*{29.5mm}
+\sum_{r=1}^\infty
\Big(
\tilde{b}^i_{-r}\tilde{b}^k_r
-\tilde{b}^k_{-r}\tilde{b}^i_r
\Big)
+\frac{1}{2}
\Big(
\tilde{b}_0^i\tilde{b}_0^k
-\tilde{b}_0^k\tilde{b}_0^i
\Big)
\bigg)
\nonumber 
\\
&&
-i\frac{2\sqrt{\pi}\phi^{\hat{+}}}{\phi^{\hat{+}}p^+-\phi^+p^{\hat{+}}}
\bigg(
\sum_{m=1}^\infty
\frac{1}{m}
\Big(
\tilde{\alpha}^i_{-m}\tilde{L}_m^{\rm tr}
-\tilde{L}_{-m}^{\rm tr}\tilde{\alpha}^i_m
\Big)
\nonumber 
\\
&&
\hspace*{29.5mm}
+\sum_{r=1}^\infty
\Big(
\tilde{b}^i_{-r}\tilde{G}_r^{\rm tr}
-\tilde{G}_{-r}^{\rm tr}\tilde{b}^i_r
\Big)
+\frac{1}{2}
\Big(
\tilde{b}_0^i\tilde{G}_0^{\rm tr}
-\tilde{G}_0^{\rm tr}\tilde{b}_0^i
\Big)
\bigg)
\nonumber 
\\
&&
+\phi^i{p_\phi}^-
-\phi^-{p_\phi}^i, 
\\
M^{i\hat{-}}
&=&
x^i
\bigg(
\frac{p^+p^{\hat{+}}}{\phi^{\hat{+}}(\phi^{\hat{+}}p^+-\phi^+p^{\hat{+}})}
\Big(\phi^+\phi^--\frac{1}{2}\phi^j\phi_j
\Big)
-\frac{p^+}{\phi^{\hat{+}}p^+-\phi^+p^{\hat{+}}}
\Big(\phi^-p^+-\phi^jp_j
\Big)
\nonumber 
\\
&&
\hspace*{6.5mm}
-\frac{2\pi\phi^+}{\phi^{\hat{+}}p^+-\phi^+p^{\hat{+}}}
\Big(
\hspace*{-1mm}
:\hspace*{-1mm}
L^{\rm tr}_0
\hspace*{-1mm}:
+
:\hspace*{-1mm}
\tilde{L}^{\rm tr}_0
\hspace*{-1mm}:
-a_0-\tilde{a}_0
\Big)
\bigg)
\nonumber 
\\
&&
-x^{\hat{-}}p^i
\nonumber 
\\
&&
-i\frac{p^+\phi_k}{\phi^{\hat{+}}p^+-\phi^+p^{\hat{+}}}
\bigg(
\sum_{m=1}^\infty\frac{1}{m}
\Big(
\alpha^i_{-m}\alpha^k_m
-\alpha^k_{-m}\alpha^i_m
\Big)
+\sum_{r=\frac{1}{2}}^\infty
\Big(
b^i_{-r}b^k_r
-b^k_{-r}b^i_r
\Big)
\bigg)
\nonumber 
\\
&&
+i\frac{2\sqrt{\pi}\phi^+}{\phi^{\hat{+}}p^+-\phi^+p^{\hat{+}}}
\bigg(
\sum_{m=1}^\infty
\frac{1}{m}
\Big(
\alpha^i_{-m}L_m^{\rm tr}-L_{-m}^{\rm tr}\alpha^i_m
\Big)
+\sum_{r=\frac{1}{2}}^\infty
\Big(b^i_{-r}G_r^{\rm tr}-G_{-r}^{\rm tr}b^i_r
\Big)
\bigg)
\nonumber 
\\
&&
-i\frac{p^+\phi_k}{\phi^{\hat{+}}p^+-\phi^+p^{\hat{+}}}
\bigg(
\sum_{m=1}^\infty\frac{1}{m}
\Big(
\tilde{\alpha}^i_{-m}\tilde{\alpha}^k_m
-\tilde{\alpha}^k_{-m}\tilde{\alpha}^i_m
\Big)
\nonumber 
\\
&&
\hspace*{30mm}
+\sum_{r=1}^\infty
\Big(
\tilde{b}^i_{-r}\tilde{b}^k_r
-\tilde{b}^k_{-r}\tilde{b}^i_r
\Big)
+\frac{1}{2}
\Big(
\tilde{b}_0^i\tilde{b}_0^k
-\tilde{b}_0^k\tilde{b}_0^i
\Big)
\bigg)
\nonumber 
\\
&&
+i\frac{2\sqrt{\pi}\phi^+}{\phi^{\hat{+}}p^+-\phi^+p^{\hat{+}}}
\bigg(
\sum_{m=1}^\infty
\frac{1}{m}
\Big(
\tilde{\alpha}^i_{-m}\tilde{L}_m^{\rm tr}
-\tilde{L}_{-m}^{\rm tr}\tilde{\alpha}^i_m
\Big)
\nonumber 
\\
&&
\hspace*{30mm}
+\sum_{r=1}^\infty
\Big(
\tilde{b}^i_{-r}\tilde{G}_r^{\rm tr}
-\tilde{G}_{-r}^{\rm tr}\tilde{b}^i_r
\Big)
+\frac{1}{2}
\Big(
\tilde{b}^i_0\tilde{G}_0^{\rm tr}
-\tilde{G}_0^{\rm tr}\tilde{b}_0^i
\Big)
\bigg)
\nonumber 
\\
&&
+\phi^i{p_\phi}^{\hat{-}}
+\frac{1}{\phi^{\hat{+}}}
\Big(
\phi^+\phi^--\frac{1}{2}\phi^j\phi_j
\Big)
{p_\phi}^i, 
\\
M^{-\hat{-}}
&=&
x^-
\bigg(
\frac{p^+p^{\hat{+}}}{\phi^{\hat{+}}(\phi^{\hat{+}}p^+-\phi^+p^{\hat{+}})}
\Big(
\phi^+\phi^--\frac{1}{2}\phi^j\phi_j
\Big)
-\frac{p^+}{\phi^{\hat{+}}p^+-\phi^+p^{\hat{+}}}
\Big(
\phi^-p^+-\phi^jp_j
\Big)
\nonumber 
\\
&&
\hspace*{8mm}
-\frac{2\pi\phi^+}{\phi^{\hat{+}}p^+-\phi^+p^{\hat{+}}}
\Big(
\hspace*{-1mm}
:\hspace*{-1mm}
L^{\rm tr}_0
\hspace*{-1mm}:
+
:\hspace*{-1mm}
\tilde{L}^{\rm tr}_0
\hspace*{-1mm}:
-a_0-\tilde{a}_0
\Big)
\bigg)
\nonumber 
\\
&&
+x^{\hat{-}}
\bigg(
\frac{(p^{\hat{+}})^2}{\phi^{\hat{+}}(\phi^{\hat{+}}p^+-\phi^+p^{\hat{+}})}
\Big(
\phi^+\phi^--\frac{1}{2}\phi^j\phi_j
\Big)
-\frac{p^{\hat{+}}}{\phi^{\hat{+}}p^+-\phi^+p^{\hat{+}}}
\Big(
\phi^-p^+-\phi^jp_j
\Big)
\nonumber 
\\
&&
\hspace*{11.5mm}
-\frac{2\pi\phi^{\hat{+}}}{\phi^{\hat{+}}p^+-\phi^+p^{\hat{+}}}
\Big(
\hspace*{-1mm}
:\hspace*{-1mm}
L^{\rm tr}_0
\hspace*{-1mm}:
+
:\hspace*{-1mm}
\tilde{L}^{\rm tr}_0
\hspace*{-1mm}:
-a_0-\tilde{a}_0
\Big)
\bigg)
\nonumber 
\\
&&
+i\frac{2\sqrt{\pi}\phi_j}{\phi^{\hat{+}}p^+-\phi^+p^{\hat{+}}}
\bigg(
\sum_{m=1}^\infty\frac{1}{m}
\Big(
\alpha^j_{-m}L^{\rm tr}_m
-L^{\rm tr}_{-m}\alpha_m^j
\Big)
+\sum_{r=\frac{1}{2}}^\infty
\Big(b_{-r}^jG_r^{\rm tr}
-G_{-r}^{\rm tr}b_r^j
\Big)
\bigg)
\nonumber 
\\
&&
+i\frac{2\sqrt{\pi}\phi_j}{\phi^{\hat{+}}p^+-\phi^+p^{\hat{+}}}
\bigg(
\sum_{m=1}^\infty\frac{1}{m}
\Big(
\tilde{\alpha}^j_{-m}\tilde{L}^{\rm tr}_m
-\tilde{L}^{\rm tr}_{-m}\tilde{\alpha}_m^j
\Big)
\nonumber 
\\
&&
\hspace*{29.5mm}
+\sum_{r=1}^\infty
\Big(
\tilde{b}_{-r}^j\tilde{G}_r^{\rm tr}
-\tilde{G}_{-r}^{\rm tr}\tilde{b}_r^j
\Big)
+\frac{1}{2}
\Big(
\tilde{b}^j_0\tilde{G}_0^{\rm tr}
-\tilde{G}_0^{\rm tr}\tilde{b}^j_0
\Big)
\bigg)
\nonumber 
\\
&&
+\phi^-{p_\phi}^{\hat{-}}
+\frac{1}{\phi^{\hat{+}}}
\Big(\phi^+\phi^--\frac{1}{2}\phi^j\phi_j
\Big){p_\phi}^-. 
\end{eqnarray}
\end{subequations}



\begin{thebibliography}{99}
\bibitem{bd}
 M.P. Blencowe and M.J. Duff, 
 Nucl.\ Phys.\ {\bf B310} (1988) 387. 
\bibitem{ov}
 H. Ooguri and C. Vafa, 
 Nucl.\ Phys.\ {\bf B367} (1991) 83. \\
 D. Kutasov and E. Martinec, 
 Nucl.\ Phys.\ {\bf B477} (1996) 652. \\
 D. Kutasov, E. Martinec and M. O'Loughlin, 
 Nucl.\ Phys.\ {\bf B477} (1996) 675.   
\bibitem{v}
 C. Vafa, 
 Nucl.\ Phys.\ {\bf B469} (1996) 403. 
\bibitem{t}
 A.A. Tseytlin, 
 Nucl.\ Phys.\ {\bf B469} (1996) 51; 
 Phys.\ Rev.\ Lett.\ {\bf 78} (1997) 1864. 
\bibitem{b}
 I. Bars, 
 Phys.\ Rev.\ {\bf D54} (1996) 5203; 
 {\it Duality and hidden dimensions} 
 Proc.\ Conf.\ in {\it Frontiers in Quantum Field Theory} 
 (Toyonaka, Japan, 1995), ed. H. Itoyama {\it et al} 
 (World Scientific, 1996). \\
 For a review, 
 see I. Bars, Class.\ Quantum\ Grav.\ {\bf 18} (2001) 3113. 
\bibitem{bdk}
 I. Bars and C. Kounnas, 
 Phys.\ Rev.\ {\bf D56} (1997) 3664. \\
 I. Bars and C. Deliduman, 
 Phys.\ Rev.\ {\bf D56} (1997) 6579. 
\bibitem{br}
 I. Bars and S.-J. Rey, 
 Phys.\ Rev.\ {\bf D64} (2001) 046005. 
\bibitem{rz}
 J.M. Romero and A. Zamora, 
 Phys.\ Rev.\ {\bf D70} (2004) 105006. 
\bibitem{ns}
 H. Nishino and E. Sezgin, 
 Phys.\ Lett.\ {\bf B388} (1996) 569. \\
 H. Nishino, 
 Phys.\ Lett.\ {\bf B426} (1998) 64; 
 Phys.\ Lett.\ {\bf B428} (1998) 85. 
\bibitem{hpnn}
 S. Hewson and M. Perry, 
 Nucl.\ Phys.\ {\bf B492} (1997) 249. \\
 H. Nishino, Phys.\ Lett.\ {\bf B437} (1998) 303. \\
 J.A. Nieto, 
 hep-th/0410003.   
\bibitem{rssu}
 I. Rudychev, E. Sezgin and P. Sundell, 
 Nucl.\ Phys.\ Proc.\ Suppl.\ {\bf 68} (1998) 285. \\
 S.F. Hewson, 
 Nucl.\ Phys.\ {\bf B534} (1998) 513. \\
 R. Manvelyan, A. Melikyan and R. Mkrtchian, 
 Mod.\ Phys.\ Lett.\ {\bf A13} (1998) 2147. \\
 T. Ueno, JHEP {\bf 0012} (2000) 006. 
\bibitem{dg}
 A.A. Deriglazov, 
 Phys.\ Lett.\ {\bf B486} (2000) 218. 
\bibitem{k}
 D. Kamani, 
 Phys.\ Lett.\ {\bf B564} (2003) 123. 
\bibitem{wata1}
 Y. Watabiki, 
 Phys.\ Rev.\ Lett.\ {\bf 62} (1989) 2907; 
 Phys.\ Rev.\ {\bf D40} (1989) 1229; 
 Mod.\ Phys.\ Lett.\ {\bf A6} (1991) 1291.  
\bibitem{wata2} 
 Y. Watabiki, 
 JHEP {\bf 0305} (2003) 001. 
\bibitem{tw1}
 T. Tsukioka and Y. Watabiki, 
 Int.\ J.\ Mod.\ Phys. {\bf A19} (2004) 1923.  
\bibitem{howe}
 P.S. Howe, 
 J.\ Phys.\ {\bf A 12} (1979) 393. 
\bibitem{tw2} 
 T. Tsukioka and Y. Watabiki, in preparation. 
\bibitem{kawata}
 N. Kawamoto and Y. Watabiki, 
 Commun.\ Math.\ Phys.\ {\bf 144} (1992) 641; 
 Mod.\ Phys.\ Lett.\ {\bf A7} (1992) 1137.
\bibitem{kostu}
 N. Kawamoto, E. Ozawa and K. Suehiro,
 Mod.\ Phys.\ Lett.\ {\bf A12} (1997) 219. \\
 N. Kawamoto, K. Suehiro, T. Tsukioka and H. Umetsu, 
 Commun.\ Math.\ Phys.\ {\bf 195} (1998) 233; 
 Nucl.\ Phys.\ {\bf B532} (1998) 429. 
\bibitem{bv}
 I.A. Batalin and G.A. Vilkovisky,
 Phys.\ Lett.\ {\bf 102B} (1981) 27; 
 Phys.\ Rev.\ {\bf D28} (1983) 2567; Errata: {\bf D30} (1984) 508.
\bibitem{koh}
 M. Kato and K. Ogawa, 
 Nucl.\ Phys.\ {\bf B212} (1983) 443. \\
 S. Hwang, 
 Phys.\ Rev.\ {\bf D28} (1983) 2614.
\bibitem{oimnu}
N. Ohta, 
 Phys.\ Rev.\ {\bf D33} (1986) 1681. \\
M. It\=o, T. Morozumi, S. Nojiri and S. Uehara, 
 Prog.\ Theor.\ Phys.\ {\bf 75} (1986) 934.
\bibitem{ggrt}
 P. Goddard, J. Goldstone, C. Rebbi and C.B. Thorn, 
 Nucl. Phys. {\bf B56} (1973) 109. 
\bibitem{gswbhp}
 M.B. Green, J.H. Schwarz and E. Witten, 
 {\it Superstring Theory} 
 (Cambridge University Press, Cambridge, 1987). \\
 L. Brink and M. Henneaux, 
 {\it Principle of String Theory} 
 (Plenum Press, New York and London, 1988). \\
 J. Polchinski, 
 {\it String Theory} (Cambridge University Press, Cambridge, 1998). 
\bibitem{ht}
 M. Henneaux and C. Teitelboim,
 {\it Quantization of Gauge Systems}
 (Princeton University Press, Princeton, 1992). 
\bibitem{gps}
 J. Gomis, J. Paris and S. Samuel, 
 Phys.\ Rep.\ {\bf 259} (1995) 1, and references therein. 
\end{thebibliography}
\end{document}